\documentclass[preprint,12pt]{elsarticle}

\usepackage[margin=2.3cm]{geometry}
\usepackage{amsmath,amssymb}
\usepackage{url}
\usepackage{tikz}
\usepackage{subcaption}
\usepackage{placeins}
\usepackage{stmaryrd}                     %
\usepackage{xcolor}
\usetikzlibrary{shapes.geometric,calc,arrows.meta,decorations.pathreplacing}
 
\tikzset{
  spmark/.style ={regular polygon, regular polygon sides=3,
                  draw=red!85!black, fill=red!85!black,
                  minimum size=6pt, inner sep=0pt},          %
  fpmark/.style ={rectangle, draw=blue!70!black, fill=blue!70!black,
                  minimum size=5pt, inner sep=0pt},          %
  reddot/.style ={circle, fill=red!85!black,  minimum size=5pt, inner sep=0pt},
  bluedot/.style={circle, fill=blue!75!black, minimum size=5pt, inner sep=0pt},
  fstar/.style  ={star, star points=5, star point ratio=2.25,
                  draw=blue!70!black, fill=blue!70!black,
                  minimum size=9pt, inner sep=0pt},          %
  lbl/.style    ={font=\scriptsize, align=center},
}

\definecolor{mplblue}{HTML}{1F77B4}
\definecolor{mplorange}{HTML}{FF7F0E}
\definecolor{mplgreen}{HTML}{2CA02C}

\definecolor{fillgray}{RGB}{170,170,170}
\definecolor{linegray}{RGB}{110,110,110}
\colorlet{hblued}{blue}          %
\colorlet{hblue}{blue}           %
\colorlet{hgreend}{green!60!black}  %
\colorlet{hgreen}{green!60!black}   %

\colorlet{hredd}{red}          %
\colorlet{hred}{red}           %
 
\usepackage[mathlines]{lineno}
\usepackage{bm}
\usepackage{pgfplots}
\pgfplotsset{compat=1.18}
\usepackage{booktabs} 
 
\usetikzlibrary{arrows.meta, positioning}
\definecolor{kblue}{HTML}{2C5AA0}
\definecolor{kblueL}{HTML}{E8F0FB}
\definecolor{kteal}{HTML}{1F7A5C}
\definecolor{ktealL}{HTML}{EAF7F3}
\definecolor{keos}{HTML}{B5651D}
\definecolor{keosL}{HTML}{FFF4E6}
\definecolor{kgray}{HTML}{555555}
\definecolor{kgrayL}{HTML}{F0F0F0}
\definecolor{kchem}{HTML}{7B2D8E}
\definecolor{kchemL}{HTML}{F5EEFC}
\definecolor{kdiff}{HTML}{A13D63}
\definecolor{kdiffL}{HTML}{FDEEF3}

\definecolor{ccomp}{HTML}{2C5AA0}
\definecolor{ccompL}{HTML}{E8F0FB}
\definecolor{ccomm}{HTML}{B5651D}
\definecolor{ccommL}{HTML}{FFF1E0}
\definecolor{cns}{HTML}{1F7A5C}
\newcommand{\stepfont}{\fontsize{5.6}{6.4}\ttfamily}
\newcommand{\lblfont}{\fontsize{6.6}{7.4}\sffamily}

\journal{Computer Physics Communications}

\begin{document}

\begin{frontmatter}

\title{Performance-portable spectral difference simulation of multi-species reacting flows on CPUs and GPUs}

\author[cerfacs]{Romaric Simo Tamou}
\author[cerfacs]{Thomas Marchal\corref{cor1}}
\ead{tmarchal@cerfacs.fr}
\author[cerfacs]{Carlos Montilla}
\author[cerfacs]{Jean-Fran\c{c}ois Boussuge}
\cortext[cor1]{Corresponding author}
\affiliation[cerfacs]{organization={CERFACS},
            addressline={42 avenue Gaspard Coriolis},
            city={Toulouse},
            postcode={31057},
            country={France}}
            
\begin{abstract}
High-order discretizations and performance-portable programming models are two
routes to cheaper simulations of turbulent reacting flows, but few solvers
combine them for multi-species chemistry on both central and graphics processing
units (CPUs and GPUs). This work extends HOPPS, a spectral difference solver
written once in C++ with Kokkos, to compressible multi-species
reacting flows with tabulated thermochemistry and a nine-species hydrogen-air
mechanism. The physical models are selected at compile time. HOPPS
on CPU and on GPU is cross-verified against the legacy Fortran solver JAGUAR on
a non-reacting and a reacting three-dimensional Taylor--Green vortex. On AMD Genoa nodes, HOPPS is 1.7 to 2.8 times faster than
JAGUAR. In strong scaling on AMD Rome nodes, for $10^8$ degrees of freedom, the
parallel efficiency at 6400 cores is 1.02 in the non-reacting case and 0.97 in
the reacting case. On GPUs, the compute kernels follow the storage of the
fields by element and keep the solution gradients in shared memory. One Nvidia
H100 and one AMD MI300A then reach the same cost per degree of freedom, 3.8 to
4.5 times lower than that of a full 192-core AMD Genoa node. In weak scaling at
$5\times10^6$ degrees of freedom per GPU, the parallel efficiency is between
0.85 and 0.95 in both cases on 512 H100 GPUs, 1024 graphics compute dies of AMD
MI250X and 256 MI300A.
On both CPUs and GPUs, activating chemistry multiplies the cost per degree of
freedom by two to 2.5, with nearly the same or a higher parallel efficiency.
\end{abstract}

\begin{highlights}
\item HOPPS extended to multi-species reacting flows from one C++/Kokkos source
\item Parallel efficiency 0.97 on 6400 CPU cores and at least 0.85 on Nvidia and AMD GPUs
\item CPU and GPU runs cross-verified against JAGUAR on a reacting Taylor--Green vortex
\item 1.7 to 2.8 times faster than the legacy Fortran solver on AMD Genoa nodes
\item Single-species RCT on one GPU 2.7 to 6 times lower than in the first HOPPS article
\end{highlights}

\begin{keyword}
Spectral difference method \sep Performance portability \sep Kokkos \sep GPU
\sep Multi-species reacting flows \sep High-order methods
\end{keyword}

\end{frontmatter}

\section{Introduction}\label{sec:introduction}

Predictive simulation of turbulent reacting flows remains one of the most
demanding workloads in computational science. Resolving the scales that control
mixing and heat release requires discretizations with low dissipation and
dispersion errors, and such simulations routinely use a large amount of
computational resources. Chemistry further increases this cost, because the
solution vector carries one transported scalar per species, the thermodynamic
closure is no longer a closed-form expression, and the reaction rates introduce
much shorter time scales. Therefore, reducing the computational cost in such
cases requires discretizations and implementations that use the available
hardware efficiently, not only more computing power.

The evolution of the hardware makes this efficiency harder to reach. Peak arithmetic
throughput has grown faster than memory bandwidth for decades, so that
mesh-based solvers of partial differential equations, which are memory-intensive,
sit on the bandwidth-limited side of the roofline \cite{Roofline}. Machines have also become
heterogeneous: an increasing fraction of the performance of the top-ranked
systems comes from accelerators, mostly graphics processing units (GPUs), sold
by several vendors with incompatible native programming models
\cite{TOP500,Leiserson2020}. A code rewritten for each new device
entails a maintenance burden that grows with the diversity of the hardware, while
the number of developers available to carry it does not. Performance
portability, that is, competitive efficiency on several classes of device from a
single maintained source \cite{Deakin2019}, has therefore become a design constraint on the same
footing as accuracy and stability.

High-order discontinuous discretizations partly answer the first of these
pressures. Because they reach a target accuracy with fewer degrees of freedom
(DOF) than low-order schemes \cite{Wang2013,GassnerBeck2013}, they move less data
per unit of resolved physics, which raises their arithmetic intensity
\cite{Klockner2009,Vermeire2017}. Their element-local structure also suits
parallel hardware: neighboring elements interact only through a common interface
flux, which limits the halo to the element faces and keeps most of the arithmetic
inside independent elements. The discontinuous Galerkin (DG) method
\cite{ReedHill1973,CockburnShu2001,Cockburn1998dg,BassiRebay1997}, the flux
reconstruction (FR) approach \cite{Huynh2007fr,Vincent2011esfr,WangGao2009,WangHuynh2016}
and the spectral difference (SD) method
\cite{KoprivaKolias1996,Liu2006sd,MayJameson2006,Wang2007sd,Sun2007sd} share
these properties but differ in how the discrete operator is built. DG solves a
weak form: the equations are multiplied by test functions and integrated by parts
over each element, which gives volume and surface integrals evaluated by
quadrature, and a mass matrix to invert. This matrix is local to each element,
and becomes diagonal when the basis is collocated with the quadrature points, as
in the DG spectral element method (DGSEM) \cite{Kopriva2009book}. FR solves the
differential form: the solution and the flux are collocated at the same points,
and the flux, discontinuous at the interfaces, is corrected by functions of
degree $p+1$ that impose the common interface flux. Particular choices of these
correction functions recover nodal DG and SD schemes in the linear case
\cite{Huynh2007fr}. SD also solves the differential form, but on two staggered
point sets: the solution is stored at the solution points (SP), the flux is evaluated
at a distinct set of flux points (FP) that includes the element boundaries, and the
derivative of the interpolating flux polynomial of degree $p+1$ gives the
divergence. Neither quadrature, nor mass matrix, nor correction function is
then required. Its kernels are small, dense and regular
tensor-product sweeps over each element, which map naturally onto both the vector
units of processors and the thread hierarchy of GPUs.

Several mature high-order frameworks exist, and they can be classified by their
strategy towards hardware diversity. A first group targets clusters of central
processing units (CPUs) through the Message Passing Interface (MPI), possibly
combined with OpenMP, without a native GPU path. It includes the DG solver FLEXI
\cite{FLEXI}, the spectral/$hp$ element framework Nektar++ \cite{Nektar} and the
high-order finite volume solver UCNS3D \cite{UCNS3D}. A second group reaches the
GPU through vendor-specific kernels. STREAmS \cite{STREAmS} accelerates direct
numerical simulations of canonical wall-bounded flows with CUDA Fortran, GALÆXI
\cite{GALAEXI,GALAEXIarch} couples a Fortran DGSEM core to hand-written CUDA and HIP
(Heterogeneous-computing Interface for Portability) kernels to support Nvidia
and AMD devices, and an FR solver has been run with CUDA on
thousands of Nvidia GPUs \cite{JourdanWang2020}. Other solvers keep
a single source annotated with OpenACC directives, such as the GPU version of
the DGSEM solver HORSES3D \cite{HORSES3D,HORSES3DGPU} and the spectral element
solver SOD2D \cite{SOD2D,SOD2Dpar}, at the price of depending on the directive
support of each compiler and vendor. A third group places a portability layer
between the numerics and the device. PyFR \cite{PyFR} generates its FR
kernels for several back-ends from a single description, NekRS
\cite{NekRS} builds its spectral element kernels on the OCCA abstraction, Neko
\cite{Neko} organizes a modern Fortran solver around interchangeable hardware
back-ends, and Trixi.jl \cite{Trixi} pursues the same goal with the Julia
language. The finite element libraries MFEM \cite{MFEM} and deal.II
\cite{dealII} expose several accelerator programming models behind matrix-free
kernels, on which application solvers such as ExaDG \cite{ExaDG} are built. Each
of these choices trades differently between the peak efficiency on a given device
and the cost of following the hardware as it evolves.

None of the discontinuous solvers above addresses multi-species reacting flows
on GPU. Combustion solvers have been ported to GPUs, but none of them uses a
high-order discontinuous discretization. S3D \cite{S3D,S3DLegion} and KARFS
\cite{KARFS} use eighth-order central finite differences, the latter with
Kokkos \cite{KOKKOS,Kokkos3}, and nekCRF \cite{nekCRF}, built on NekRS,
continuous spectral elements for low-Mach reacting flows.
The DG formulations developed for detailed chemistry~\cite{LvIhme2014,Keeton2025} and the FR formulations developed for multi-species reacting flows~\cite{SimoFR2025} have both been demonstrated on CPUs.
To the authors'
knowledge, the combination of a high-order discontinuous discretization,
multi-species chemistry and GPU execution has hardly been explored, although the
trade-offs change in this regime.
Adding $N_s$ transported species widens every field and every halo message.
Evaluating a kinetic mechanism adds local arithmetic, dominated by
transcendental functions and free of communication. Replacing a constant heat
capacity by tabulated thermochemical data replaces a closed-form temperature
evaluation by a table search that depends on the local state. A search loop
whose trip count varies from point to point is cheap on a CPU, where branch
prediction and caches absorb it. On a GPU, it forces every thread of a warp to
wait for the slowest one, and this control divergence is largest at interfaces,
where neighboring threads handle different states. A design adequate for a
single-species perfect gas may therefore behave differently once chemistry is
active, and this has to be measured rather than assumed.

The solver studied here, HOPPS \cite{HOPPS}, addresses this gap. It is a C++20 SD
solver whose shared-memory parallelism is written once with the Kokkos library
\cite{KOKKOS} and compiled for OpenMP, CUDA or HIP back-ends, with MPI for the
distributed layer. It succeeds the CPU-only Fortran solver JAGUAR
\cite{JAGUAR,Cassagne2015}, which provides a reference point to assess the
rewrite. Its first release was restricted to flows without chemical species
\cite{HOPPS}. The present article extends it to multi-species compressible flows,
reacting or not, and quantifies the cost and the gain of this extension on current
hardware. The contributions are the following.

\begin{itemize}
  \item The compressible Navier--Stokes equations for a reacting mixture of $N_s$
        species, closed by tabulated thermochemistry \cite{JANAF} and a
        hydrogen-air mechanism \cite{Boivin}, are implemented in the SD
        framework. The equation set, the equation of state, the diffusive
        closure, the chemistry model, the Riemann solver and the time integrator
        are template parameters resolved at compile time, so that no runtime
        branch on the physics reaches the compute kernels.
  \item An \emph{auxiliary variable}, which carries the primitive state and a
        few derived thermodynamic quantities, is built at the FP from
        the interpolated primitive variables. The table inversion is then performed once per SP
        and removed from all the interface operations, including the MPI halo
        exchange. This design trades a slightly wider exchanged vector against
        the removal of a divergent search loop.
  \item The extended solver is cross-verified on a three-dimensional
        Taylor--Green vortex (TGV) carrying hydrogen, in a non-reacting
        three-species and a reacting nine-species configuration: HOPPS on CPU
        and HOPPS on GPU reproduce the solution of JAGUAR, whose formulation for
        reacting flows was validated previously \cite{Marchal2023}.
  \item On GPU, the compute kernels follow the storage of the fields by
        element, and the gradients of the solution stay in shared memory, which
        makes the same source efficient on Nvidia and AMD GPUs.
  \item A performance study covers four CPU generations from two vendors,
        four Nvidia GPU generations and three AMD GPUs. It reports kernel
        profiling, the cost per DOF and per Runge--Kutta (RK) stage as a
        function of the local problem size, the gain over JAGUAR, and weak and
        strong scaling up to 6400 CPU cores, 512 Nvidia H100 GPUs, 1024
        graphics compute dies (GCD) of AMD MI250X and 256 AMD MI300A
        accelerated processing units (APUs). Reacting and
        non-reacting configurations are reported side by side, to separate the
        cost of chemistry from the cost of the discretization.
\end{itemize}

Section~\ref{sec:NSequationNumMethods} gives the governing equations, the
closures and the SD discretization. Section~\ref{sec:hopps-arch-impl} describes
the code architecture, the Kokkos portability layer and the auxiliary variable.
Section~\ref{sec:mpi} details the distributed-memory strategy and the overlap of
communication with computation on CPU and GPU. Section~\ref{sec:validationTGV}
presents the cross-verification on the TGV, Section~\ref{sec:perfoAnalysis} the
performance analysis, and Section~\ref{sec:conclusion} concludes.

\section{Governing equations and numerical methods}\label{sec:NSequationNumMethods}
\subsection{Navier--Stokes equations for a reacting gas mixture}\label{sec:GoverningEquations}
The compressible Navier--Stokes equations for a mixture of $N_s$ species read,
in conservative form \cite{Poinsot},
\begin{equation}
\frac{\partial \mathbf{U}}{\partial t}
+ \frac{\partial \mathbf{E}}{\partial x}
+ \frac{\partial \mathbf{F}}{\partial y}
+ \frac{\partial \mathbf{G}}{\partial z}
= \mathbf{S},
\label{eq:conservative-form}
\end{equation}
where $\mathbf{U} = (\rho, \rho u, \rho v, \rho w, \rho E, \rho Y_1, \dots ,\rho Y_{N_s})^{\mathrm{T}}$ is the vector of conservative variables,
$\mathbf{E} = \mathbf{E}_c + \mathbf{E}_d$, $\mathbf{F} = \mathbf{F}_c + \mathbf{F}_d$
and $\mathbf{G} = \mathbf{G}_c + \mathbf{G}_d$ are the sums of the convective and
diffusive fluxes along $x$, $y$ and $z$, and $\mathbf{S}$ is the chemical source
vector. The convective fluxes read
\begin{equation}
\mathbf{E}_c = \begin{pmatrix} \rho u \\ \rho u^2 + P \\ \rho uv \\ \rho uw \\ u\left(P+\rho E\right) \\ \rho u Y_1 \\ \vdots \\ \rho u Y_{N_s} \end{pmatrix},\quad
\mathbf{F}_c = \begin{pmatrix} \rho v \\ \rho uv \\ \rho v^2 + P \\ \rho vw \\ v\left(P+\rho E\right) \\ \rho v Y_1 \\ \vdots \\ \rho v Y_{N_s} \end{pmatrix},\quad
\mathbf{G}_c = \begin{pmatrix} \rho w \\ \rho uw \\ \rho vw \\ \rho w^2 + P \\ w\left(P+\rho E\right) \\ \rho w Y_1 \\ \vdots \\ \rho w Y_{N_s} \end{pmatrix},
\label{eq:convective-fluxes}
\end{equation}
and the diffusive fluxes and the source vector
\begin{equation}
\begin{gathered}
\mathbf{E}_d = \begin{pmatrix} 0 \\ -\tau_{11} \\ -\tau_{12} \\ -\tau_{13} \\ q_x - u\tau_{11} - v\tau_{12} - w\tau_{13} \\ -M_{1,x} \\ \vdots \\ -M_{N_s,x} \end{pmatrix},\quad
\mathbf{F}_d = \begin{pmatrix} 0 \\ -\tau_{21} \\ -\tau_{22} \\ -\tau_{23} \\ q_y - u\tau_{21} - v\tau_{22} - w\tau_{23} \\ -M_{1,y} \\ \vdots \\ -M_{N_s,y} \end{pmatrix},\\[1ex]
\mathbf{G}_d = \begin{pmatrix} 0 \\ -\tau_{31} \\ -\tau_{32} \\ -\tau_{33} \\ q_z - u\tau_{31} - v\tau_{32} - w\tau_{33} \\ -M_{1,z} \\ \vdots \\ -M_{N_s,z} \end{pmatrix},\quad
\mathbf{S} = \begin{pmatrix} 0 \\ 0 \\ 0 \\ 0 \\ \dot{\omega}_T \\ \dot{\omega}_1 \\ \vdots \\ \dot{\omega}_{N_s} \end{pmatrix}.
\end{gathered}
\label{eq:diffusive-fluxes}
\end{equation}
In Eqs.~\eqref{eq:convective-fluxes} and \eqref{eq:diffusive-fluxes}, $\rho$ is
the density, ${\bf{u}}=(u,v,w)$ the velocity, $P$ the pressure,
$E$ the total energy without chemical contribution, that is, the sum of the
sensible internal energy and the kinetic energy, and $Y_k$ the mass fraction of
species $k$. The closures of the
viscous stress $\boldsymbol{\tau}$, the heat flux ${\bf{q}}=(q_x,q_y,q_z)$, the species
diffusion flux ${\bf{M}}_k=(M_{k,x},M_{k,y},M_{k,z})$, the source terms and the
equation of state are given below.

Assuming negligible bulk viscosity, the viscous stress tensor reads
\begin{equation}
    {\tau}_{ij}=\mu\left(\partial_{x_i}u_j+\partial_{x_j}u_i-\tfrac{2}{3}\,\delta_{ij}\nabla\cdot{\bf{u}}\right),
\end{equation}
where $\mu$ is the dynamic viscosity and $\delta_{ij}$ the Kronecker symbol. The
viscosity depends on temperature only, through the power law
$\mu=\mu_\text{ref}\left(T/T_\text{ref}\right)^{\beta}$, with
$\mu_\text{ref}=1.8405\times10^{-5}\,\mathrm{Pa\,s}$, $T_\text{ref}=300\,\mathrm{K}$
and $\beta=0.6759$.

The heat flux combines Fourier conduction and the sensible enthalpy transported by
species diffusion \cite{Poinsot},
\begin{equation}
    {\bf{q}} = -\lambda \nabla T +\rho \sum_{k=1}^{N_s}h_{k}Y_k{\bf{V}}_k,
\end{equation}
where $\lambda$ is the thermal conductivity, $h_k$ the sensible enthalpy of species $k$
and ${\bf{V}}_k$ its diffusion velocity, defined by
$\rho Y_k {\bf{V}}_k = -{\bf{M}}_k$. Binary diffusion coefficients are costly to
evaluate, and constant Schmidt numbers $\text{Sc}_k$ and a constant Prandtl
number $\text{Pr}$ are assumed. This gives $D_k=\mu/(\rho\,\text{Sc}_k)$ and
$\lambda=\mu c_p/\text{Pr}$, with $\text{Pr}=0.75$ and the Schmidt numbers of
Table~\ref{tab:schmidt}.
The species diffusion flux follows the Hirschfelder--Curtiss
approximation with a correction velocity \cite{Poinsot},
\begin{equation}
    {\bf{M}}_k= \rho\left(D_k\dfrac{W_k}{W}\nabla X_k-Y_k {\bf{V}}^c\right),
    \qquad
    {\bf{V}}^c=\sum_{k=1}^{N_s}D_k\dfrac{W_k}{W}\nabla X_k,
\end{equation}
where $W_k$ is the molar mass of species $k$, $W$ that of the mixture and $X_k$
the mole fraction. The correction velocity ensures
$\sum_k {\bf{M}}_k=0$, hence global mass conservation.
\begin{table}[htbp]
\centering
\small
\begin{tabular}{@{}lccccccccc@{}}
\toprule
Species & $\mathrm{H_2}$ & $\mathrm{O_2}$ & $\mathrm{N_2}$ & $\mathrm{H}$ & $\mathrm{OH}$ & $\mathrm{O}$ & $\mathrm{H_2O}$ & $\mathrm{HO_2}$ & $\mathrm{H_2O_2}$ \\
\midrule
$\text{Sc}_k$ & 0.247 & 0.953 & 1.370 & 0.167 & 0.621 & 0.610 & 0.806 & 0.944 & 0.950 \\
\bottomrule
\end{tabular}
\caption{Schmidt numbers of the nine species, used with the hydrogen-air mechanism of Boivin et al.~\cite{Boivin}. The non-reacting case uses the values of $\mathrm{H_2}$, $\mathrm{O_2}$ and $\mathrm{N_2}$.}
\label{tab:schmidt}
\end{table}

The chemical source terms $\dot\omega_k$ and $\dot\omega_T$ of
$\mathbf{S}$ couple species and energy. For a
mechanism of $N_r$ reactions with stoichiometric coefficients $\nu'_{kj}$ and
$\nu''_{kj}$, the mass production rate of species $k$ and the heat release rate
in the energy equation are
\begin{equation}
    \dot\omega_k = W_k\sum_{j=1}^{N_r}(\nu''_{kj}-\nu'_{kj})\,\mathcal{Q}_j,
    \qquad
    \dot\omega_T = -\sum_{k=1}^{N_s}\Delta h_{f,k}^{\,0}\,\dot\omega_k,
    \label{eq:source-terms}
\end{equation}
where $\Delta h_{f,k}^{\,0}$ is the standard enthalpy of formation of species $k$
and $\mathcal{Q}_j$ is the rate of progress of reaction $j$,
\begin{equation}
    \mathcal{Q}_j = k_{f,j}\prod_{k=1}^{N_s}\left(\frac{\rho Y_k}{W_k}\right)^{\nu'_{kj}}
                  - k_{r,j}\prod_{k=1}^{N_s}\left(\frac{\rho Y_k}{W_k}\right)^{\nu''_{kj}}.
\end{equation}
The forward and reverse rate constants both follow Arrhenius laws,
\begin{equation}
    k_{f,j}=A_{f,j}\,T^{\beta_{f,j}}\exp\left(-\frac{E_{a,f,j}}{\overline{R}\,T}\right),
    \qquad
    k_{r,j}=A_{r,j}\,T^{\beta_{r,j}}\exp\left(-\frac{E_{a,r,j}}{\overline{R}\,T}\right),
\end{equation}
whose parameters are given by the mechanism. In the present work, it is the
skeletal hydrogen-air mechanism of Boivin et al.~\cite{Boivin}, with 9 species
and 12 reactions.

Thermochemical data are taken from the Joint Army Navy Air Force (JANAF) tables
\cite{JANAF}, which give, for each species $k$, the enthalpy and entropy at
temperatures spaced by $100\,\text{K}$. They are used to tabulate the sensible
internal energy $e_{sk}$ and the heat capacities $C_{pk}$ and $C_{vk}$ on each
interval $[T^n,T^{n+1}]$, with $T^n=100\,n$~K. Since the mixture energy depends
on the composition, the temperature cannot be read directly from the table. It
is recovered from the specific internal energy $e$ and the mass fractions $Y_k$
by searching the interval $n$ such that
\begin{equation}
    e(T^n)=\sum_{k=1}^{N_s}e_{sk}(T^n)Y_k \leq e \leq \sum_{k=1}^{N_s}e_{sk}(T^{n+1})Y_k=e(T^{n+1}),
    \label{energyInterval}
\end{equation}
after which $T$ follows by linear interpolation,
\begin{equation}
    T = 100\left(n+\dfrac{e - e(T^n)}{e(T^{n+1}) - e(T^n)}\right).\label{tempfromEnergy}
\end{equation}
The pressure follows from Dalton's law for ideal gases,
$P=\rho R T$, where $R=\overline{R}\sum_{k=1}^{N_s}Y_k/W_k$ is the gas constant
of the mixture and $\overline{R} = 8.314\,\mathrm{J\,mol^{-1}\,K^{-1}}$ the
universal gas constant.

System~\eqref{eq:conservative-form} is redundant, since the sum of the $N_s$
species equations gives the mass equation. In HOPPS, the $N_s$ species equations
are solved and the density is obtained as $\rho = \sum_{k=1}^{N_s} \rho Y_k$.

\subsection{Numerical methods}
\subsubsection{General principle}
\label{subsec:sd-principle}
 
The SD method is a high-order, element-based discretization of hyperbolic
conservation laws \cite{Liu2006sd,Wang2007sd,Sun2007sd}. Like the DG method, it
represents the solution by piecewise polynomials that may be discontinuous
across element interfaces. Unlike DG, it uses a purely differential (collocation)
formulation: the scheme operates on nodal values connected by Lagrange
interpolation, so that neither variational projection, nor quadrature, nor mass
matrix inversion is required. The method is only recalled here, and the reader
is referred to Sun et al.~\cite{Sun2007sd} for its derivation and to Dutka et
al.~\cite{HOPPS} for its implementation in HOPPS.

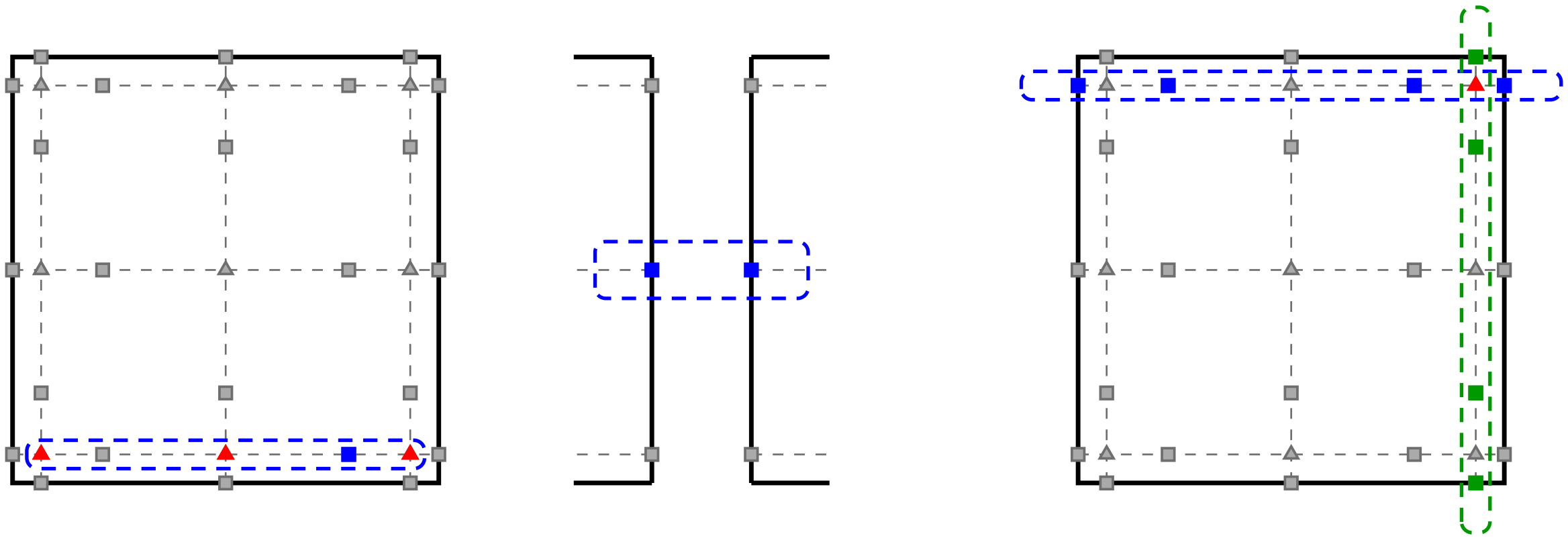
\begin{figure}[htbp]
  \centering
\begin{tikzpicture}[x=0.7cm,y=0.7cm,
  gsq/.style={rectangle, draw=linegray, fill=fillgray, minimum size=3.5pt, inner sep=0pt, line width=.7pt},
  gtr/.style={regular polygon, regular polygon sides=3, draw=linegray, fill=fillgray, minimum size=4.5pt, inner sep=0pt, line width=.7pt},
  ksq/.style={rectangle, draw=black, fill=black, minimum size=3.5pt, inner sep=0pt, line width=.7pt},
  ktr/.style={regular polygon, regular polygon sides=3, draw=hredd, fill=hred, minimum size=4.5pt, inner sep=0pt, line width=.7pt},
  bsq/.style={rectangle, draw=hblued, fill=hblue, minimum size=3.5pt, inner sep=0pt, line width=.7pt},
  osq/.style={rectangle, draw=hgreend, fill=hgreen, minimum size=3.5pt, inner sep=0pt, line width=.7pt},
  hgrid/.style={linegray, dashed, line width=.5pt},
  border/.style={black, line width=1.3pt},
  bell/.style={hblued, dashed, line width=1pt, dash pattern=on 4pt off 3pt},
  oell/.style={hgreend, dashed, line width=1pt, dash pattern=on 4pt off 3pt},
  bell/.style={hblued, dashed, line width=1pt, dash pattern=on 4pt off 3pt, rounded corners=3pt},
  oell/.style={hgreend, dashed, line width=1pt, dash pattern=on 4pt off 3pt, rounded corners=3pt},
]

\def\posarr{{0, 0.401924, 1.267949, 3, 4.732051, 5.598076, 6}}
\pgfmathsetmacro{\pA}{\posarr[0]}\pgfmathsetmacro{\pB}{\posarr[1]}
\pgfmathsetmacro{\pC}{\posarr[2]}\pgfmathsetmacro{\pD}{\posarr[3]}
\pgfmathsetmacro{\pE}{\posarr[4]}\pgfmathsetmacro{\pF}{\posarr[5]}
\pgfmathsetmacro{\pG}{\posarr[6]}

\def\drawmesh{%
  \foreach \k in {1,3,5}{%
    \pgfmathsetmacro{\pk}{\posarr[\k]}%
    \draw[hgrid] (\pk,0)--(\pk,6);
    \draw[hgrid] (0,\pk)--(6,\pk);
  }
  \draw[border] (0,0) rectangle (6,6);
  \foreach \i in {0,...,6}{%
    \pgfmathsetmacro{\px}{\posarr[\i]}%
    \foreach \j in {0,...,6}{%
      \pgfmathtruncatemacro{\jj}{6-\j}%
      \pgfmathsetmacro{\py}{\posarr[\jj]}%
      \pgfmathtruncatemacro{\io}{mod(\i,2)}%
      \pgfmathtruncatemacro{\jo}{mod(\j,2)}%
      \ifnum\io=1
        \ifnum\jo=1
          \node[gtr] at (\px,\py){}; %
        \else
          \node[gsq] at (\px,\py){}; %
        \fi
      \else
        \ifnum\jo=1
          \node[gsq] at (\px,\py){}; %
        \fi
      \fi
    }
  }
}

\begin{scope}
  \drawmesh
  \node[ktr] at (\pB,\pB){};
  \node[ktr] at (\pD,\pB){};
  \node[ktr] at (\pF,\pB){};
  \node[bsq] at (\pE,\pB){};
\draw[bell] (\pD-2.8,\pF-5.396152) rectangle (\pD+2.8,\pF-4.9961519999999995);
\end{scope}

\begin{scope}[shift={(9,0)}]
  \def\yl{\pF}\def\ym{\pD}\def\yb{\pB}
  \def\xa{0}
  \draw[border] (\xa,0)--(\xa,6);
  \draw[border] (\xa,6)--(\xa-1.1,6);
  \draw[border] (\xa,0)--(\xa-1.1,0);
  \foreach \yy in {\yl,\ym,\yb}{\draw[hgrid] (\xa,\yy)--(\xa-1.1,\yy);}
  \def\xb{1.4}
  \draw[border] (\xb,0)--(\xb,6);
  \draw[border] (\xb,6)--(\xb+1.1,6);
  \draw[border] (\xb,0)--(\xb+1.1,0);
  \foreach \yy in {\yl,\ym,\yb}{\draw[hgrid] (\xb,\yy)--(\xb+1.1,\yy);}
  \node[gsq] at (\xa,\yl){}; \node[gsq] at (\xa,\yb){}; \node[bsq] at (\xa,\ym){};
  \node[gsq] at (\xb,\yl){}; \node[gsq] at (\xb,\yb){}; \node[bsq] at (\xb,\ym){};
  \draw[bell] (0.7-1.5,\ym-0.4) rectangle (0.7+1.5,\ym+0.4);
\end{scope}

\begin{scope}[shift={(15,0)}]
  \drawmesh
  \foreach \x in {\pA,\pC,\pE,\pG}{\node[bsq] at (\x,\pF){};}   %
  \foreach \y in {\pA,\pC,\pE,\pG}{\node[osq] at (\pF,\y){};}   %
  \node[ktr] at (\pF,\pF){};                                    %
  \draw[bell] (\pD-3.8,\pB+4.9961519999999995) rectangle (\pD+3.8,\pB+5.396152);
  \draw[oell] (\pF-0.2,\pD-3.7) rectangle (\pF+0.2,\pD+3.7);
\end{scope}

\end{tikzpicture}
\caption{Stencils of the SD operations on a quadrilateral element with $p = 2$, in the reference frame $(\xi, \eta)$. SP are triangles and FP squares. Left: extrapolation to an FP (step 2). Middle: common interface flux (step 4). Right: flux derivative at an SP (step 6). Adapted from Dutka et al.~\cite{HOPPS}.}
\label{fig:sd-stencil-2d}
\end{figure}

The principle is given for the one-dimensional conservation law
\begin{equation}
  \frac{\partial {\mathbf U}}{\partial t}
  + \frac{\partial {\mathbf E}}{\partial x} = 0,
  \label{eq:conservation}
\end{equation}
where ${\mathbf U}$ contains the conservative variables and ${\mathbf E} = {\mathbf E}({\mathbf U})$ the associated flux.
Each element of the mesh is mapped onto the reference element $[0,1]$ by an
isoparametric map of Jacobian $J=\partial x/\partial \xi$, and the scheme is
written for the transformed variables $\hat{\mathbf U} = |J|{\mathbf U}$ and
$\hat{\mathbf E} = |J|J^{-1}{\mathbf E}$. For a solution of degree $p$, each
element carries two staggered sets of points: $p+1$ SP,
where the unknowns are stored, and $p+2$ FP, which include
the two element boundaries. In HOPPS, the SP are the Gauss--Chebyshev points and
the interior FP the Gauss--Legendre points, a robust choice for stability
\cite{Abeele2008sd,Jameson2010sd}. Each stage of the time integration performs
six operations:
\begin{enumerate}
  \item \emph{Solution polynomial.} A Lagrange polynomial of degree $p$ is
        built from the values $\hat{\mathbf U}_s$ at the SP,
        \begin{equation}
          \hat{\mathbf U}_h(\xi) = \sum_{s=1}^{p+1} \hat{\mathbf U}_s\, l_s^{SP}(\xi),
          \label{eq:sol-poly}
        \end{equation}
        where $\{l_s^{SP}\}$ are the Lagrange basis functions associated with
        the SP. It is generally discontinuous at element interfaces.
  \item \emph{Extrapolation.} $\hat{\mathbf U}_h$ is evaluated at the FP.
  \item \emph{Flux evaluation.} The flux $\hat{\mathbf E}$ is computed at each
        FP. At an interface, the two adjacent elements give two different values.
  \item \emph{Common interface flux.} These two values are replaced by a common
        flux given by an approximate Riemann solver,
        \begin{equation}
          \hat{\mathbf E}^{\,\mathrm{com}} = \mathcal{F}\!\left({\mathbf U}_L, {\mathbf U}_R\right),
          \label{eq:riemann}
        \end{equation}
        where ${\mathbf U}_L$ and ${\mathbf U}_R$ are the left and right states.
        Among the classical solvers (Rusanov \cite{LaxFriedrichs1,LaxFriedrichs2},
        Roe \cite{Roe}, Harten--Lax--van Leer--Contact (HLLC) \cite{HLLC}), all results of this article use HLLC.
  \item \emph{Flux polynomial.} A Lagrange polynomial of degree $p+1$,
        continuous across the interfaces, is built from the $p+2$ flux values,
        \begin{equation}
          \hat{\mathbf E}_h(\xi) = \sum_{f=1}^{p+2} \hat{\mathbf E}_f\, l_f^{FP}(\xi).
          \label{eq:flux-poly}
        \end{equation}
  \item \emph{Flux divergence.} $\hat{\mathbf E}_h$ is differentiated
        analytically at the SP.
\end{enumerate}
On hexahedral elements, the SP and FP are tensor products of the
one-dimensional sets, and each operation acts along lines of points in one
direction (Fig.~\ref{fig:sd-stencil-2d}). The scheme is of order $p+1$. Its
stencil is compact: an element is coupled only to its direct neighbors, through
the common interface fluxes of step~4 and of Section~\ref{subsec:sd-parabolic}.
For linear problems, the SD method is a particular case of the FR
framework \cite{Huynh2007fr,Jameson2010sd}.
\subsubsection{Semi-discrete form and time integration}
\label{subsec:sd-time}

 Collecting the previous operations, the SD discretization of
Eq.~\eqref{eq:conservation} reduces, at each SP, to the
semi-discrete ordinary differential equation
\begin{equation}
  \frac{\mathrm{d} \hat{\mathbf U}_s}{\mathrm{d} t}
  = - \left.\frac{\partial \hat{\mathbf E}_h}{\partial \xi}\right|_{\xi_s^{\mathrm{SP}}} ,
  \qquad s = 1,\dots,p+1 .
  \label{eq:semidiscrete}
\end{equation}
This system is advanced in time with explicit RK schemes. All results
of this article use the five-stage, fourth-order strong-stability-preserving
scheme of Spiteri and Ruuth \cite{ruuthspi,TVD}, denoted RK(5,4).

The explicit time step is $\Delta t = \min(\Delta t^{\mathrm{conv}}, \Delta t^{\mathrm{diff}}, \Delta t^{\mathrm{chem}})$
(see Table~\ref{tab:timesteps}). The convective and diffusive limits are controlled
by the Courant--Friedrichs--Lewy (CFL) and Fourier (Fo) numbers, and the chemical
limit is the characteristic time of the source terms. The first two use the characteristic size
$\Delta_{SP} = |J|_{SP}^{1/d}/(p+1)$ at each SP, with $d$ the space
dimension, the speed of sound $c$ and the kinematic viscosity $\nu=\mu/\rho$.
The values CFL~$=0.4$ and Fo~$=1$ are used in this article. The chemical time scale does not limit stability for the kinetic schemes considered here.
\begin{table}[htbp]
  \centering
  \caption{Time-step contributions governing explicit integration.}
  \label{tab:timesteps}
  \renewcommand{\arraystretch}{2.2}
  \begin{tabular}{@{}llc@{}}
    \toprule
    Contribution & Controlling number & Time step \\
    \midrule
    Convective & CFL &
      $\displaystyle \Delta t^{\mathrm{conv}}
        = \frac{\mathrm{CFL}\times \Delta_{SP}}{\|\mathbf{u}\|_{2} + c}$ \\
    Diffusive & Fo &
      $\displaystyle \Delta t^{\mathrm{diff}}
        = \frac{\mathrm{Fo}\times \Delta_{SP}^{2}}
               {\max\!\left(\nu,\tfrac{\nu}{\mathrm{Pr}},\tfrac{\nu}{\mathrm{Sc}_{k}}\right)}$ \\
    Chemical & none &
      $\displaystyle \Delta t^{\mathrm{chem}}
        = \min_{k}\!\left(\rho Y_{k}/|\dot{\omega}_{k}|\right)$ \\
    \bottomrule
  \end{tabular}
\end{table}

\subsubsection{Extension to diffusive fluxes}
\label{subsec:sd-parabolic}
The discretization of Section~\ref{subsec:sd-principle} applies to hyperbolic
equations, where the flux depends on $\mathbf{U}$ only. Diffusive fluxes also
depend on the solution gradient, as for the heat equation, where
$\mathbf{E} = -\lambda\,\partial\mathbf{U}/\partial x$. The diffusive flux must
then be defined at the interfaces, either with the centered scheme of Sun et
al.~\cite{Sun2007sd} or with the SDLIFT approach of Marchal et
al.~\cite{Marchal2023}. This work uses the centered scheme of Sun et
al.~\cite{Sun2007sd}, equivalent to the first scheme of Bassi and Rebay (BR1)
\cite{BassiRebay1997}. The solution gradient is obtained by two successive
differentiation steps, each enforcing continuity at the interfaces through the
arithmetic average of the two adjacent states. These gradients are then used to
evaluate the diffusive flux at each FP.

\section{Code architecture and implementation design}\label{sec:hopps-arch-impl}
\subsection{HOPPS architecture}\label{subsec:hopps-arch}
HOPPS \cite{HOPPS} is a C++ library implementing the SD method for the
compressible Euler and Navier--Stokes equations. Its first version was restricted
to flows without chemical species, and the present article extends it to
multi-species flows, reacting or not.

HOPPS succeeds the in-house Fortran solver JAGUAR \cite{JAGUAR}, designed for
CPUs only. Its rewrite was motivated by a common observation in high-performance
computing (HPC): the number of developers available to port and maintain a
computational fluid dynamics code remains limited, while the hardware becomes
more diverse and evolves faster. The design objective of HOPPS is therefore to
sustain good performance across hardware generations without duplicating or
rewriting the numerical logic.

\begin{figure}[htbp]
  \centering
\begin{tikzpicture}[
  font=\scriptsize,
  every node/.style={align=center},
  box/.style={draw=kblue, line width=0.9pt, fill=kblueL, rounded corners=2pt,
              text width=#1, minimum height=0.85cm, font=\fontsize{6.7}{7.5}\ttfamily},
  box/.default=2.0cm,
  leaf/.style={draw, line width=0.7pt, dashed, rounded corners=2pt,
               text width=#1, minimum height=0.7cm, font=\fontsize{6.2}{7}\ttfamily},
  leaf/.default=2.0cm,
  desc/.style={font=\fontsize{5.6}{6.4}\itshape\sffamily},
  arr/.style={-{Latex[length=2mm,width=1.4mm]}, line width=0.8pt},
  ]

  \node[box=6.0cm] (kernelsd) at (6.8,-2.2) {KernelSD<MeshType, ModelType,\\ RiemannSolver, Integrator>};

  \node[box=1.75cm] (mesh)     at (2.75,-4.) {MeshType};
  \node[box=1.75cm, draw=kteal, fill=ktealL, font=\fontsize{6.7}{7.5}\ttfamily\bfseries] (model) at (5.3,-4.) {ModelType};
  \node[box=1.75cm] (riemann)  at (7.8,-4.) {RiemannSolver};
  \node[box=1.75cm] (integ)    at (10.3,-4.) {TimeIntegrator};

  \draw[arr] (kernelsd.south) -- ++(0,-0.35) -| (mesh.north);
  \draw[arr] (kernelsd.south) -- ++(0,-0.35) -| (model.north);
  \draw[arr] (kernelsd.south) -- ++(0,-0.35) -| (riemann.north);
  \draw[arr] (kernelsd.south) -- ++(0,-0.35) -| (integ.north);

  \node[box=4.6cm, draw=kteal, fill=ktealL] (euler) at (2.7,-6.3) {Euler<dim, EOS,\\ InterpolateField, KineticChemistry>};
  \node[box=5.8cm, draw=kteal, fill=ktealL] (ns)    at (10.6,-6.3) {NavierStokes<dim, EOS, DiffusiveClosure,\\ InterpolateField, KineticChemistry>};

  \draw[arr, kteal] (ns.west) -- (euler.east);

  \draw[arr] (model.south) -- ++(0,-0.6) -| (ns.north);
  \draw[arr] (model.south) -- ++(0,-0.6) -| (euler.north);

  \node[box=2.5cm, draw=keos, fill=keosL] (eos) at (1.6,-8.1) {EOS};
  \node[box=2.7cm, draw=kgray, fill=kgrayL] (ifld) at (4.7,-8.1) {InterpolateField (IF)};
  \node[box=3.0cm, draw=kchem, fill=kchemL] (chem) at (7.9,-8.1) {KineticChemistry};
  \node[box=3.6cm, draw=kdiff, fill=kdiffL] (diff) at (11.9,-8.1) {DiffusiveClosure};

  \draw[arr, kteal] (euler.south) -- ++(0,-0.3) -| (eos.north);
  \draw[arr, kteal] (euler.south) -- ++(0,-0.3) -| (ifld.north);
  \draw[arr, kteal] (euler.south) -- ++(0,-0.3) -| (chem.north);

  \draw[arr, kblue] (ns.south) -- ++(0,-0.55) -| (eos.north);
  \draw[arr, kblue] (ns.south) -- ++(0,-0.55) -| (ifld.north);
  \draw[arr, kblue] (ns.south) -- ++(0,-0.55) -| (chem.north);
  \draw[arr, kblue] (ns.south) -- ++(0,-0.55) -| (diff.north);

  \node[leaf=2.5cm, draw=keos] (eos1) at (1.6,-9.4) {PerfectGasConstantCp};
  \node[leaf=2.5cm, draw=keos] (eos2) at (1.6,-10.25) {JanafTableCp};
  \draw[keos, dashed] (eos.south) -- (eos1.north);

  \node[leaf=2.3cm, draw=kgray] (if1) at (4.7,-9.35) {Conservative};
  \node[leaf=2.3cm, draw=kgray] (if2) at (4.7,-10.1) {PrimitiveT};
  \node[leaf=2.3cm, draw=kgray] (if3) at (4.7,-10.85) {PrimitiveP};
  \draw[kgray, dashed] (ifld.south) -- (if1.north);

  \node[leaf=3.4cm, draw=kchem] (c2) at (7.9,-9.35) {ClassicalChemistry (Arrhenius)};
  \node[leaf=3.4cm, draw=kchem] (c3) at (7.9,-10.1) {AnalyticChemistry (hard-coded)};
  \node[leaf=3.4cm, draw=kchem] (c4) at (7.9,-10.85) {NoSourceTerms (non-reacting)};
  \draw[kchem, dashed] (chem.south) -- (c2.north);

  \node[leaf=3.6cm, draw=kdiff] (nf) at (11.9,-9.35) {NewtonianFluidMultiSpecies<dim, Power, IF>};
  \draw[kdiff, dashed] (diff.south) -- (nf.north);

  \node[leaf=3.6cm, draw=kdiff] (nfms) at (11.9,-10.15) {NewtonianFluid<dim, ConstantViscosity, IF>};

  \draw[gray!60, dashed] (-0.3,-11.7) -- (14.9,-11.7);

  \node[box=1.0cm, minimum height=0.4cm] at (1.3,-12.05) {};
   \node[anchor=west, font=\fontsize{6.5}{7.5}\sffamily] at (2.1,-12.05) {composite type / template parameter};

\node[leaf=1.0cm, minimum height=0.4cm] at (7.0,-12.05) {};
\node[anchor=west, font=\fontsize{6.5}{7.5}\sffamily, text width=4.2cm] at (7.2,-12.05) {interchangeable concrete implementation (compile-time choice)};

  \draw[arr] (11.8,-12.05) -- (12.4,-12.05);
  \node[anchor=west, font=\fontsize{6.5}{7.5}\sffamily] at (12.5,-12.05) {composition};

\end{tikzpicture}
\caption{HOPPS architecture: compile-time assembly of the SD kernel and of the physical model. Each box is a template parameter, and each dashed box a concrete implementation selected at compile time.}
  \label{fig:hopps_architecture}
\end{figure}
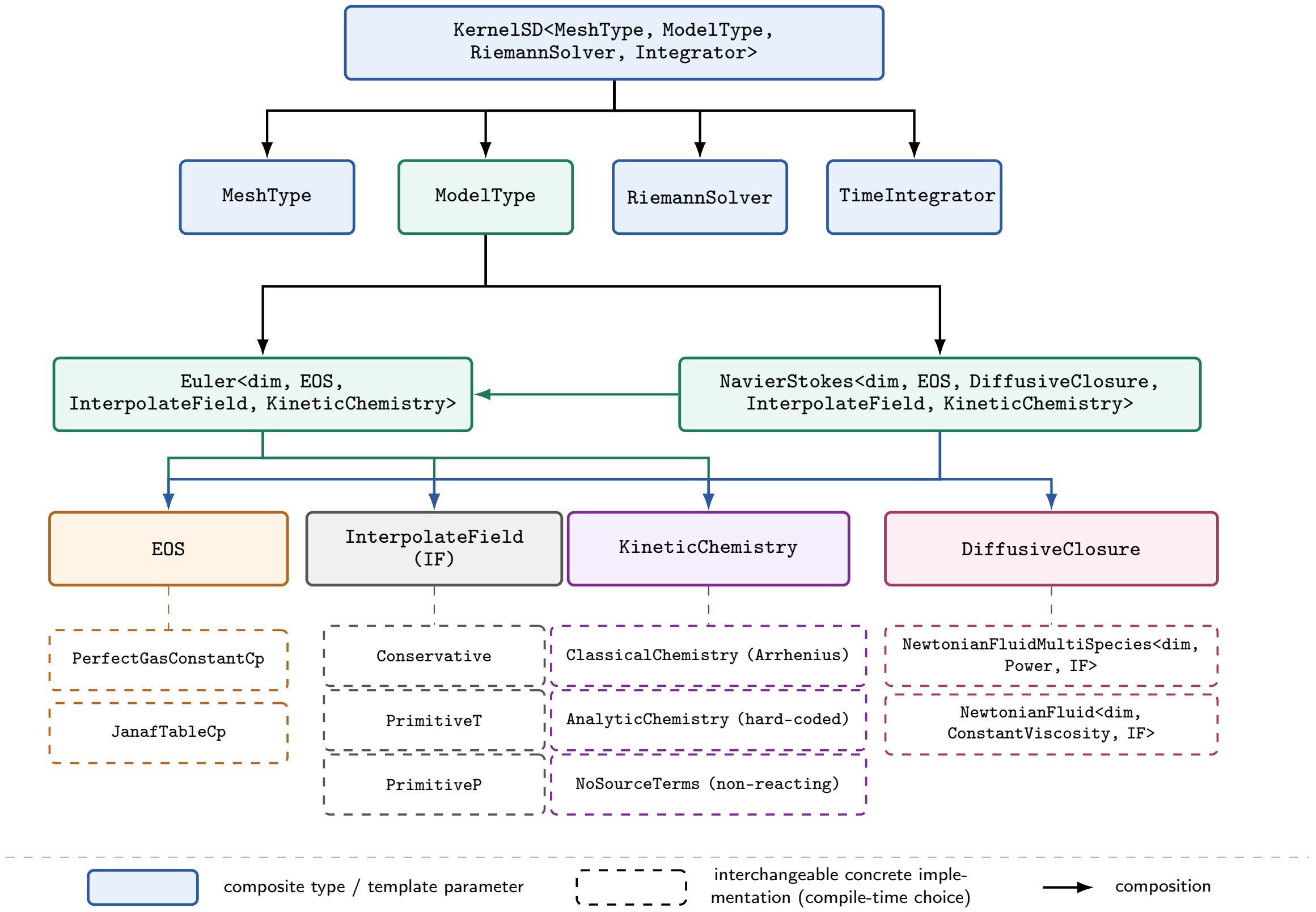

The numerical core is written in C++20. It uses the Kokkos library
\cite{KOKKOS,Kokkos3} to abstract shared-memory parallelism, combined with MPI for the
distributed domain decomposition. A Python layer built with pybind11
\cite{pybind11} drives the solver from scripts without recompilation. Each
combination of physical models (dimension, equation of state, diffusive closure,
chemical mechanism) is exposed as a distinct Python class, which favors ease of
use and type checking at the language boundary. Figure~\ref{fig:hopps_architecture}
shows this compile-time assembly. The SD kernel \texttt{KernelSD} is templated
on a mesh type, a physical model (Euler or Navier--Stokes), a Riemann solver and a
time integrator. The physical model is itself parameterized at compile time by an
equation of state (perfect gas with constant heat capacity, or JANAF tables), the
variables interpolated to the FP (see Section~\ref{sec:Auxvariable}), a
chemistry model (generic Arrhenius kinetics, or a mechanism hard-coded in
analytic form, as for analytically reduced chemistry, ARC
\cite{Pepiot2008,Felden2018}, and for the mechanism of this article) and, for
Navier--Stokes, a diffusive closure (single- or multi-species Newtonian fluid).

The core data structures are multidimensional \texttt{Kokkos::View} objects
that store all DOF of a mesh, indexed by element and by SP
or FP. Before any computation, the elements are sorted by geometric type
and polynomial degree, so that each homogeneous group is processed by regular
loops without conditional branching inside the compute kernels.

The target hardware covers x86 (Intel, AMD) and ARM processors, and Nvidia
(CUDA) and AMD (HIP) GPUs. Both GPU vendors were assessed in the first HOPPS
article \cite{HOPPS}.

\subsection{Implementation: programming model and portability layer (Kokkos)}
\label{subsec:hopps-kokkos}

The performance portability of HOPPS rests on Kokkos \cite{KOKKOS,Kokkos3}, a
C++ library that separates the numerical algorithm from the execution hardware.
A single annotated source is compiled for a multi-core processor through the
OpenMP back-end, for an Nvidia GPU through CUDA, or for an AMD GPU through HIP,
without any hardware-dependent branch in the numerical logic. The reasons for
this choice over directive-based and vendor-specific programming models, and its
implementation, are detailed in the first HOPPS article \cite{HOPPS}, and only the
features used in the rest of this article are recalled here. The compute
functions are annotated so that a single implementation is compiled for both
host and device. The physical fields are stored in multidimensional
\texttt{Kokkos::View} objects with the same row-major layout on all back-ends:
the values of all the points and variables of an element form one contiguous
block of memory, and the element index is the slowest. On CPU, this layout lets
each thread work on contiguous data and the compiler vectorize the inner loops.
On GPU, the kernels are organized around these blocks
(see Section~\ref{subsec:intragpu}). The buffers used for halo exchanges impose a
layout contiguous per point, because the linear packing of MPI messages relies
on it.
Finally, the sorting of the elements by type and polynomial degree
(see Section~\ref{subsec:hopps-arch}) is exploited by a compile-time loop that
generates, for each group, a specialized kernel whose inner loop size is known to
the compiler, which enables full loop unrolling and vectorization on CPU, and
suitable thread-block sizes on GPU, for any polynomial degree.

\subsection{Auxiliary variable}\label{sec:Auxvariable}
Rather than the conservative or primitive variables alone, HOPPS reconstructs at
the FP an \emph{auxiliary variable}, whose size depends on the
thermodynamic closure and is fixed at compile time. For a perfect gas with
constant heat capacity, it reduces to the usual conservative or primitive state.
For the tabulated closure required by multi-species mixtures with variable heat
capacities, here the JANAF tables \cite{JANAF}, it is enriched with the local gas
constant $R$ of the mixture, the heat capacity $C_p$ and the total energy $\rho E$. Its
$N_s+8$ components are therefore
$(T, u, v, w, P, Y_1,\dots,Y_{N_s}, R, C_p, \rho E)$ for the interpolation field \texttt{IF = PrimitiveT} (Fig.~\ref{fig:hopps_architecture}) used in this article, that is, 11 values for 3 species and 17 for 9 species, against
8 and 14 for the conservative state. The rest of this section explains why this
enriched variable is used for multi-species flows.

Computing the temperature from the internal energy, needed for the pressure, the speed of
sound and the viscosity, is immediate for a perfect gas with constant heat
capacity. For a tabulated closure, it requires the table search and interpolation
of Eqs.~\eqref{energyInterval} and \eqref{tempfromEnergy}, whose cost depends on
the local state. If only the conservative variables were reconstructed, this
inversion would be repeated wherever a derived thermodynamic quantity is needed:
on each side of every Riemann problem, during gradient reconstruction, and again
for the viscous flux. With the auxiliary variable, the inversion is performed
once per SP, when the conservative state is converted to the
primitive state $(T,u,v,w,P,Y_k)$. The primitive variables are then interpolated
to the FP, where the auxiliary variable is built from them: $R$ from
the interpolated mass fractions, $C_p$ from the interpolated temperature, and
$\rho=P/(RT)$, so that no table search is needed at the FP and the
equation of state holds by construction. Both steps form the kernel labeled
\texttt{ConsToAuxVar} in Sections~\ref{sec:mpi} and~\ref{sec:profiling}.
The result is available and used by all subsequent operations of the RK stage, including the Riemann problems on the faces shared between partitions, since the auxiliary variable itself is what crosses the first halo (\texttt{AuxVar\_halo}, see Section~\ref{sec:mpi}).
Interpolating the pressure rather than the total energy also avoids the
spurious pressure oscillations that conservative interpolation can produce at
material interfaces of multi-species flows \cite{Marchal2023}. JAGUAR follows
the same approach.

This choice has a cost: the size of the FP vector, and hence the volume
of data read, written and exchanged through MPI, grows with the number of added
quantities. It is a classical trade-off between computation and memory traffic,
resolved here in favor of reusing an expensive computation at the price of a
slightly wider vector. The rationale is that the avoided table search, with its
indirect memory accesses, costs more than moving a few additional contiguous
floating-point values.

The avoided operation is not only redundant, it is a data-dependent search loop
(a \texttt{while} loop over the table until the right interval is found). On a
CPU, branch prediction and cache reuse absorb much of its cost. On a GPU, threads execute in
lockstep warps, and a search loop of variable length makes the whole warp wait
for its slowest thread. This divergence is worse at interfaces, where neighboring
threads handle different states. Reading contiguous values already in memory
removes this divergence, which is the main motivation for the auxiliary
variable on GPU.

\section{MPI parallelization strategy}
\label{sec:mpi}

The distributed parallelism of HOPPS relies on a classical domain
decomposition: the mesh is split into subdomains of comparable size, each
assigned to an MPI rank, and the fields of each rank are stored as
\texttt{Kokkos::View} objects over its elements. The partitioning is delegated to
a tool based on t8code \cite{t8code,t8codeSW}, an open-source library for parallel
adaptive mesh refinement, which keeps this task outside the solver. Each rank
receives a contiguous range of the elements, in the order in which they are
stored in the mesh file: the shape of the subdomains, and thus the number of
faces exchanged, follows from this order (see Sections~\ref{sec:weakstrong-cpu}
and \ref{sec:weakstrong-gpu}).

Thanks to the compact stencil of the SD method
(see Section~\ref{subsec:sd-principle}), neighboring ranks exchange only
the values at the FP lying on partition interfaces. Two halos are
exchanged at each RK stage: the auxiliary variable for the convective
flux, and the averaged flux for the diffusive scheme. Both follow a non-blocking
protocol. The non-blocking communications (\texttt{MPI\_Irecv},
\texttt{MPI\_Isend}) are posted as early as possible, the interfaces internal to
the partition are processed while the exchange is in progress, and the
synchronization (\texttt{MPI\_Waitall}) is placed as late as possible, just
before the processing of the interfaces coupled through MPI. Communication is
thus overlapped with the processing of the internal interfaces which require no remote data.

\subsection{CPU parallelization}
\label{subsec:cpu}

On a CPU node, two levels of parallelism are combined: MPI ranks, and OpenMP
threads managed by Kokkos within each rank. Their distribution is not fixed in
the code but set by the compilation mode and the launch script, and the
configuration used in this article is given in Section~\ref{sec:perfoAnalysis}.

Within a rank, the element kernels traverse a two-index iteration space
(element, point) with \texttt{Kokkos::parallel\_for}, whereas the GPU organizes the same
operations differently (see Section~\ref{subsec:intragpu}). Under the OpenMP back-end, Kokkos tiles this iteration space and assigns contiguous blocks of tiles to the threads of a persistent thread pool. On the host, the default row-major layout stores the points and variables of an element contiguously, with the element index slowest, so that each thread works on a contiguous memory segment. Inside each iteration, the reconstruction loop, whose size $p+1$ is known at compile time (see Section~\ref{subsec:hopps-kokkos}), is auto-vectorized. The static dispatch over
homogeneous element groups therefore benefits both architectures, through vectorization on CPU and through thread occupancy on GPU.

The kernel that computes the gradients and the fluxes is the exception: CPU and
GPU run the same fused kernel (see Section~\ref{subsec:intragpu}). On CPU, a team
has a single thread, which processes one element, loops over its points with a
step known at compile time, and keeps the gradients of the element in a buffer
of the team, small enough to stay in the cache of the core. On the CPUs of
Section~\ref{sec:perfoAnalysis}, this fused kernel is in general as fast as a
gradient kernel followed by a flux kernel, or faster, with smaller gains than
on GPU: the organization chosen for the GPU does not penalize the CPU, and
often improves it.

\subsection{Inter-GPU parallelization}
\label{subsec:intergpu}

Between GPUs, on the same node or not, communication follows the same protocol:
one MPI rank per GPU, a non-blocking halo exchange limited to the interface FP, and an overlap with the processing of the internal interfaces. Two
aspects are specific to the GPU.

The first one is the location of the exchanged data. By default, HOPPS relies
on a GPU-aware MPI: the buffers passed to the non-blocking calls are device
pointers, and the MPI library handles the device-to-device transfer, with
hardware support such as GPUDirect RDMA where available, without going through
host memory. The simulation data thus remain in GPU memory during the whole
computation, as in other GPU compressible solvers \cite{STREAmS,GALAEXI}. A
second path, enabled at compile time, serves MPI stacks or hardware without
device-to-device transfers: the buffers are copied to host mirrors before sending,
and back to the device after reception.

\begin{figure}[htbp]
  \centering
\begin{tikzpicture}[
  seg/.style={draw=ccomp, line width=0.9pt, fill=ccompL, rounded corners=1.5pt,
              minimum height=1.05cm, align=center, font=\lblfont, anchor=west},
  commseg/.style={draw=ccomm, line width=0.9pt, fill=ccommL, rounded corners=1.5pt,
              minimum height=1.05cm, align=center, font=\lblfont, anchor=west},
  stepnote/.style={font=\stepfont, text=black!55, align=center},
  sync/.style={line width=1.1pt, draw=black},
  launch/.style={line width=0.9pt, draw=black!65, dash pattern=on 3pt off 2pt},
  ]

  \fill[black!4] (0,-0.55) rectangle (15.4,-1.65);
  \fill[black!4] (0,-2.55) rectangle (15.4,-3.65);

  \node[seg, minimum width=1.8cm, text width=1.3cm] (segA) at (0,-1.1) {\texttt{Element}\\\texttt{loop}};
  \node[stepnote, text width=2.2cm] at (0.75,-0.2) {\texttt{ConsToAuxVar}\\ \texttt{AuxVar\_halo}};

  \draw[launch] (1.5,-0.55) -- (1.5,-3.65);
  \node[stepnote, fill=white, inner sep=1.5pt]
        at (1.5,-3.9) {\texttt{MPI\_Irecv}\\\texttt{MPI\_Isend}};

  \node[seg, minimum width=2.8cm, text width=2.6cm] (segB) at (1.5,-1.1) {\texttt{Internal faces}\\\texttt{loop}};
  \node[commseg, minimum width=2.8cm, text width=2.5cm] (segBn) at (1.5,-3.1) {communication\\ in progress};
  \node[stepnote, text width=3.0cm] at (2.9,-0.35) {\texttt{RiemannPb\_intern}};

  \draw[sync] (4.3,-0.55) -- (4.3,-3.65);
  \node[stepnote, text width=2.0cm] at (4.3,-3.9) {\texttt{MPI\_Waitall}};

  \node[seg, minimum width=2.0cm, text width=1.8cm] (segC) at (4.3,-1.1) {\texttt{MPI faces}\\\texttt{loop}};
  \node[stepnote, text width=2.4cm] at (5.3,-0.35) {\texttt{RiemannPb\_mpi}};

  \node[seg, minimum width=1.5cm, text width=1.3cm] (segD) at (6.4,-1.1) {\texttt{Element}\\\texttt{loop}};
  \node[stepnote, text width=2.0cm] at (7.15,-0.1) {\texttt{GradViscIntConvFlx}\\ \texttt{flux\_halo}};

  \draw[launch] (8.0,-0.55) -- (8.0,-3.65);
  \node[stepnote, fill=white, inner sep=1.5pt]
        at (8.0,-3.9) {\texttt{MPI\_Irecv}\\\texttt{MPI\_Isend}};

  \node[seg, minimum width=2.6cm, text width=2.4cm] (segE) at (8.0,-1.1) {\texttt{Internal faces}\\\texttt{loop}};
  \node[commseg, minimum width=2.6cm, text width=2.4cm] (segEn) at (8.0,-3.1) {communication\\ in progress};
  \node[stepnote, text width=2.8cm] at (9.3,-0.35) {\texttt{AvgNormalFlx\_intern}};

  \draw[sync] (10.7,-0.55) -- (10.7,-3.65);
  \node[stepnote, text width=2.0cm] at (10.8,-3.9) {\texttt{MPI\_Waitall}};

  \node[seg, minimum width=2.0cm, text width=1.8cm] (segF) at (10.7,-1.1) {\texttt{MPI faces}\\\texttt{loop}};
  \node[stepnote, text width=2.4cm] at (11.7,-0.35) {\texttt{AvgNormalFlx\_mpi}};

  \node[seg, minimum width=2.cm, text width=2.4cm] (segG) at (12.8,-1.1) {\texttt{Element}\\\texttt{loop}};
  \node[stepnote, text width=2.8cm] at (14.1,-0.35) {\texttt{FlxDivSrcTerms}};

  \draw[decorate, decoration={brace, amplitude=4pt}, line width=0.7pt]
        (0.05,0.5) -- node[above=4pt, font=\lblfont] {convective flux} (7.85,0.5);
  \draw[decorate, decoration={brace, amplitude=4pt}, line width=0.7pt]
        (6.45,1.15) -- node[above=4pt, font=\lblfont] {diffusive flux (Navier--Stokes only)} (12.65,1.15);
  \draw[decorate, decoration={brace, amplitude=4pt}, line width=0.7pt]
        (12.85,0.5) -- node[above=4pt, font=\lblfont, align=center] {divergence and\\ source terms} (15.15,0.5);

  \node[seg, minimum width=0.8cm, minimum height=0.4cm] at (0,-5.1) {};
  \node[font=\lblfont, anchor=west] at (0.95,-5.1) {Kokkos kernel (compute)};

  \node[commseg, minimum width=0.8cm, minimum height=0.4cm] at (4.9,-5.1) {};
  \node[font=\lblfont, anchor=west] at (5.85,-5.1) {MPI transfer in progress};

  \draw[launch] (9,-4.95) -- (9,-5.25);
  \node[font=\lblfont, anchor=west] at (9.2,-5.1) {non-blocking launch};

  \draw[sync] (12.4,-4.95) -- (12.4,-5.25);
  \node[font=\lblfont, anchor=west] at (12.6,-5.1) {synchronization (\texttt{wait})};
\end{tikzpicture}
\caption{Pipeline of \texttt{KernelSD::compute\_rhs} for one RK stage: overlap of computation and MPI communication. Top lane: Kokkos kernels. Bottom lane: MPI transfers.}
  \label{fig:hopps_pipeline}
\end{figure}
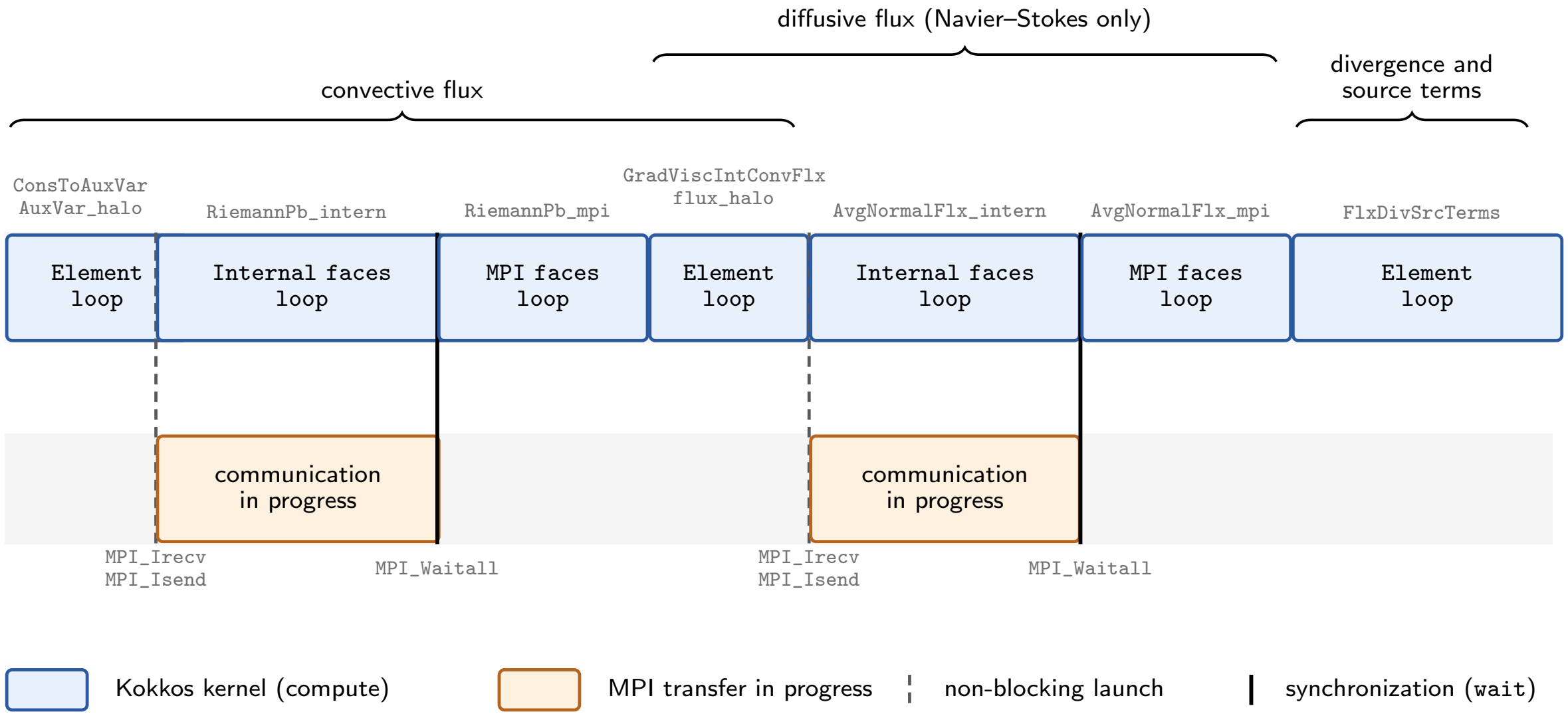

The second aspect is the overlap itself. Its mechanism is the same as on CPU,
but it matters more on GPU, where the latency of a network exchange, of the order
of microseconds, is comparable to the execution time of a whole Kokkos kernel.
Figure~\ref{fig:hopps_pipeline} details the pipeline of one RK stage,
split into a convective phase and a diffusive phase, followed by the
flux divergence.

The convective phase has three steps. First, an element loop reconstructs the
auxiliary variable at the FP (steps~1 and~2 of
Section~\ref{subsec:sd-principle}, \texttt{ConsToAuxVar},
Section~\ref{sec:Auxvariable}) and packs its interface values into the buffer
\texttt{AuxVar\_halo}, whose non-blocking exchange starts immediately. Second,
while the exchange is in progress, the Riemann problem (step~4) is solved on the internal
faces (\texttt{RiemannPb\_intern}). Third, after \texttt{MPI\_Waitall}, the
Riemann problem is solved on the faces shared between partitions
(\texttt{RiemannPb\_mpi}) with the received data.

The diffusive phase follows the same pattern. An element loop,
\texttt{GradViscIntConvFlx}, computes the solution gradient
(see Section~\ref{subsec:sd-parabolic}) and, from it, the interior convective and
viscous fluxes (step~3), while packing the buffer \texttt{flux\_halo} for a second non-blocking exchange. The
averaged normal flux is computed on the internal faces
(\texttt{AvgNormalFlx\_intern}) during this exchange, and on the faces shared
between partitions (\texttt{AvgNormalFlx\_mpi}) after the second
synchronization.

A last element loop, \texttt{FlxDivSrcTerms}, differentiates the flux
polynomial (steps~5 and~6 of Section~\ref{subsec:sd-principle}) and adds the chemical source
terms (see Section~\ref{sec:GoverningEquations}), which gives the right-hand side of
Eq.~\eqref{eq:semidiscrete}.

\subsection{Intra-GPU parallelization}
\label{subsec:intragpu}

Within a GPU, the element kernels called by the right-hand side
(reconstruction of the auxiliary variable, gradient, interior convective and
diffusive fluxes, divergence) follow the storage of the fields by element
(see Section~\ref{subsec:hopps-kokkos}). Each kernel launches one team of threads
per element, with one thread per point, so that consecutive threads process
consecutive points (SP or FP) of the same element. To reconstruct an FP, a
thread reads the $p+1$ SP of the corresponding line
(Fig.~\ref{fig:sd-stencil-2d}, left). The team as a whole therefore reads the
contiguous block of its element, and every cache line brought from memory is
used by its threads. The team size follows from the polynomial degree, known at
compile time ($216$ threads for the SP at $p=5$), and no parameter depends on
the GPU model.

The gradient, the diffusive fluxes and the interior convective fluxes are
computed in a single kernel. The team first computes the gradients at the SP of
its element and keeps them in the scratch memory of the team, then evaluates
the fluxes at the FP from them: the gradients are never written to global
memory and read back. Kokkos maps the scratch memory to the shared memory of a
multiprocessor, or, at a second level, to a block of global memory reserved per
team. In the multi-species case, the time loop computes only the gradients that
the diffusive fluxes need, those of the temperature, of the velocity components
and of the mass fractions, that is, $3(4+N_s)$ components per SP. At $p=5$, they
take 35\,KiB per element with three species and 66\,KiB with nine. The first
size fits within the 48\,KiB that Kokkos allows per team in shared memory. For
the second one, the kernel keeps in shared memory as many components as fit on
AMD GPUs, and places all of them in the global scratch memory on Nvidia GPUs,
where this choice was measured faster. On AMD GPUs, when two teams do not fit in
the shared memory of a compute unit, the team size is doubled, so that each
SIMD unit still alternates between two wavefronts.

This organization replaces one in which the element index was the fastest in
memory and each warp processed one point of consecutive elements, relying on
the L2 cache to keep the data of the elements in progress. Its effect is
measured in Section~\ref{sec:singlegpuperf}.

The interface kernels (Riemann solver and averaged viscous flux) iterate over
the pairs of FP shared by two elements. Each pair holds two handles to the FP values of the neighboring elements, so the kernels read and write the FP field in place, without intermediate face arrays: only non-conforming faces go through a projection onto mortar points, and the remote side of an MPI pair is read directly from the halo receive buffer, without an unpacking kernel. The pairs internal to a GPU are sorted by the memory address of their first FP, so that neighboring threads also access neighboring addresses. Since the
variables of an FP are contiguous in memory, a thread that solves one Riemann
problem would write its flux at addresses far from those of the neighboring
threads. The Riemann kernel therefore processes the pairs by teams of 64: each
thread writes its flux into shared memory, and the team then copies the fluxes
of its pairs to the FP field, with consecutive threads writing consecutive
addresses. The averaged viscous flux assigns one thread to each variable of a
pair, which gives contiguous accesses directly.

\section{Cross-verification on a three-dimensional reacting Taylor--Green vortex}\label{sec:validationTGV}
This section checks that the multi-species extension of HOPPS reproduces the
solution of JAGUAR, whose SD formulation for reacting flows was validated
previously \cite{Marchal2023}, and that the CPU and GPU executions of HOPPS give
the same solution. It is a cross-verification between codes and back-ends, not
a validation of the physical models, which are those of JAGUAR. The performance
study of Section~\ref{sec:perfoAnalysis} relies on this equivalence.

\subsection{Benchmark presentation}
The TGV is a standard test case for numerical schemes: a smooth, analytically
defined velocity field decays into small-scale structures in a periodic cube,
without walls or inflow and outflow boundaries. The discretization error is thus
isolated from any boundary treatment, and chemistry can be added without
geometric complexity. Following Abdelsamie et al.~\cite{AbdelsamieTGVReacting},
who proposed this case to validate low-Mach reacting solvers, non-uniform
temperature and composition fields are superimposed on the classical TGV
velocity field, in a non-premixed configuration. The fuel, hydrogen diluted in
nitrogen, fills a central region bounded by two planes normal to $x$, and air
fills the rest of the domain. The temperature peaks at the two fuel-air
interfaces. Temperature and composition vary along $x$ only, through hyperbolic
tangent profiles, and the reacting case uses profiles that reproduce the
equilibrium composition expected at the fuel-air interfaces.
The case is used here as a convenient unsteady multi-species configuration: the
reference solution of Abdelsamie et al.\ was computed with low-Mach solvers
($u_0=4$~m/s, $L_0=1$~mm, $\mathrm{Re}=u_0L_0/\nu_\mathrm{min}=251.62$ with
$\nu_\mathrm{min}$ the kinematic viscosity of air at 300~K,
$\mathrm{Ma}\approx 0.01$), whereas JAGUAR and HOPPS are both
compressible explicit solvers, and the comparison is made between them only.

Both variants use the same mesh of $N^3$ hexahedral elements, with $N=32$
elements per direction and $p=3$, that is, $N^3(p+1)^3=128^3$ DOF, and the numerical settings of Section~\ref{sec:NSequationNumMethods}.
Both codes use the RK(5,4) scheme with CFL~$=0.4$. The profiles along the lines
$y=z=L/2$ and $x=z=L/2$ are compared at $t=2\tau_\mathrm{ref}=0.5$~ms, where
$\tau_\mathrm{ref}=0.25$~ms is the reference time of the vortex
(\ref{app:tgv-ic}), once the vortices have folded the initial fuel region. The
maximum temperature in the domain is compared over the whole computation, up to
$t=10\tau_\mathrm{ref}$.

\paragraph{Non-reacting configuration} A three-species mixture
($\mathrm{H_2}$, $\mathrm{O_2}$, $\mathrm{N_2}$) is transported with the JANAF
closure but without chemical source terms, so that only convective transport
and molecular diffusion modify the initial temperature and composition fields.

\paragraph{Reacting configuration} The velocity field is the same and the
initial temperature and composition fields are similar, but the mixture contains
nine species ($\mathrm{H_2}$,
$\mathrm{O_2}$, $\mathrm{N_2}$, $\mathrm{H}$, $\mathrm{OH}$, $\mathrm{O}$,
$\mathrm{H_2O}$, $\mathrm{HO_2}$, $\mathrm{H_2O_2}$) and reacts according to the
mechanism of Boivin et al.~\cite{Boivin} (see Section~\ref{sec:GoverningEquations}).
The species production rates $\dot\omega_k$ and the heat release rate
$\dot\omega_T$ (see Eq.~\eqref{eq:source-terms}) then act in addition to transport
and diffusion.

\begin{figure}[htbp]
  \centering
  \begin{subfigure}[b]{0.48\textwidth}\centering
\includegraphics[width=\linewidth]{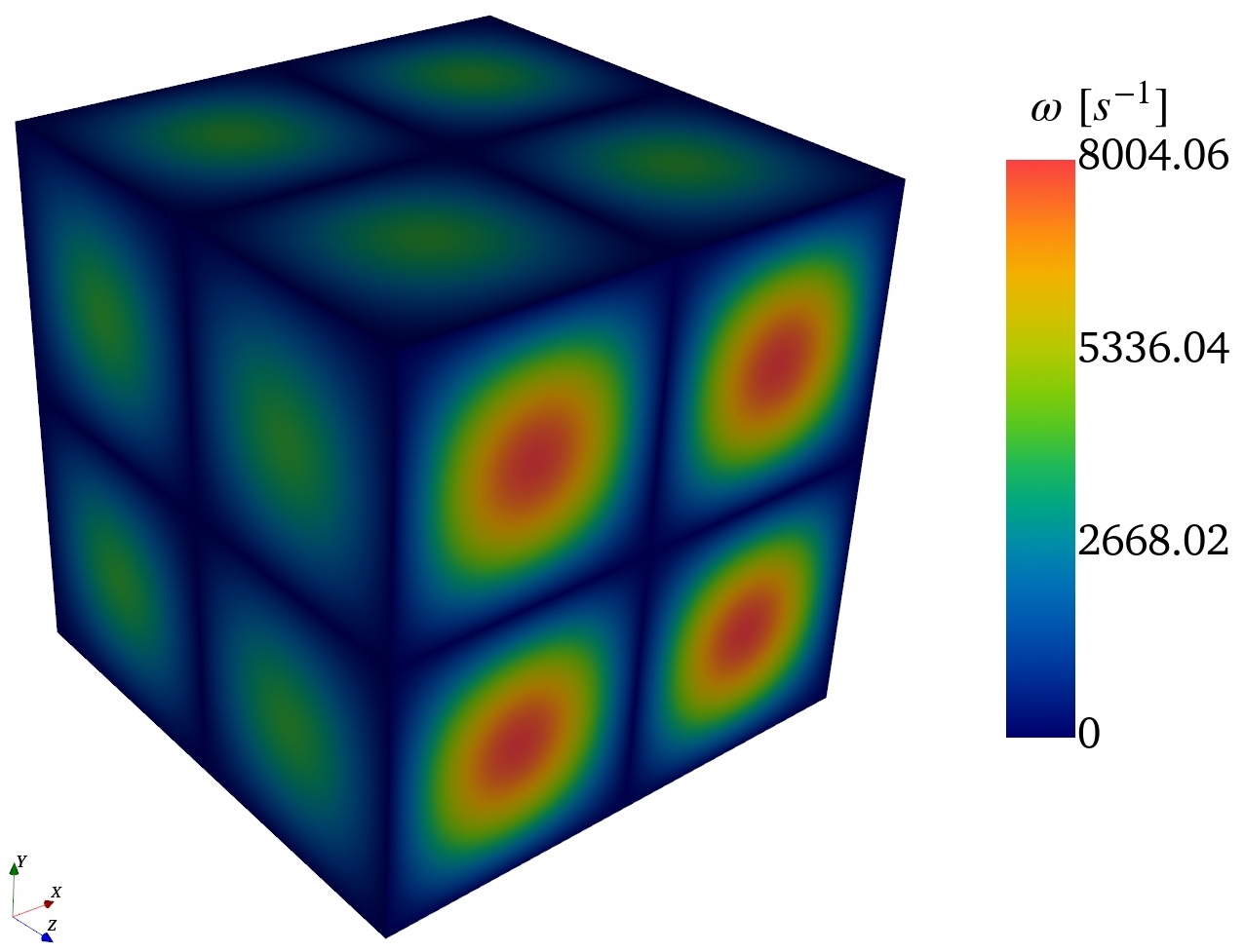}

\end{subfigure}
  \begin{subfigure}[b]{0.48\textwidth}\centering
\includegraphics[width=\linewidth]{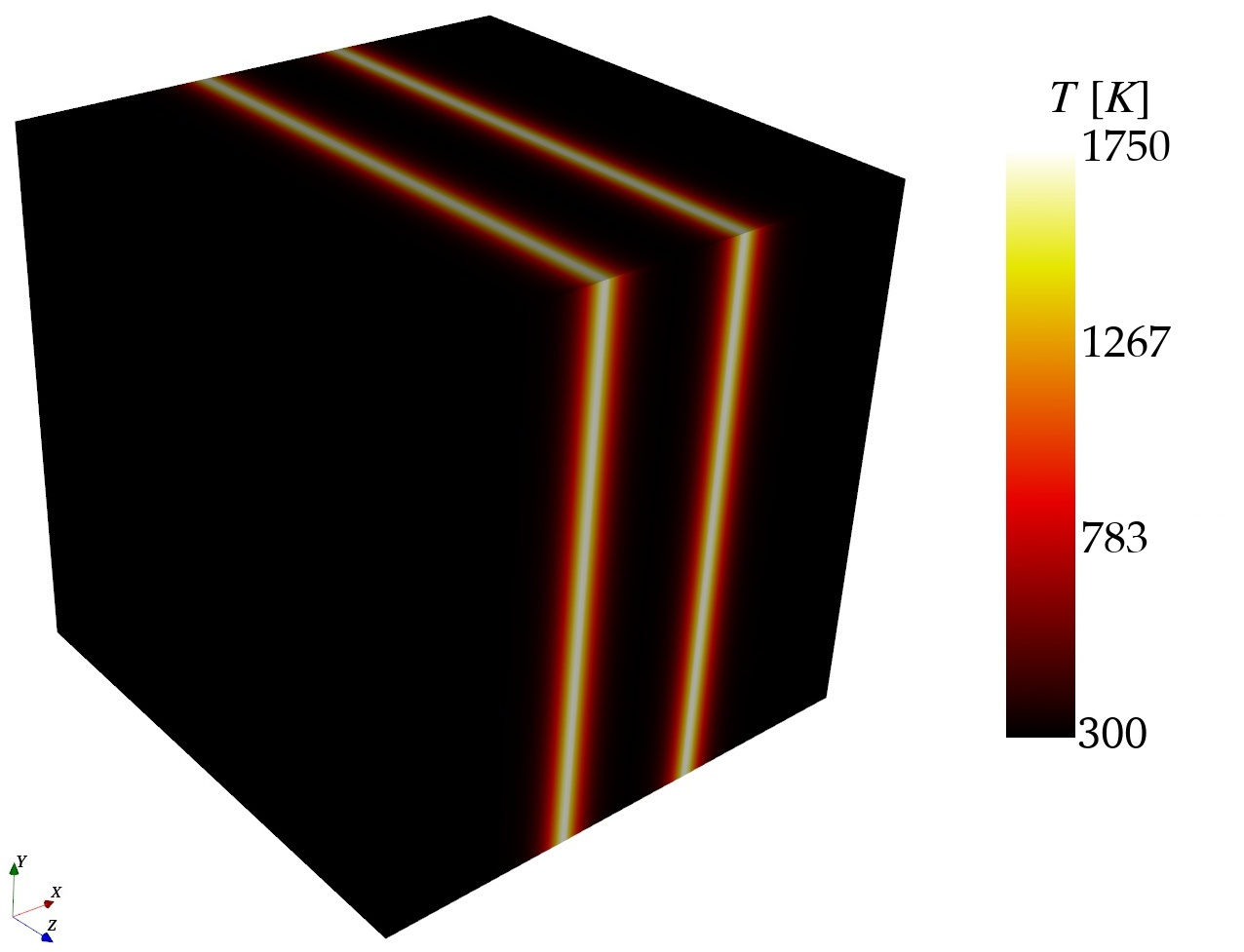}

\end{subfigure}
  \begin{subfigure}[b]{0.48\textwidth}\centering
\includegraphics[width=\linewidth]{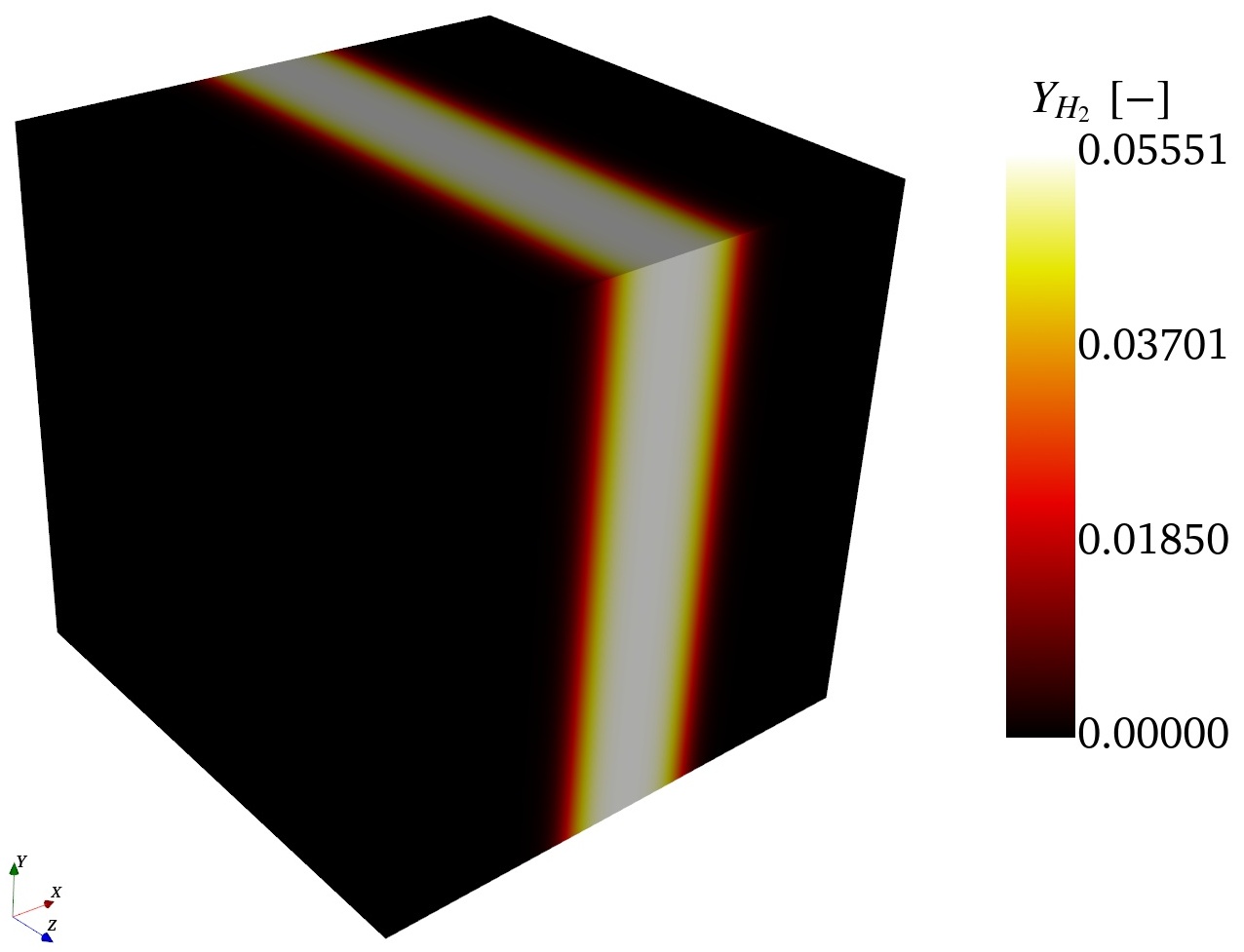}

\end{subfigure}
  \begin{subfigure}[b]{0.48\textwidth}\centering
\includegraphics[width=\linewidth]{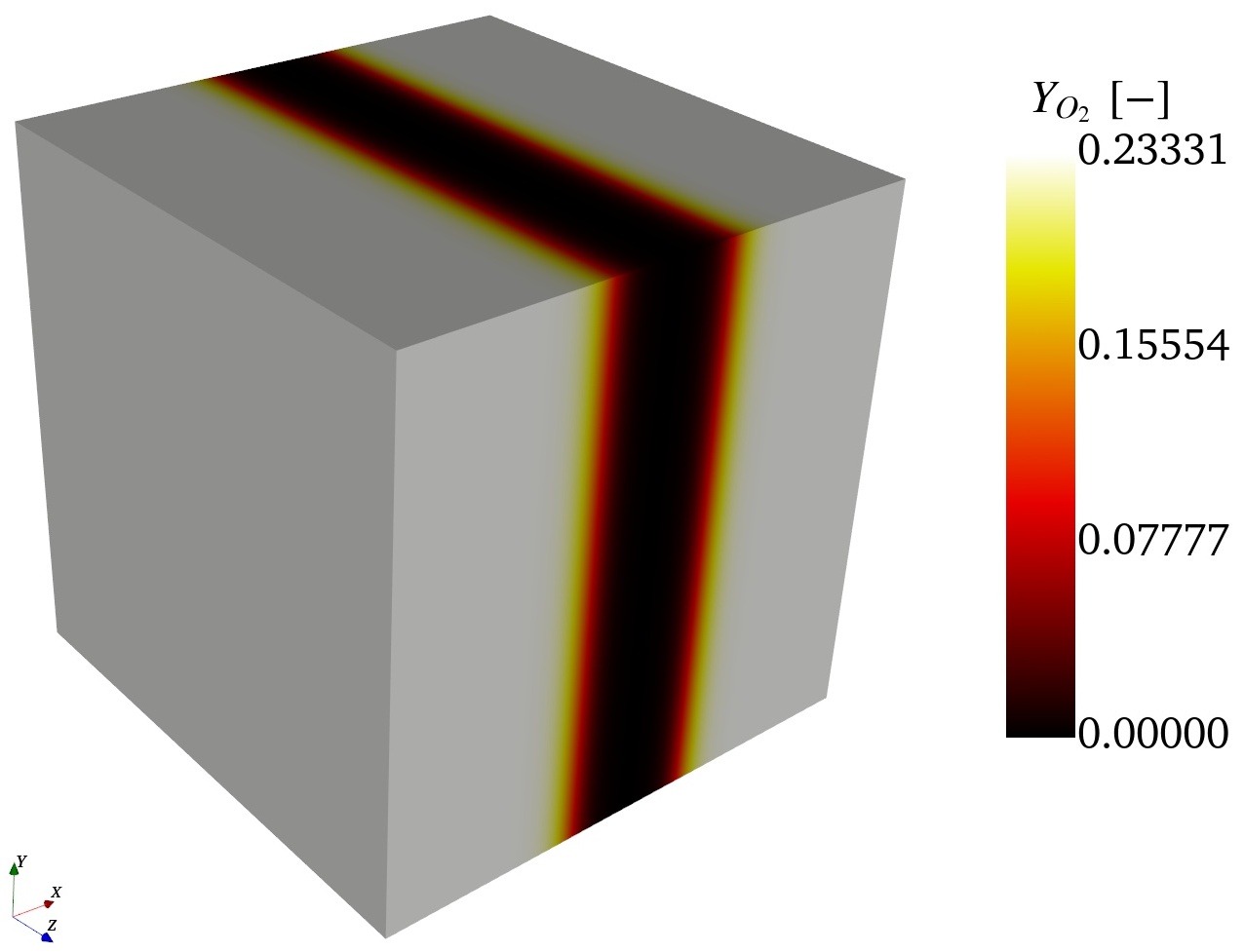}

\end{subfigure}
\caption{Initial fields ($t=0$) of the non-reacting case: vorticity magnitude, temperature, and mass fractions of $\mathrm{H_2}$ and $\mathrm{O_2}$.}
  \label{fig:bench_init_vort_temp_h2_o2_3D}
\end{figure}

The exact expressions of the initial velocity, pressure, temperature and species
fields, as implemented in HOPPS, are given in \ref{app:tgv-ic}
(Fig.~\ref{fig:bench_nonreact_init_1D}).
Figure~\ref{fig:bench_init_vort_temp_h2_o2_3D} shows the initial vorticity,
temperature and $\mathrm{H_2}$ and $\mathrm{O_2}$ mass fraction fields.

\subsection{Non-reacting case}\label{sec:nonreactingcase}
Without chemistry, only convective transport by the TGV velocity field and
molecular diffusion (see Section~\ref{sec:GoverningEquations}) modify the initial
temperature and composition fields.

\begin{figure}[htbp]
  \centering
  \begin{subfigure}[b]{0.48\textwidth}\centering
\includegraphics[width=\linewidth]{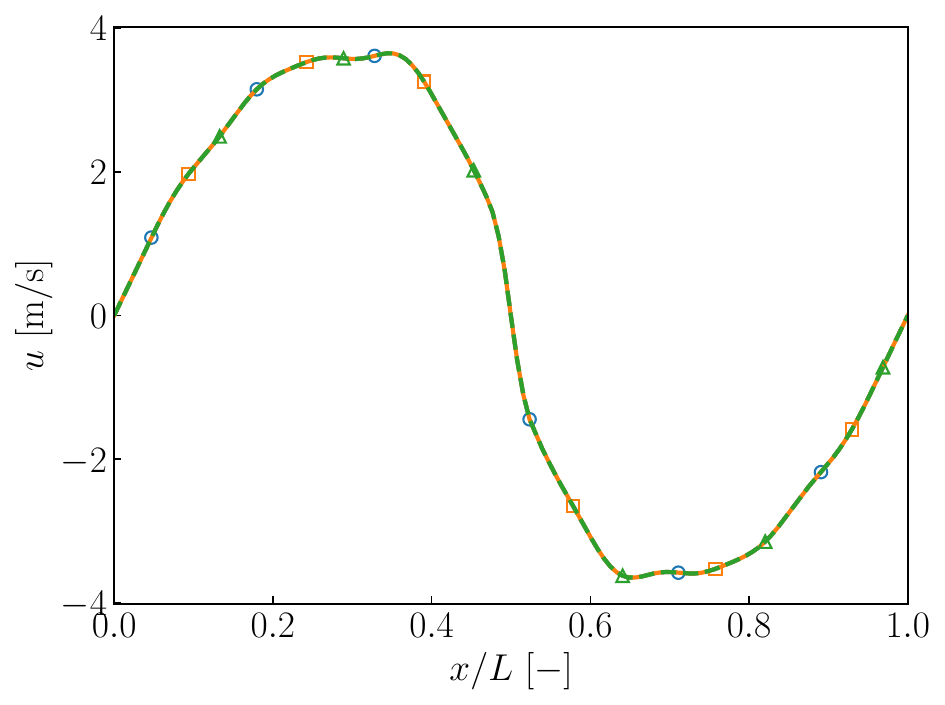}

\end{subfigure}
  \begin{subfigure}[b]{0.48\textwidth}\centering
\includegraphics[width=\linewidth]{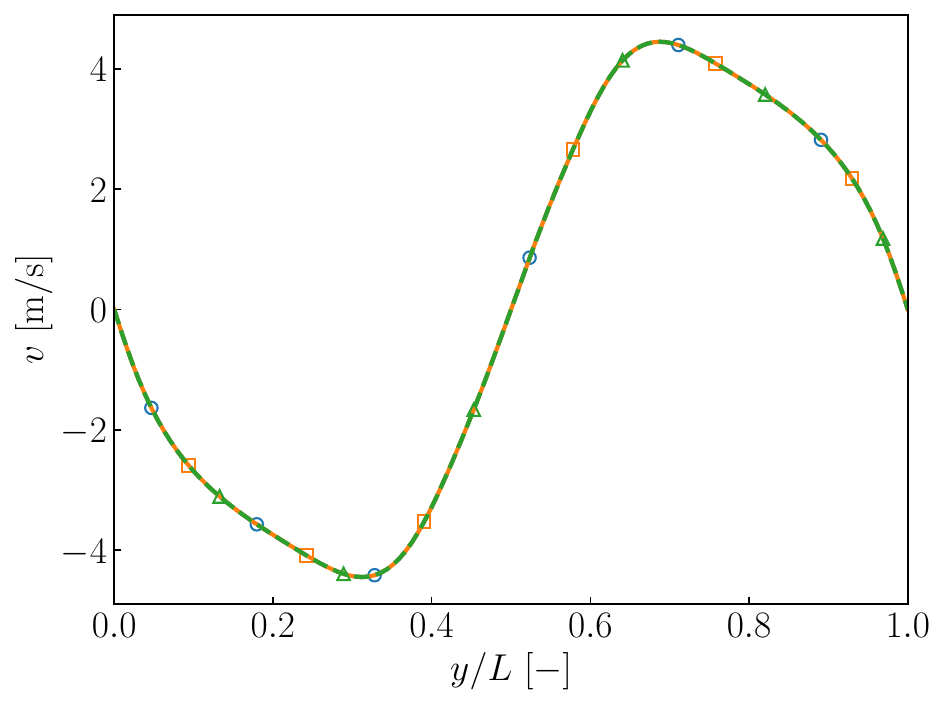}

\end{subfigure}
\caption{Non-reacting case: velocity component $u$ along the line $y = z = 0.5L$ (left) and $v$ along the line $x = z = 0.5L$ (right) at $t = 2\tau_\mathrm{ref} = 0.5$~ms.
Legend:
  \textcolor{mplblue}{$-\!\circ\!-$} JAGUAR on AMD Genoa CPU,
  \textcolor{mplorange}{$-\!\square\!-$} HOPPS on AMD Genoa CPU,
  \textcolor{mplgreen}{$-\!\triangle\!-$} HOPPS on Nvidia H100 GPU.}
\label{fig:bench_nonreact_vx_vy_1D}
\end{figure}

\begin{figure}[htbp]
  \centering
  \begin{subfigure}[b]{0.48\textwidth}\centering
\includegraphics[width=\linewidth]{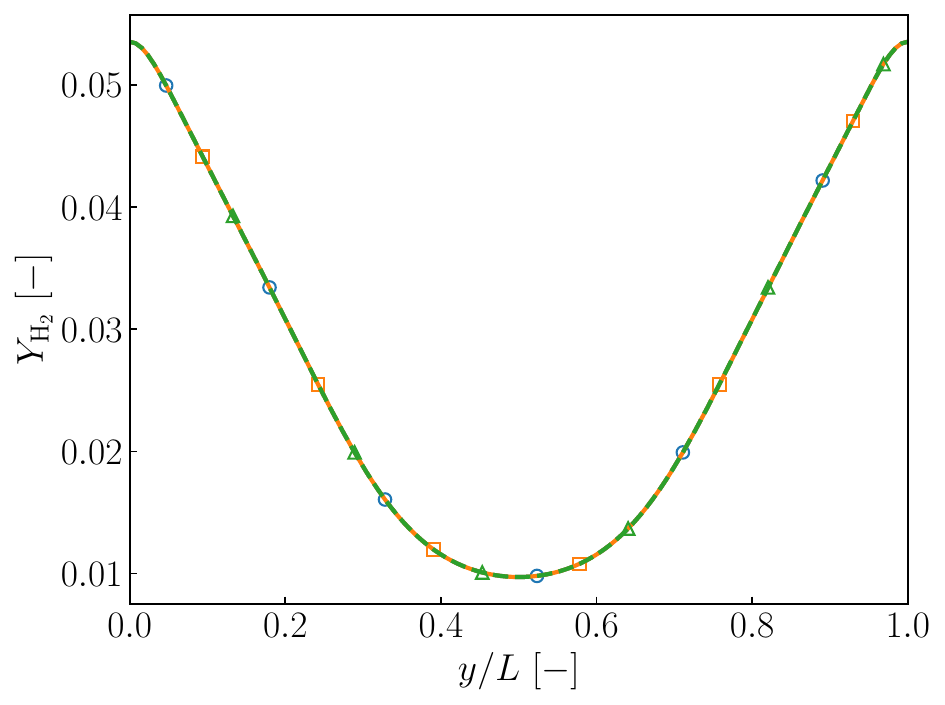}

\end{subfigure}
  \begin{subfigure}[b]{0.48\textwidth}\centering
\includegraphics[width=\linewidth]{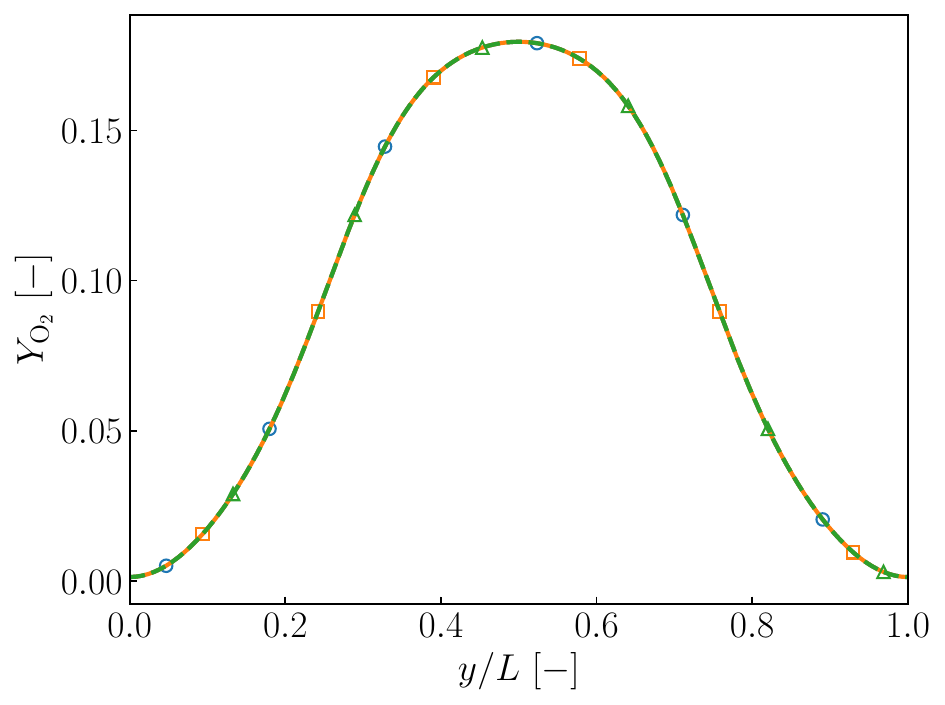}

\end{subfigure}
  \begin{subfigure}[b]{0.48\textwidth}\centering
\includegraphics[width=\linewidth]{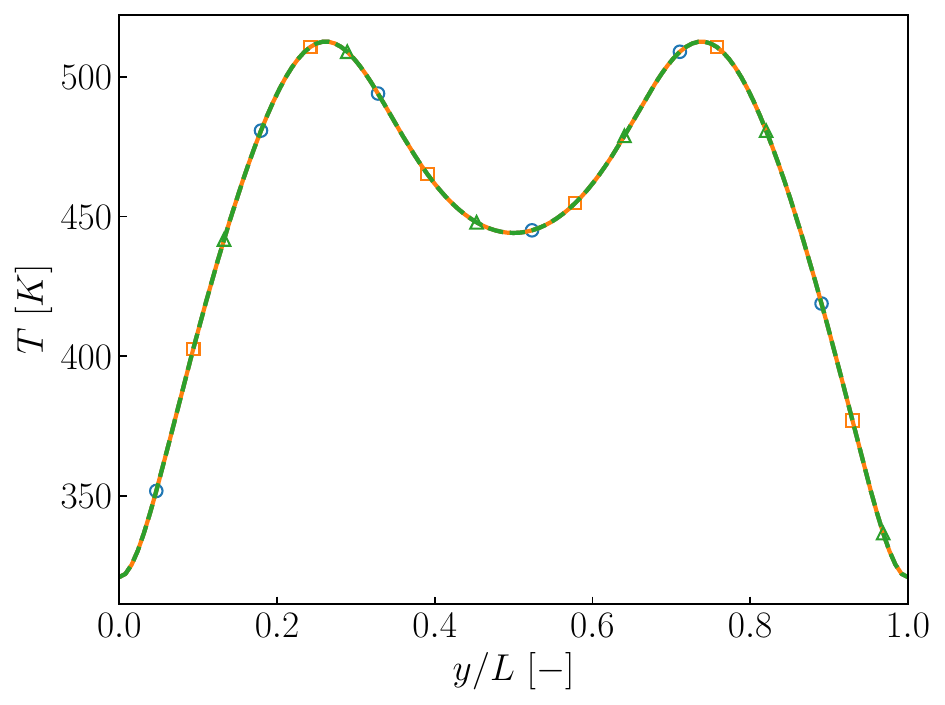}

\end{subfigure}
\caption{Non-reacting case: mass fractions of $\mathrm{H_2}$ and $\mathrm{O_2}$, and temperature, along the line $x = z = 0.5L$ at $t = 2\tau_\mathrm{ref} = 0.5$~ms.
  Legend:
  \textcolor{mplblue}{$-\!\circ\!-$} JAGUAR on AMD Genoa CPU,
  \textcolor{mplorange}{$-\!\square\!-$} HOPPS on AMD Genoa CPU,
  \textcolor{mplgreen}{$-\!\triangle\!-$} HOPPS on Nvidia H100 GPU.}
\label{fig:bench_nonreact_h2_o2_temp_1D}
\end{figure}

\begin{figure}[htbp]
  \centering
  \begin{subfigure}[b]{0.48\textwidth}\centering
\includegraphics[width=\linewidth]{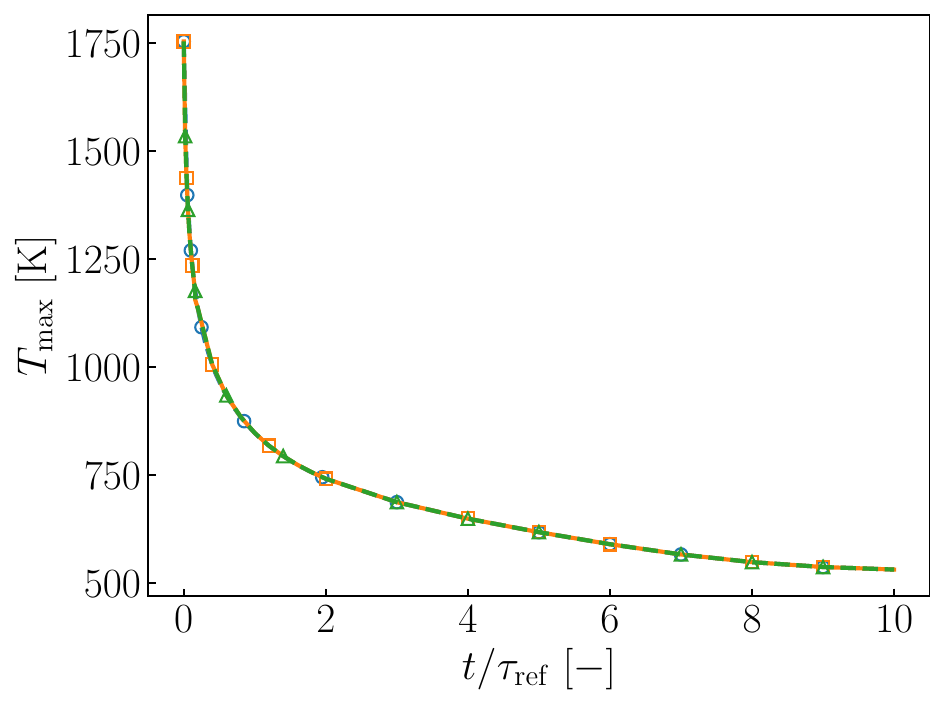}

\end{subfigure}
\caption{Non-reacting case: maximum temperature in the domain versus time.
  Legend:
  \textcolor{mplblue}{$-\!\circ\!-$} JAGUAR on AMD Genoa CPU,
  \textcolor{mplorange}{$-\!\square\!-$} HOPPS on AMD Genoa CPU,
  \textcolor{mplgreen}{$-\!\triangle\!-$} HOPPS on Nvidia H100 GPU.}
\label{fig:bench_nonreact_tmax_1D}
\end{figure}

Three computations are compared along lines through the domain center: the
Fortran solver JAGUAR on AMD Genoa CPUs, HOPPS on the same CPUs, and HOPPS on an
Nvidia H100 GPU. The velocity components (Fig.~\ref{fig:bench_nonreact_vx_vy_1D}),
the $\mathrm{H_2}$ and $\mathrm{O_2}$ mass fractions and the temperature
(Fig.~\ref{fig:bench_nonreact_h2_o2_temp_1D}) of the three computations are
superimposed over the whole line, including the two off-center temperature maxima
and the central minimum produced by the folding of the initial fuel region. The maximum
temperature in the domain (Fig.~\ref{fig:bench_nonreact_tmax_1D}) decreases
monotonically from its initial value, as expected for pure mixing, and the three
histories are also superimposed.

The multi-species extension of HOPPS with JANAF thermochemistry therefore
reproduces JAGUAR \cite{Marchal2023} on all the quantities examined, and the CPU
and GPU executions of HOPPS agree with each other as closely as with JAGUAR.

\subsection{Reacting case}\label{sec:reactingcase}

With chemistry, heat is released where the folded interface reaches a
composition and a temperature that allow significant reaction rates. The
associated dilatation accelerates the flow locally.

\begin{figure}[htbp]
  \centering
  \begin{subfigure}[b]{0.48\textwidth}\centering
\includegraphics[width=\linewidth]{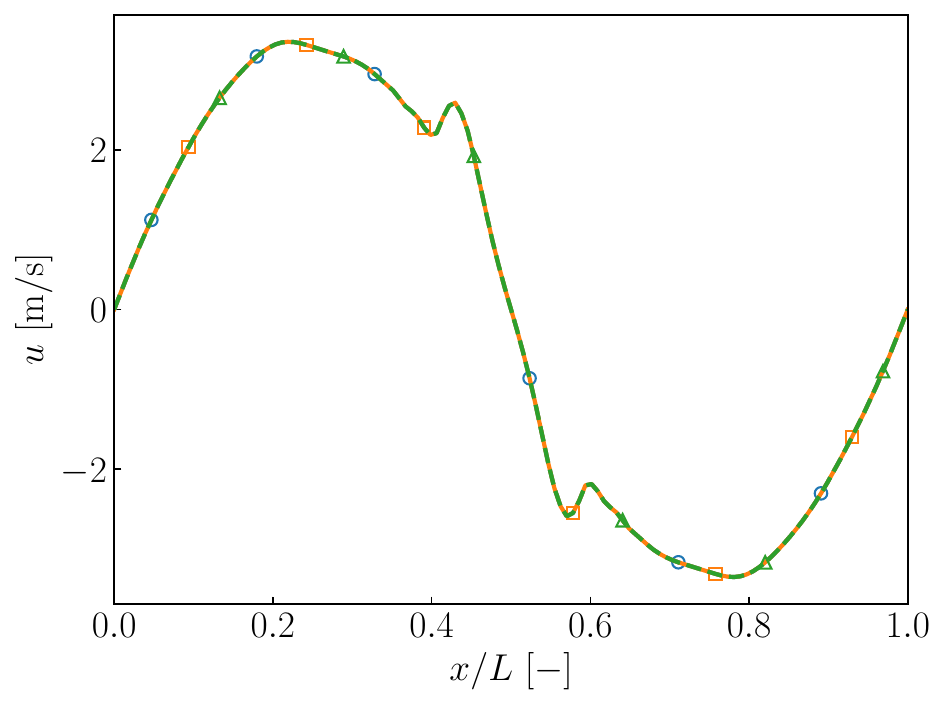}

\end{subfigure}
  \begin{subfigure}[b]{0.48\textwidth}\centering
\includegraphics[width=\linewidth]{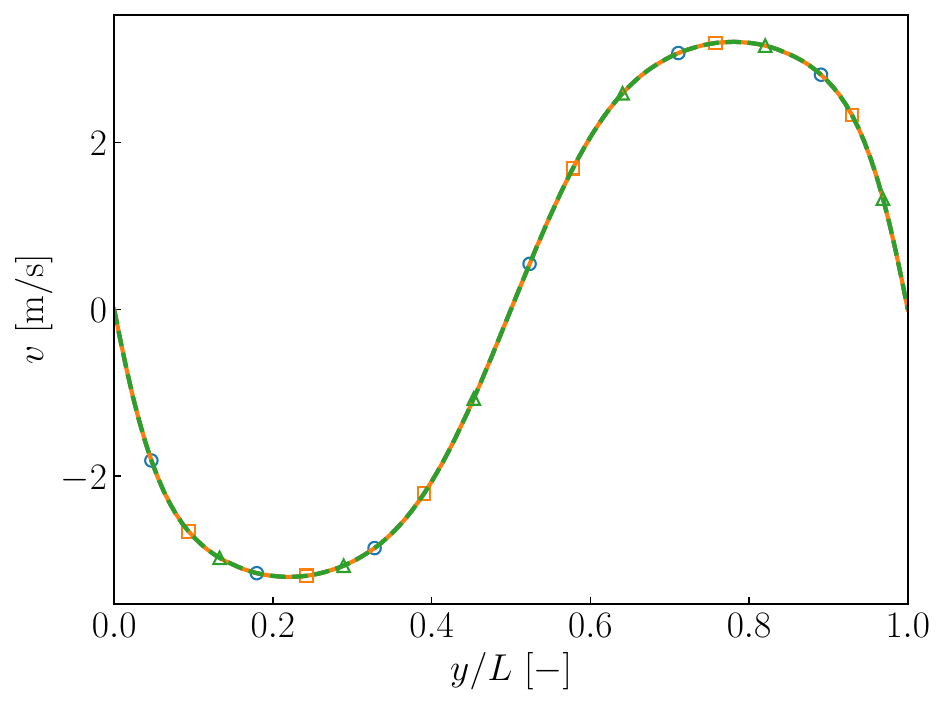}

\end{subfigure}
\caption{Reacting case: velocity component $u$ along the line $y = z = 0.5L$ (left) and $v$ along the line $x = z = 0.5L$ (right) at $t = 2\tau_\mathrm{ref} = 0.5$~ms.
  Legend:
  \textcolor{mplblue}{$-\!\circ\!-$} JAGUAR on AMD Genoa CPU,
  \textcolor{mplorange}{$-\!\square\!-$} HOPPS on AMD Genoa CPU,
  \textcolor{mplgreen}{$-\!\triangle\!-$} HOPPS on Nvidia H100 GPU.}
  \label{fig:bench_vx_vy_1D}
\end{figure}

\begin{figure}[htbp]
  \centering
  \begin{subfigure}[b]{0.48\textwidth}\centering
\includegraphics[width=\linewidth]{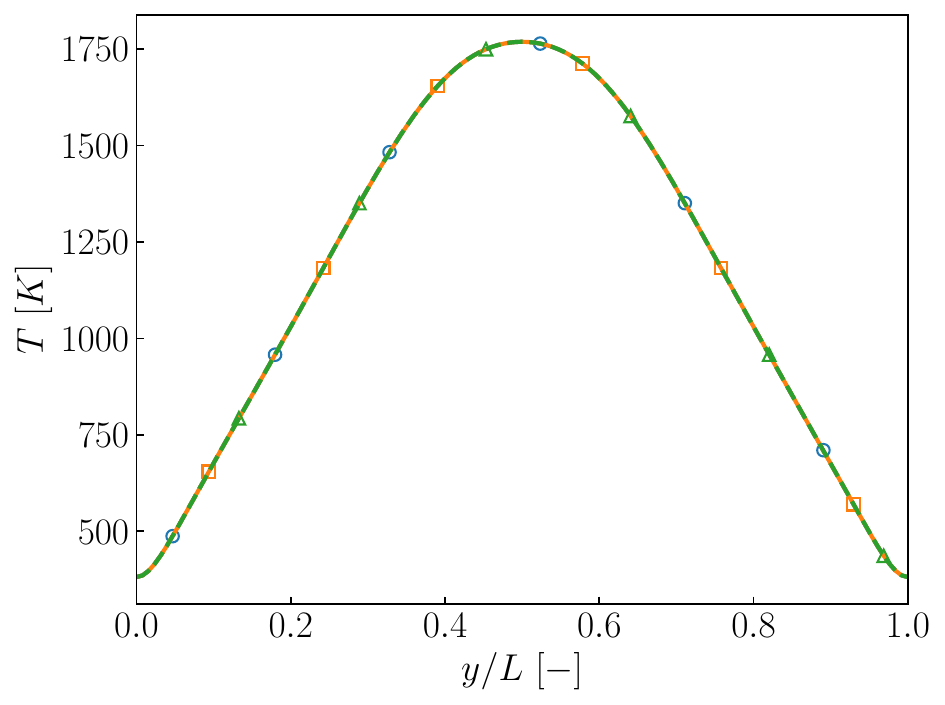}
\end{subfigure}
  \begin{subfigure}[b]{0.48\textwidth}\centering
\includegraphics[width=\linewidth]{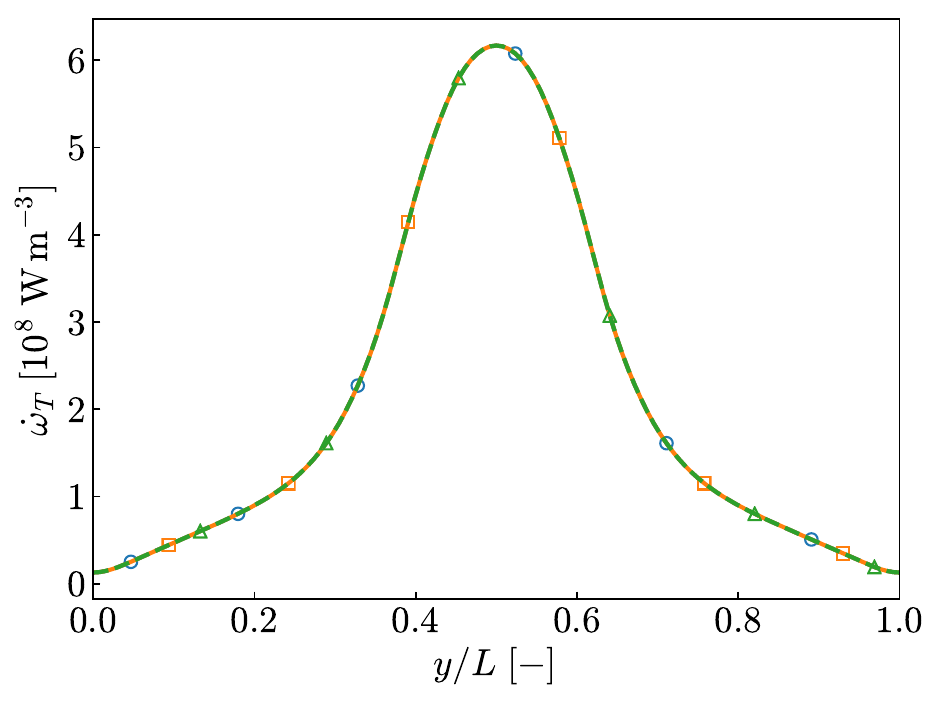}
\end{subfigure}
\caption{Reacting case: temperature (left) and heat release rate (right) along the line $x = z = 0.5L$ at $t = 2\tau_\mathrm{ref} = 0.5$~ms.
  Legend:
  \textcolor{mplblue}{$-\!\circ\!-$} JAGUAR on AMD Genoa CPU,
  \textcolor{mplorange}{$-\!\square\!-$} HOPPS on AMD Genoa CPU,
  \textcolor{mplgreen}{$-\!\triangle\!-$} HOPPS on Nvidia H100 GPU.}
  \label{fig:bench_temp_hr_1D}
\end{figure}

\begin{figure}[htbp]
  \centering
  \begin{subfigure}[b]{0.48\textwidth}\centering
\includegraphics[width=\linewidth]{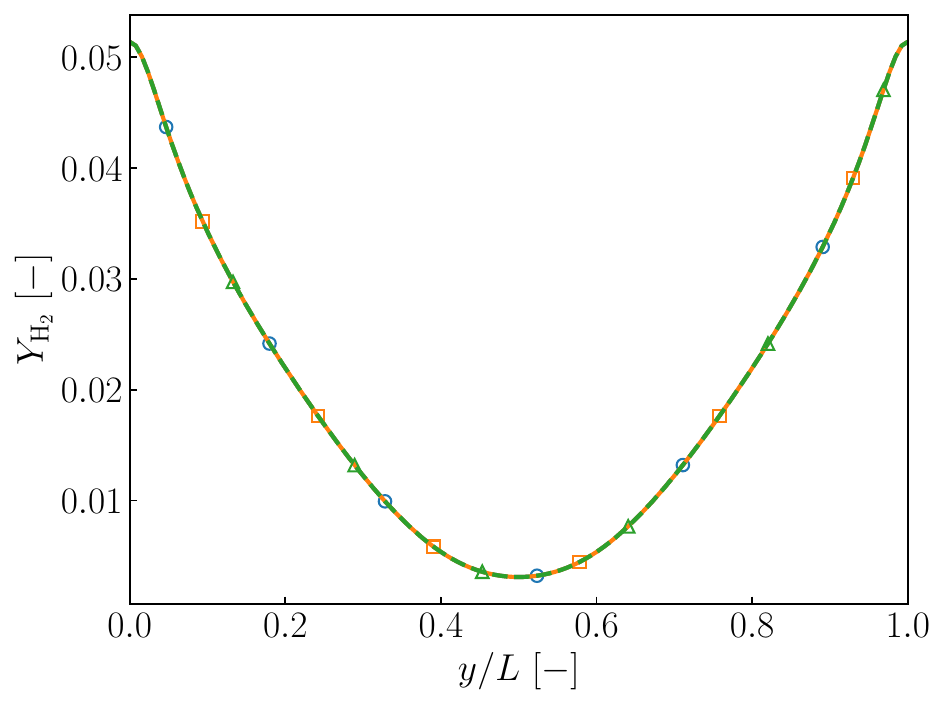}
\end{subfigure}
  \begin{subfigure}[b]{0.48\textwidth}\centering
\includegraphics[width=\linewidth]{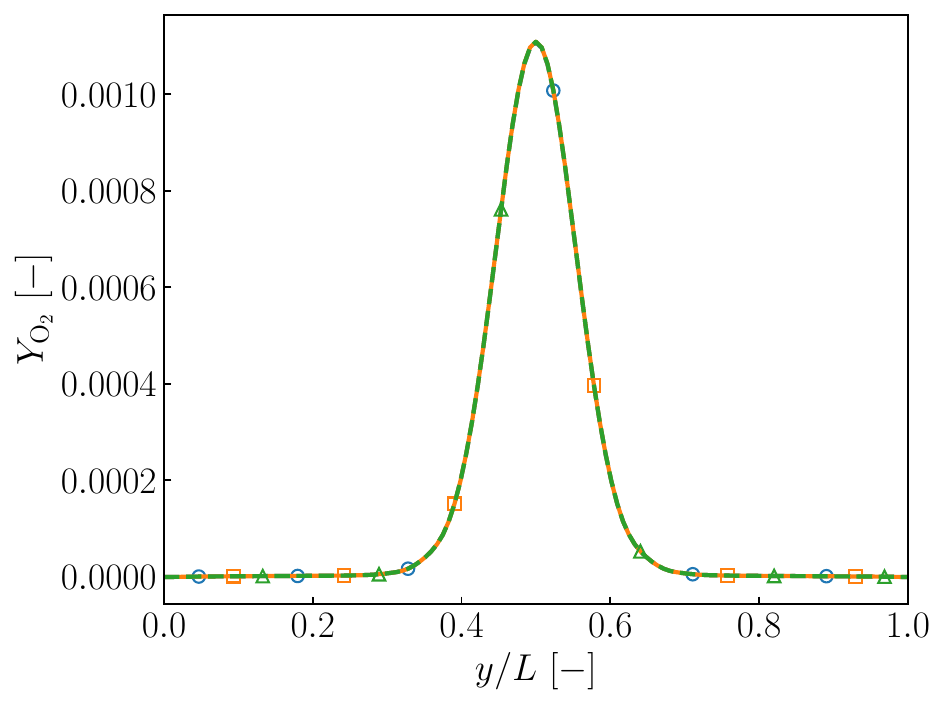}
\end{subfigure}
  \begin{subfigure}[b]{0.48\textwidth}\centering
\includegraphics[width=\linewidth]{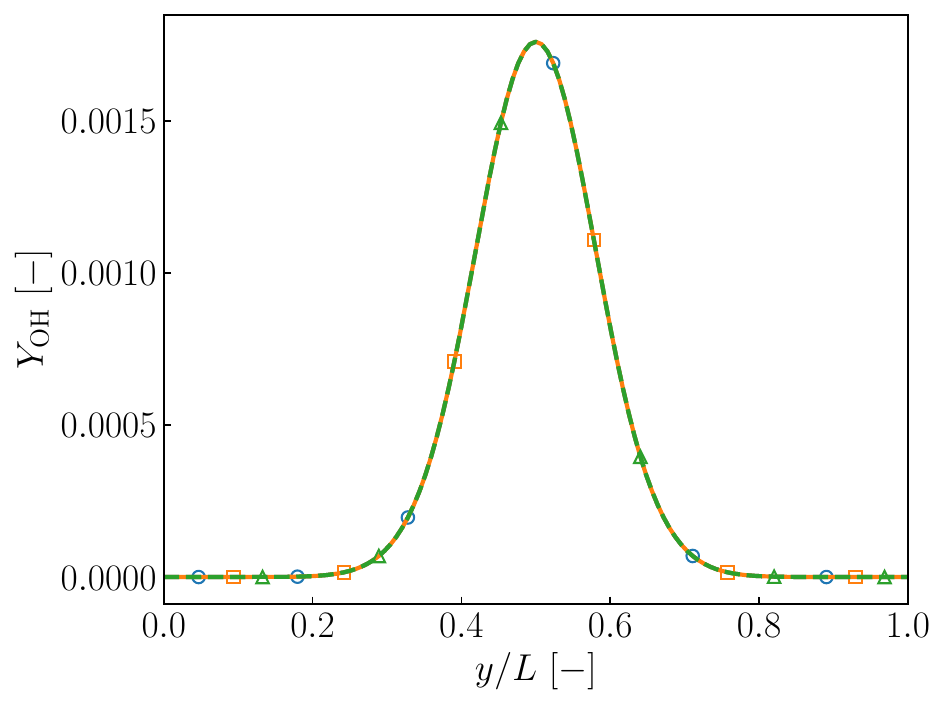}
\end{subfigure}
  \begin{subfigure}[b]{0.48\textwidth}\centering
\includegraphics[width=\linewidth]{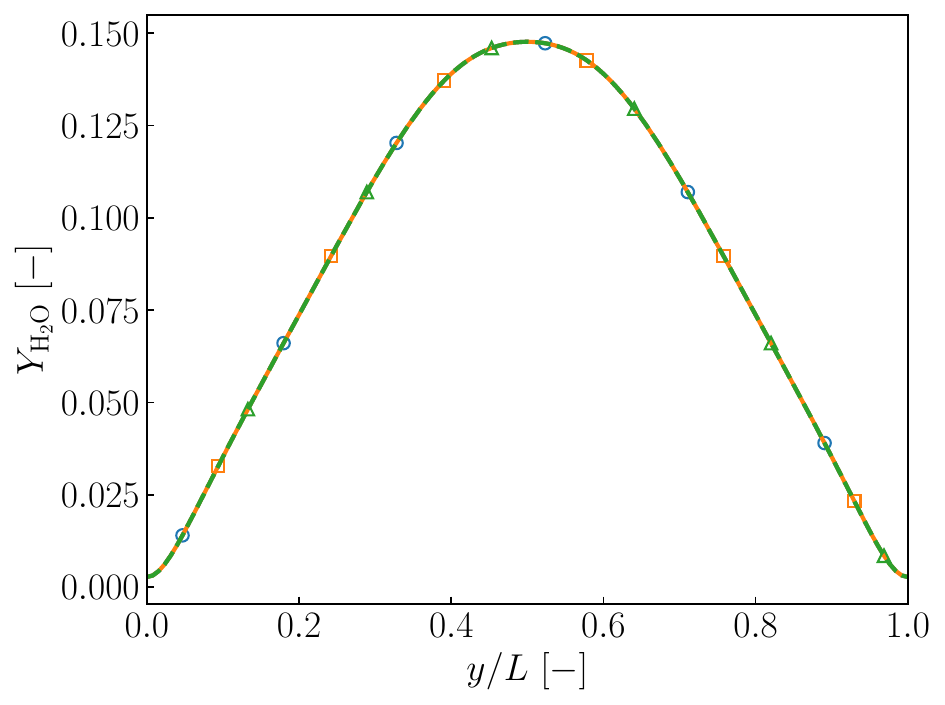}
\end{subfigure}
\caption{Reacting case: mass fractions of $\mathrm{H_2}$, $\mathrm{O_2}$, $\mathrm{OH}$ and $\mathrm{H_2O}$ along the line $x = z = 0.5L$ at $t = 2\tau_\mathrm{ref} = 0.5$~ms.
  Legend:
  \textcolor{mplblue}{$-\!\circ\!-$} JAGUAR on AMD Genoa CPU,
  \textcolor{mplorange}{$-\!\square\!-$} HOPPS on AMD Genoa CPU,
  \textcolor{mplgreen}{$-\!\triangle\!-$} HOPPS on Nvidia H100 GPU.}
  \label{fig:bench_h2_o2_oh_1D}
\end{figure}

\begin{figure}[htbp]
  \centering
  \begin{subfigure}[b]{0.48\textwidth}\centering
\includegraphics[width=\linewidth]{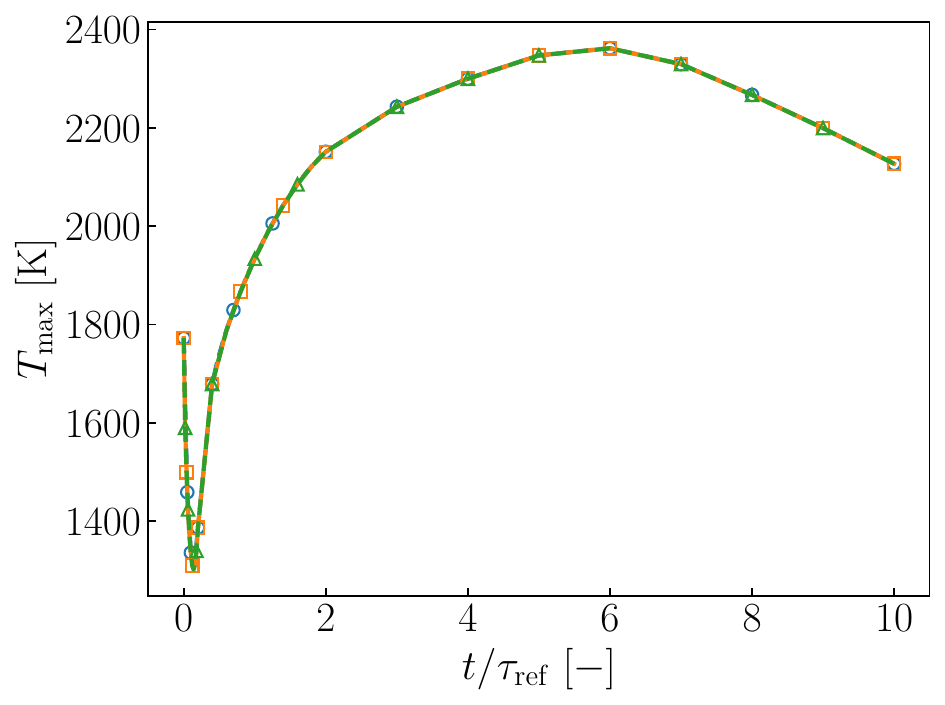}

\end{subfigure}
\caption{Reacting case: maximum temperature in the domain versus time.
  Legend:
  \textcolor{mplblue}{$-\!\circ\!-$} JAGUAR on AMD Genoa CPU,
  \textcolor{mplorange}{$-\!\square\!-$} HOPPS on AMD Genoa CPU,
  \textcolor{mplgreen}{$-\!\triangle\!-$} HOPPS on Nvidia H100 GPU.}
\label{fig:bench_react_tmax_1D}
\end{figure}

The same three computations are compared for this case. The velocity profiles
(Fig.~\ref{fig:bench_vx_vy_1D}) differ from the non-reacting ones: $u$ has two
additional local extrema, and the extrema of $v$ are lower. The three computations remain
superimposed. The temperature and the
heat release rate (Fig.~\ref{fig:bench_temp_hr_1D}) agree as well, including the
narrow peak of the heat release rate, the stiffest quantity of the comparison.
The mass fractions of $\mathrm{H_2}$, $\mathrm{O_2}$, $\mathrm{OH}$ and
$\mathrm{H_2O}$ (Fig.~\ref{fig:bench_h2_o2_oh_1D}) are also superimposed,
including the narrow peaks of $\mathrm{OH}$ and $\mathrm{O_2}$. The maximum
temperature (Fig.~\ref{fig:bench_react_tmax_1D}) is no longer monotonic. It first
decreases, as heat diffusion lowers the temperature peak faster than the heat
release rate raises it.
It then rises above its initial value when the vortex brings fresh reactants into
contact with the hot products, and decreases again as these reactants are
consumed. This ignition competing with mixing is consistent with the observations
of Abdelsamie et al.~\cite{AbdelsamieTGVReacting}.
The initial maximum, about 1750~K, is lower than $T_\mathrm{ad}=1945.5$~K
(\ref{app:tgv-ic}) because the initial temperature profile reaches
$T_\mathrm{ad}$ only at a sharp peak, at $|x-\pi L_0|=0.875$~mm, where no
SP lies.

The addition of chemical source terms therefore introduces no visible difference
between the CPU and GPU executions, nor between HOPPS and JAGUAR.


\section{Performance analysis}\label{sec:perfoAnalysis}
The performance analysis of HOPPS for flows without species \cite{HOPPS} is
extended here in two directions: to the multi-species TGV configurations of
Section~\ref{sec:validationTGV}, and to a comparison with JAGUAR \cite{JAGUAR}.
The optimizations made for the multi-species reacting solver also improve the
single-species solver, whose performance on one GPU is given in
\ref{app:mono}.
Tables~\ref{tab:cpu-hw} and \ref{tab:gpu-hw} describe the CPU nodes and the
GPUs employed.
HOPPS is built with Kokkos~4.7.00 \cite{KokkosSW}, with the architecture flag of each device
(\texttt{Kokkos\_ARCH\_ZEN4}, \texttt{ZEN2}, \texttt{ICX}, \texttt{SKX} on CPU,
\texttt{VOLTA70}, \texttt{AMPERE80}, \texttt{HOPPER90}, \texttt{AMD\_GFX90A},
\texttt{AMD\_GFX942} on GPU), so that the compiler generates code for the
instruction set of each processor. The software stacks are given in
Table~\ref{tab:sw-stack}. CPU runs use pure MPI, one rank per core, bound to
cores, with one OpenMP thread per rank. GPU runs use one MPI rank per GPU and a
GPU-aware MPI. An MI250X card contains two GCD, which
the software sees as two GPUs: each GCD is driven by its own MPI rank and counts
as one GPU in the following. An MI300A is an
APU, whose CPU and GPU share the same HBM memory. It is used
here as a GPU, with one MPI rank, and counts as one GPU. The single-GPU
results of Section~\ref{sec:singlegpuperf} also give the reduced computational time (RCT) of a full MI250X
card, a metric defined in Section~\ref{sec:perfometrics}.

\begin{table}[htbp]
\centering
\small
\begin{tabular}{@{}lll@{}}
\toprule
Device & Compiler & MPI \\
\midrule
Genoa & GCC 14.3.0 & Open MPI 4.1.8 \\
Rome (single node), Icelake & GCC 11.2.0 & Open MPI 4.1.1 \\
Rome (scaling) & GCC 14.1 & Open MPI 4.1.5.4 \\
Skylake & GCC 14.3.0 & Open MPI 4.1.7 \\
V100, A30, A100 & GCC 12.3.0 and CUDA 12.4 (\texttt{nvcc}) & Open MPI 5.0.10 \\
H100 & GCC 12.2.0 and CUDA 12.4.1 (\texttt{nvcc}) & Open MPI 4.1.6 \\
MI210 & ROCm 6.3.1 (\texttt{amdclang++}) & none (single GPU) \\
MI250X & ROCm 6.1.2 (\texttt{amdclang++}), Cray PE 24.07 & Cray MPICH 8.1.30 \\
MI300A & ROCm 6.2.1 (\texttt{amdclang++}), Cray PE 24.07 & Cray MPICH 8.1.30 \\
\bottomrule
\end{tabular}

\vspace{2pt}
\caption{Software stack of HOPPS for each device.}
\label{tab:sw-stack}
\end{table}

\begin{table}[htbp]
\centering
\small
\resizebox{\textwidth}{!}{\begin{tabular}{@{}lllll@{}}
\toprule
Label & CPU model & Cores/node & Memory/node & Bandwidth/node \\
\midrule
Skylake (SKX) & Intel Xeon Gold 6140 & $2\times18=36$ & 96\,GB DDR4 & 256\,GB/s \\
Icelake (ICX) & Intel Xeon Platinum 8368 & $2\times38=76$ & 256\,GB DDR4 & 410\,GB/s \\
Rome (Zen2), single node & AMD EPYC 7702 & $2\times64=128$ & 512\,GB DDR4 & 410\,GB/s \\
Rome (Zen2), scaling & AMD EPYC 7742 & $2\times64=128$ & 256\,GB DDR4-3200 & 410\,GB/s$^{\dagger}$ \\
Genoa (Zen4) & AMD EPYC 9654 & $2\times96=192$ & 384\,GB DDR5 & 920\,GB/s \\
\bottomrule
\end{tabular}}

\vspace{2pt}
\caption{CPU nodes used in Section~\ref{sec:perfoAnalysis}. Bandwidth: theoretical peak, from \cite{HOPPS}, $^{\dagger}$computed from 8 DDR4-3200 channels per socket. The Rome nodes labeled ``single node'' are used for the single-node measurements (see Section~\ref{sec:singlenodecpuperf}), those labeled ``scaling'' for the scaling study (see Section~\ref{sec:weakstrong-cpu}). The Rome ``scaling'' nodes are those of Belenos (Météo-France), the Skylake, Icelake and Rome ``single node'' are those of Kraken (CERFACS), and the Genoa nodes are those of Calypso (CERFACS).}
\label{tab:cpu-hw}
\end{table}

\begin{table}[htbp]
\centering
\small
\resizebox{\textwidth}{!}{\begin{tabular}{@{}llllll@{}}
\toprule
Label & Architecture & Memory & Bandwidth & L2 cache & Host node \\
\midrule
V100 & Volta & 16\,GB HBM2 & 897\,GB/s & 6\,MB & bi-socket node, 192\,GB RAM \\
A30 & Ampere & 24\,GB HBM2 & 933\,GB/s & 24\,MB & 4$\times$A30/node, 264\,GB RAM \\
A100 & Ampere & 40\,GB HBM2 & 1.56\,TB/s & 40\,MB & AMD Rome node, 512\,GB RAM \\
H100 & Hopper (SXM5) & 80\,GB HBM3 & 3.35\,TB/s & 50\,MB & 4$\times$H100/node, 4$\times$400\,Gb/s InfiniBand \\
MI210 & CDNA2 & 64\,GB HBM2e & 1.6\,TB/s & 8\,MB & 2$\times$MI210/node, AMD EPYC 9124, 384\,GB RAM \\
MI250X (1 GCD) & CDNA2 & 64\,GB HBM2e & 1.6\,TB/s & 8\,MB & 4$\times$MI250X/node, 4$\times$200\,Gb/s Slingshot \\
MI300A & CDNA3 (APU) & 128\,GB HBM3$^{\ddagger}$ & 5.3\,TB/s & 4\,MB$^{\ast}$ & 4$\times$MI300A/node, 4$\times$200\,Gb/s Slingshot \\
\bottomrule
\end{tabular}}

\vspace{2pt}
\caption{GPUs used in Section~\ref{sec:perfoAnalysis}. Bandwidth: theoretical peak, from \cite{HOPPS} and from the vendor specifications. MI250X: values per GCD (two GCD per card). $^{\ddagger}$Shared between the CPU and GPU parts of the APU. $^{\ast}$Per accelerator complex die (six per APU), backed by a 256\,MB Infinity Cache. The H100 nodes are those of Jean Zay (IDRIS), the MI250X and MI300A nodes those of Adastra (CINES), the MI210 nodes those of Calypso and the V100, A30 and A100 nodes those of Kraken (CERFACS).}
\label{tab:gpu-hw}
\end{table}

\subsection{Profiling}\label{sec:profiling}
\begin{figure}[htbp]
  \centering
  \includegraphics[width=\linewidth]{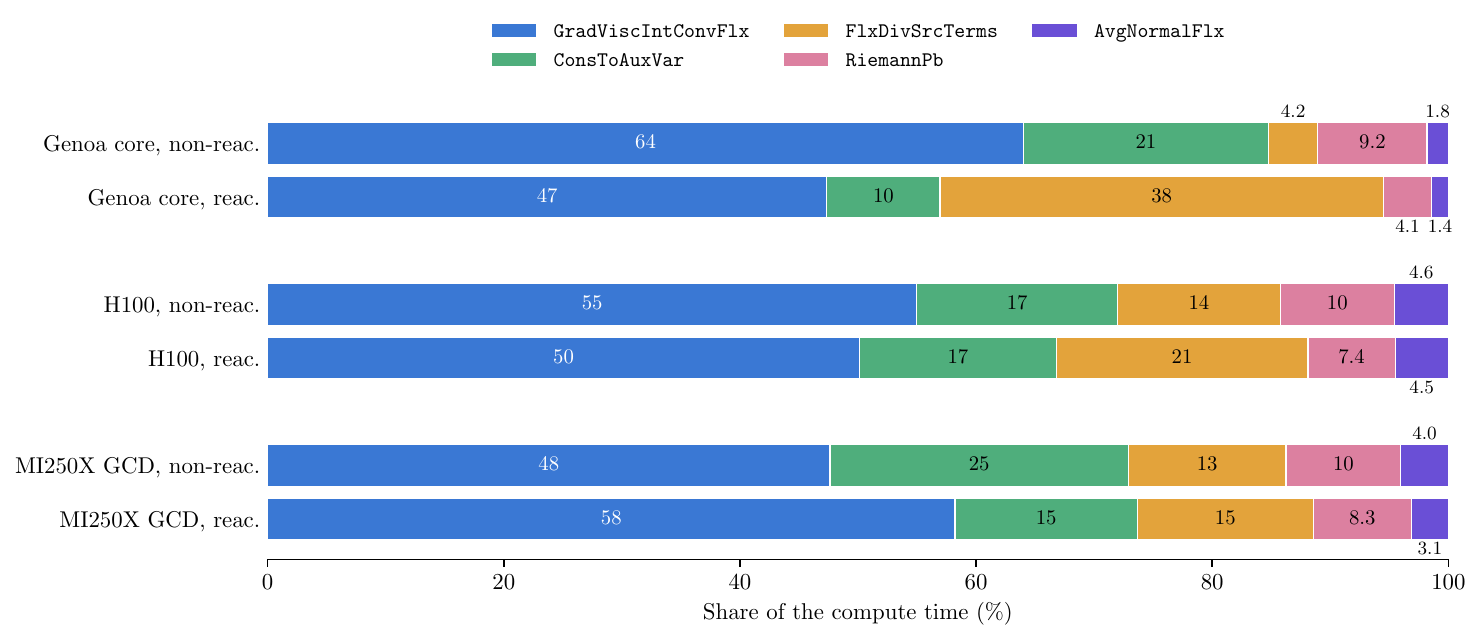}
  \caption{Share of the compute time spent in each group of kernels on one AMD Genoa core, on one Nvidia H100 GPU and on one GCD of an AMD MI250X, $p = 5$, $4.7 \times 10^6$ DOF. The three element-local groups come first, the two interface kernels last. Values in percent. Non-reac.: non-reacting case, reac.: reacting case.}
  \label{fig:prof}
\end{figure}

Figure~\ref{fig:prof} gives the share of the compute time spent in each group
of kernels at $p=5$ with $4.7\times10^{6}$ DOF, on one AMD Genoa core, on one
Nvidia H100 GPU and on one GCD of an AMD MI250X. The kernels form two groups.
The element-local kernels are the gradient with the interior convective and
viscous fluxes (\texttt{GradViscIntConvFlx}), which form a single kernel on CPU
and GPU (see Section~\ref{subsec:intragpu}), the reconstruction of the auxiliary variable (\texttt{ConsToAuxVar}) and
the flux divergence with the source terms (\texttt{FlxDivSrcTerms}). The
interface kernels are the Riemann solver (\texttt{RiemannPb}) and the averaged
viscous flux (\texttt{AvgNormalFlx}). The kernel times are measured with the
\texttt{simple-kernel-timer} of Kokkos Tools, over 100 iterations on CPU and
1000 iterations on GPU, where the profiler changes the RCT by about 1\,\%. The
update of the RK stages and the conversion to primitive variables, outside
these groups, take 4.7 to 5.5\,\% of the time loop on the Genoa core, 6.5\,\%
on H100 and 11 to 13\,\% on MI250X.

On CPU (Fig.~\ref{fig:prof}), the element-local kernels take 88.9\,\% of the
time in the non-reacting case and 94.5\,\% in the reacting case.
\texttt{GradViscIntConvFlx} dominates the non-reacting case with 64.0\,\%. With
chemistry, the share of \texttt{FlxDivSrcTerms}, which evaluates the chemical
source terms, rises from 4.2\,\% to 37.5\,\%: the evaluation of the reaction
rates becomes the largest addition to the right-hand side.

On GPU, the element-local kernels also dominate, with 85.8\,\% and 88.0\,\% of
the time on H100 and 86.3\,\% and 88.5\,\% on MI250X, in the non-reacting and
reacting cases. \texttt{GradViscIntConvFlx} alone takes about half of the time.
The interface kernels take 14.2\,\% (H100) and 13.7\,\% (MI250X) of the time in
the non-reacting case, against 11.1\,\% on CPU, and 11.9\,\% and 11.4\,\% in
the reacting case, against 5.5\,\%. These kernels follow the indirect
addressing of the interfaces and weigh more on GPU than on one CPU core, even
though the sorting of the FP pairs (see Section~\ref{subsec:intragpu}) keeps
the accesses of neighboring threads close in memory. With chemistry, the share of \texttt{FlxDivSrcTerms} rises from
13.8\,\% to 21.3\,\% on H100 and from 13.4\,\% to 14.9\,\% on MI250X, far less
than on CPU.

The absolute kernel times (see Table~\ref{tab:kernel-times}) separate the two
sources of the extra cost of the reacting case: the evaluation of the reaction
rates, in \texttt{FlxDivSrcTerms}, and the transport of 14 variables instead of
8, in all the other kernels. On one Genoa core, the reacting case costs 2.5
times the non-reacting one. \texttt{FlxDivSrcTerms} becomes 23 times more
expensive and accounts for 59\,\% of the extra time, while the other kernels
together cost 1.7 times more. On one H100, the reacting case costs 2.0 times
the non-reacting one, \texttt{FlxDivSrcTerms} costs 3.1 times more and accounts
for 29\,\% of the extra time, and the other kernels together cost 1.9 times
more. On one GCD of MI250X, the reacting case costs 2.4 times the non-reacting
one, \texttt{FlxDivSrcTerms} costs 2.7 times more and accounts for 16\,\% of
the extra time, and the other kernels together cost 2.4 times more. On GPU, the
extra cost therefore comes mostly from the larger number of transported
variables, whereas on a single CPU core the chemistry remains a large part of
the cost. On GPU, the transport
kernels cost more than the ratio $14/8=1.75$ of the number of variables, since
the species diffusion flux and the correction velocity add work per species.
On MI250X, \texttt{GradViscIntConvFlx} costs 3.0 times more in the reacting
case, against 1.8 times on H100. With nine species, the gradients of this kernel
exceed the shared memory available per team (see Section~\ref{subsec:intragpu}).

\begin{table}[htbp]
\centering
\small
\begin{tabular}{@{}lrrrrrr@{}}
\toprule
 & \multicolumn{2}{c}{Genoa, 1 core (s)} & \multicolumn{2}{c}{H100 (s)} & \multicolumn{2}{c}{MI250X, 1 GCD (s)} \\
Kernel & non-reac. & reac. & non-reac. & reac. & non-reac. & reac. \\
\midrule
\texttt{GradViscIntConvFlx} & 1256.0 & 2362.2 & 20.39 & 37.69 & 45.02 & 133.98 \\
\texttt{ConsToAuxVar} & 406.9 & 478.6 & 6.31 & 12.52 & 23.89 & 35.52 \\
\texttt{FlxDivSrcTerms} & 82.0 & 1872.9 & 5.11 & 16.03 & 12.63 & 34.30 \\
\texttt{RiemannPb} & 181.4 & 204.8 & 3.58 & 5.55 & 9.18 & 19.11 \\
\texttt{AvgNormalFlx} & 35.5 & 70.1 & 1.70 & 3.39 & 3.81 & 7.21 \\
\midrule
Total & 1961.9 & 4988.6 & 37.08 & 75.18 & 94.52 & 230.13 \\
\bottomrule
\end{tabular}

\vspace{2pt}
\caption{Time spent in each group of kernels, $p=5$, $4.7\times10^6$ DOF, 100 iterations on CPU and 1000 iterations on GPU.}
\label{tab:kernel-times}
\end{table}

\subsection{Performance metrics}\label{sec:perfometrics}
Following the methodology of the first HOPPS article \cite{HOPPS}, the main metric
is the RCT, denoted $\mathcal{R}$,
\begin{equation}
\mathcal{R} = \frac{T_\mathrm{solve}\,N_\mathrm{cu}}{N_\mathrm{dof}\,N_\mathrm{it}\,N_\mathrm{st}},
\label{eq:rct}
\end{equation}
where $T_\mathrm{solve}$ is the wall-clock time of the time loop, without
initialization and without output, measured over 100 iterations on CPU and 1000
iterations on GPU, $N_\mathrm{dof}$ the total number of DOF, $N_\mathrm{it}$ the
number of iterations of the loop, $N_\mathrm{st}$ the number of RK stages per
iteration and $N_\mathrm{cu}$ the number of compute units (CU), that
is, CPU cores or GPUs. The RCT is the cost of updating one DOF for one stage on
one CU, and the
lower, the better. It depends on the local load and on the hardware, which is
what the following sections analyze. For scaling studies, the speedup $S$ and
the parallel efficiency $E$ are computed with respect to a reference run on
$N_\mathrm{ref}$ CU,
\begin{equation}\label{eq:speedup-efficiency}
    S(N_\mathrm{cu}) = \dfrac{N_\mathrm{ref}\,T_\mathrm{solve}(N_\mathrm{ref})}{T_\mathrm{solve}(N_\mathrm{cu})}, \qquad E(N_\mathrm{cu}) = \dfrac{S(N_\mathrm{cu})}{N_\mathrm{cu}}.
\end{equation}
The RCT is used in Sections~\ref{sec:perfometrics} to \ref{sec:singlegpuperf},
and $S$ and $E$ in the scaling studies of Sections~\ref{sec:weakstrong-cpu} and
\ref{sec:weakstrong-gpu}.

\begin{figure}[htbp]
  \centering
  \begin{subfigure}[b]{0.48\textwidth}\centering
\includegraphics[width=\linewidth]{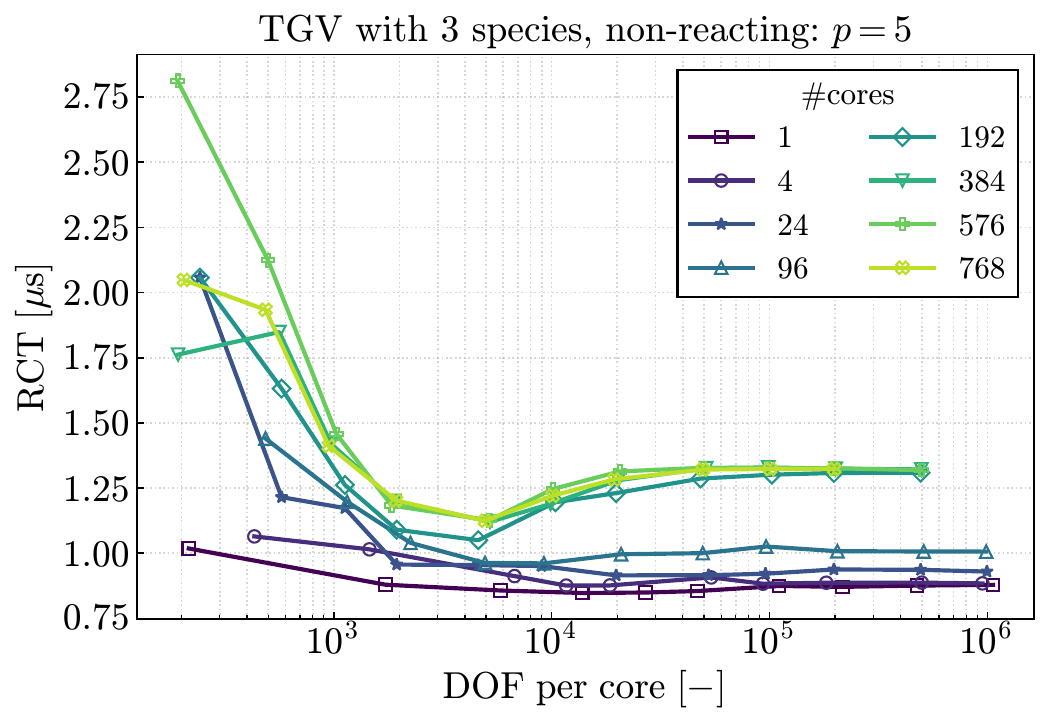}

\end{subfigure}
  \begin{subfigure}[b]{0.48\textwidth}\centering
\includegraphics[width=\linewidth]{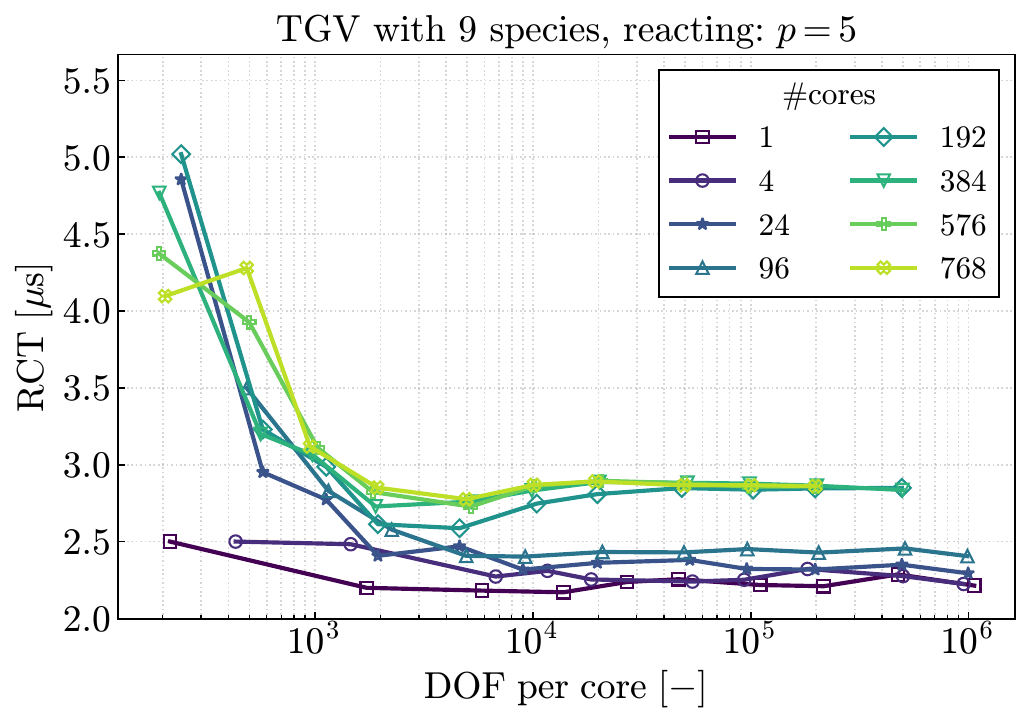}

\end{subfigure}
\caption{RCT versus DOF per core on AMD Genoa nodes, $p=5$. Left: non-reacting case. Right: reacting case.}
  \label{fig:rct_cpu_zen4}
\end{figure}

\begin{figure}[htbp]
  \centering
  \begin{subfigure}[b]{0.48\textwidth}\centering
\includegraphics[width=\linewidth]{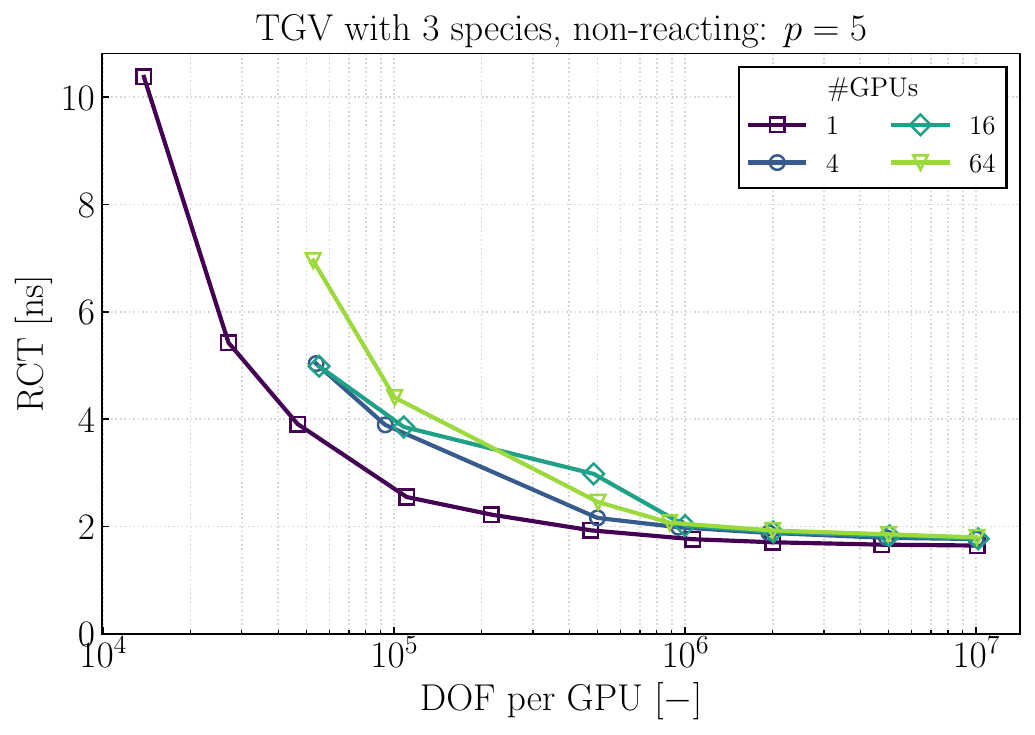}

\end{subfigure}
  \begin{subfigure}[b]{0.48\textwidth}\centering
\includegraphics[width=\linewidth]{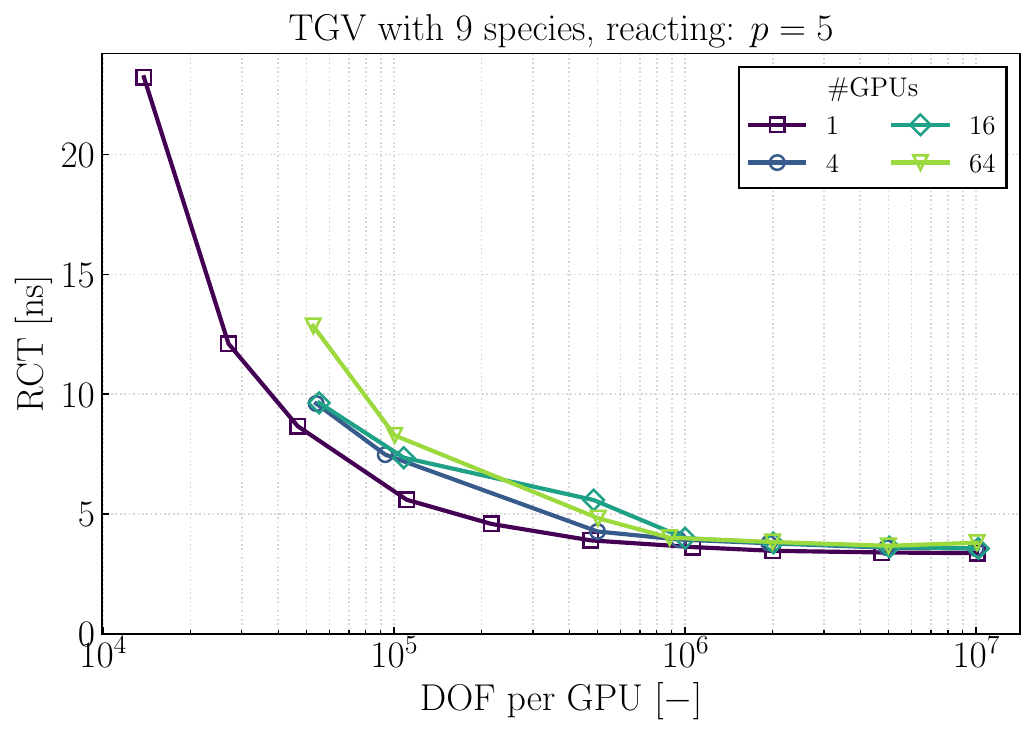}

\end{subfigure}
\caption{RCT versus DOF per GPU on 1 to 64 Nvidia H100 GPUs, $p=5$. Left: non-reacting case. Right: reacting case.}
  \label{fig:rct_gpu_h100_multi}
\end{figure}

Figures~\ref{fig:rct_cpu_zen4} and \ref{fig:rct_gpu_h100_multi} show the RCT versus
the number of DOF per CU (DOF/CU) for the non-reacting and reacting cases at $p=5$, on
AMD Genoa nodes and on 1 to 64 Nvidia H100 GPUs. On CPU, the RCT is high at low DOF/CU,
where each core has too little work to hide latencies. It then decreases to a
minimum, rises again, and reaches a plateau beyond about $10^5$ DOF/CU for all
core counts. These two regimes are analyzed on a single node in
Section~\ref{sec:singlenodecpuperf}.

On GPU, the RCT also decreases from a starvation regime at low DOF/GPU. On one
H100, it then decreases slowly with the load, and reaches a plateau beyond about
$10^6$ DOF (see Section~\ref{sec:singlegpuperf}). The multi-GPU curves lie above the
single-GPU one, and below $10^6$ DOF/GPU the gap grows with the number of GPUs:
it measures the cost of the communications, analyzed with the scaling studies
(see Section~\ref{sec:weakstrong-gpu}). At $10^7$ DOF/GPU, the RCT on 4 to 64 GPUs
is $1.77$ to $1.80\,\mathrm{ns}$ in the non-reacting case and $3.56$ to
$3.80\,\mathrm{ns}$ in the reacting case, against $1.65$ and $3.37\,\mathrm{ns}$
on one GPU.

On Genoa nodes, beyond $10^5$ DOF per core, the RCT is about
$1.3\,\mu\mathrm{s}$ in the non-reacting case and $2.85\,\mu\mathrm{s}$ in the
reacting case from 192 to 768 cores. On one H100 GPU, it ranges from $1.65$ to
$2.22\,\mathrm{ns}$ in the non-reacting case and from $3.37$ to $4.59\,\mathrm{ns}$
in the reacting case between $2\times10^5$ and $10^7$ DOF. The ratio of about two between the
two cases combines the transport of 14 variables instead of 8 and the chemistry
itself, in proportions that differ between CPU and GPU
(see Section~\ref{sec:profiling}).

\subsection{Gain over JAGUAR}\label{sec:gainjaguar}
\begin{figure}[htbp]
  \centering
  \begin{subfigure}[b]{0.48\textwidth}\centering
\includegraphics[width=\linewidth]{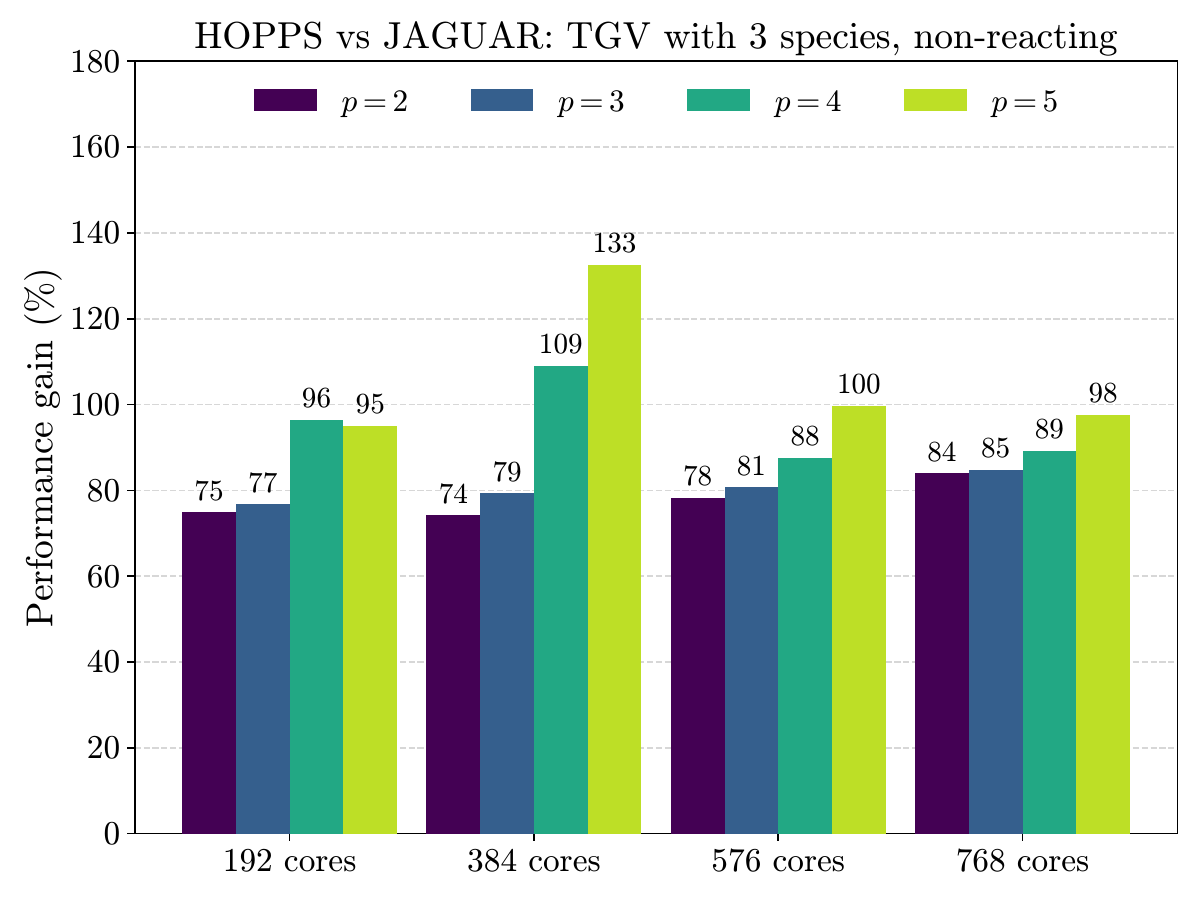}
\end{subfigure}
  \begin{subfigure}[b]{0.48\textwidth}\centering
\includegraphics[width=\linewidth]{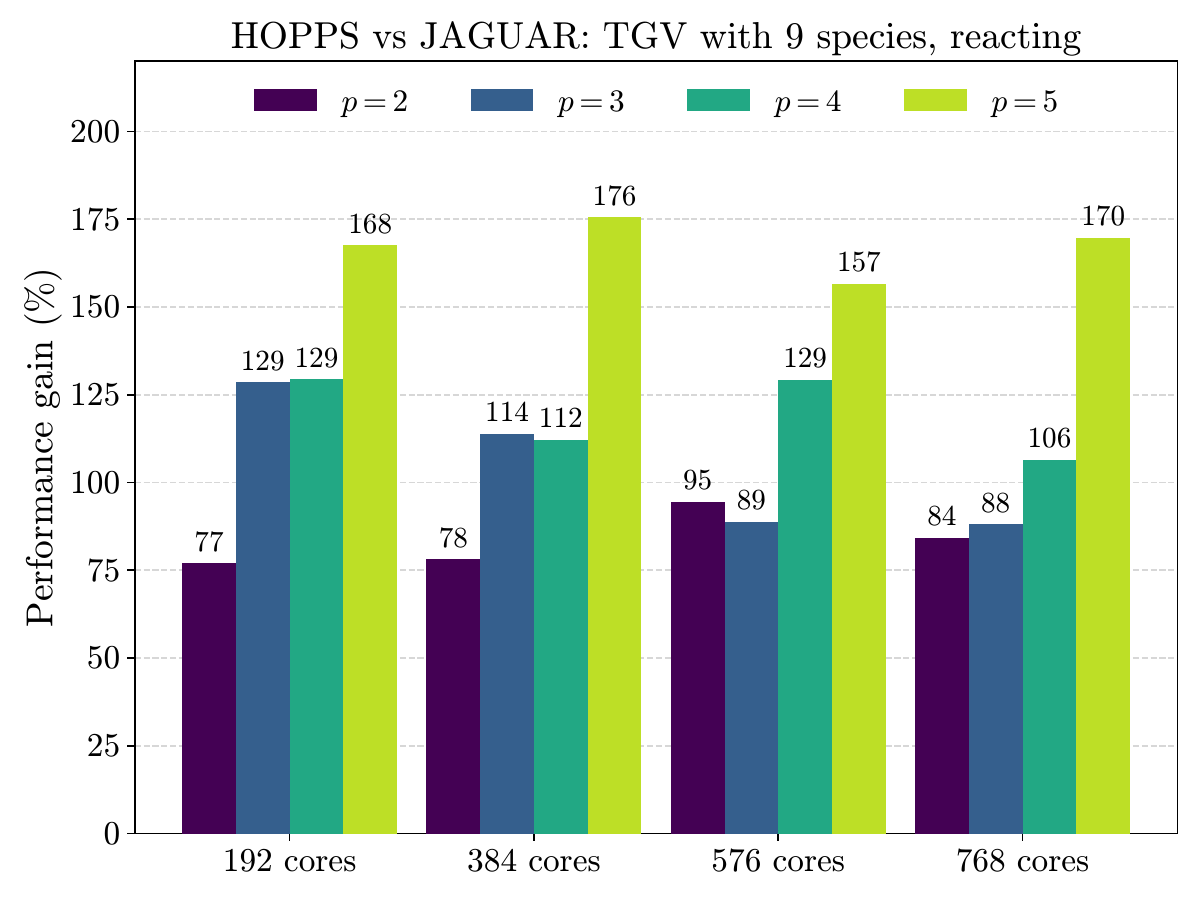}
\end{subfigure}
\caption{Gain of HOPPS over JAGUAR, $(\mathcal{R}_\mathrm{JAGUAR}/\mathcal{R}_\mathrm{HOPPS}-1)\times 100$, on AMD Genoa nodes. Left: non-reacting case. Right: reacting case.}
  \label{fig:gain_hopps_vs_jaguar}
\end{figure}

Figure~\ref{fig:gain_hopps_vs_jaguar} gives the gain of HOPPS over JAGUAR,
$(\mathcal{R}_\mathrm{JAGUAR}/\mathcal{R}_\mathrm{HOPPS} - 1)\times100$, on Genoa
nodes, for four core counts (192 to 768) and four polynomial degrees ($p=2$ to
5). The mesh is the same for all $p$ ($N=64$): the number of DOF per core
grows by a factor of 8 from $p=2$ to $p=5$. HOPPS is faster in all cases. The
gain is at least 74\,\% in the non-reacting case and 77\,\% in the reacting
case, and HOPPS is up to 2.8 times faster than JAGUAR. The gain is larger in the reacting case for
all 16 configurations, and the largest gains are obtained at $p=5$ in the
reacting case (168\,\%, 176\,\%, 157\,\% and 170\,\% for the four core counts).
The gain grows with $p$ because the RCT of HOPPS decreases faster with $p$ than
that of JAGUAR in the non-reacting case, whereas in the reacting case the RCT of
JAGUAR increases with $p$. At a given $p$, the RCT of HOPPS varies by at most
3\,\% between 192 and 768 cores, except at $p=2$ in the non-reacting case, where
it decreases with the load per core (9\,\%), and at $p=3$ on 768 cores in the
reacting case (8\,\%). That of JAGUAR varies by up to 20\,\%, without trend
with the core count. The departures from a monotonic growth of the gain with $p$,
such as the gain of 133\,\% at $p=5$ on 384 cores in the non-reacting case, come
from this variability of the JAGUAR runs.
Both codes solve the same equations with the same numerical choices: RK(5,4)
time integration, HLLC Riemann solver, BR1 viscous scheme and JANAF
thermochemical tables. Both run in pure MPI with one rank per core,
and JAGUAR reports the same metric as Eq.~\eqref{eq:rct} up to the factor
$N_\mathrm{st}=5$: its time loop without output, per iteration, per SP and per process. The
runs of Fig.~\ref{fig:gain_hopps_vs_jaguar} use JAGUAR compiled with GCC 11.2.0 and
\texttt{-O3}, without architecture flag, and HOPPS compiled with GCC 14.3.0 and \texttt{-O3} for
the Zen4 architecture (see Table~\ref{tab:sw-stack}). With JAGUAR compiled with the
same compiler and options as HOPPS, the gains of
Fig.~\ref{fig:gain_hopps_vs_jaguar} change by less than 2\,\%: the gain does not come from the compiler.

Several differences in the implementation probably explain this gain. In
JAGUAR, the number of species is fixed at compile time, as in HOPPS, but the
polynomial degree is not: it may vary from one element to the next, and the loops
read the degree and the data offsets of each element at run time. In HOPPS, the
degree is also known at compile time (see Section~\ref{sec:hopps-arch-impl}). JAGUAR copies the solution at the
interface FP into face arrays before the Riemann solver, and copies the fluxes
back afterwards, whereas the interface kernels of HOPPS read and write the FP
arrays directly through the FP pairs. At each interface FP, JAGUAR also
evaluates the thermochemical tables for the left and right states, which HOPPS
avoids by interpolating the auxiliary variable (see Section~\ref{sec:Auxvariable}).
Most of these differences reduce the memory traffic per DOF, which limits the
RCT on a full node (see Section~\ref{sec:singlenodecpuperf}).

\subsection{Single-node CPU performance}\label{sec:singlenodecpuperf}
\begin{figure}[p]
  \centering
  \begin{subfigure}[b]{\textwidth}\centering\phantomcaption\label{fig:rct_cpu_zen4_one_node}%
\rotatebox{90}{\hspace{1.2cm}\small (a) AMD Genoa}\hspace{0.01\textwidth}%
\includegraphics[width=0.45\textwidth]{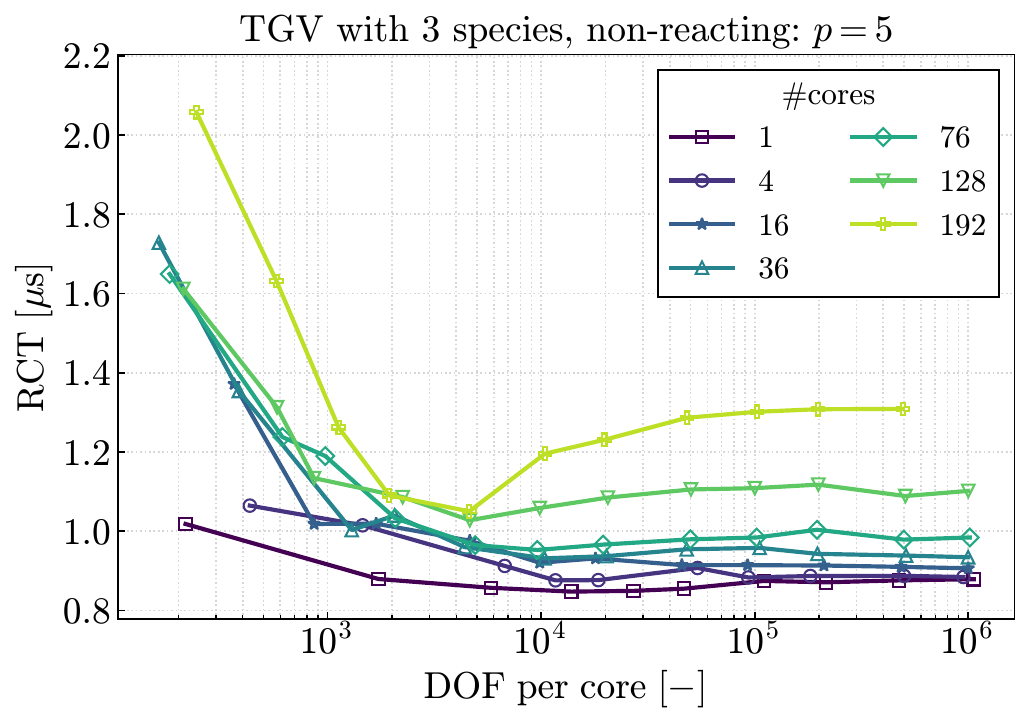}\hspace{0.01\textwidth}%
\includegraphics[width=0.45\textwidth]{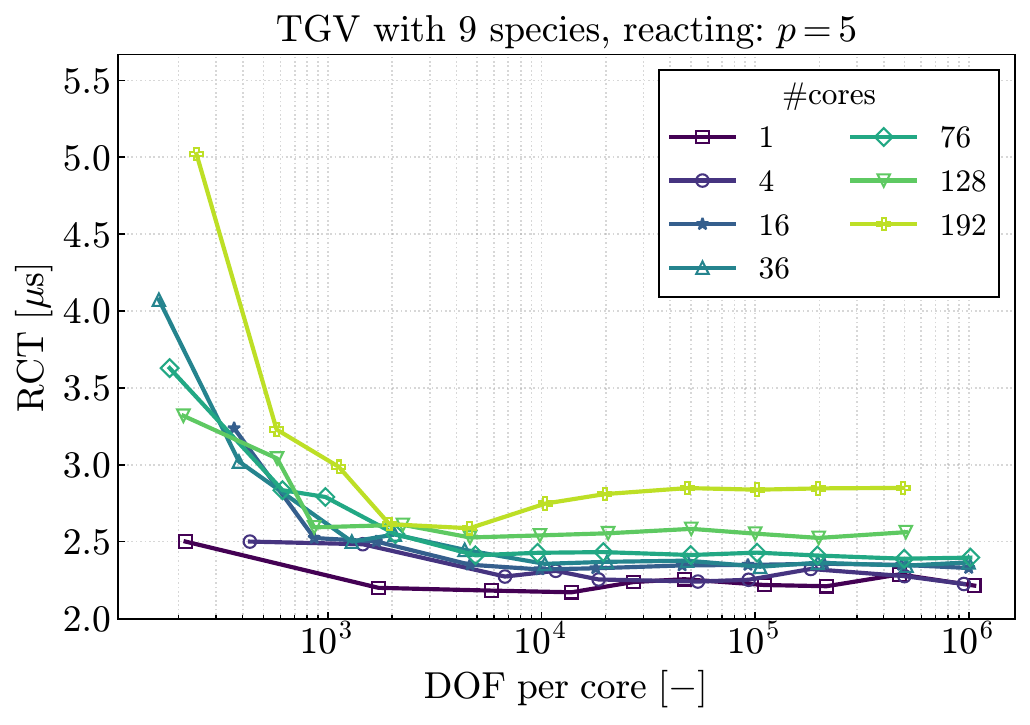}
  \end{subfigure}

  \begin{subfigure}[b]{\textwidth}\centering\phantomcaption\label{fig:rct_cpu_icx}%
\rotatebox{90}{\hspace{1.2cm}\small (b) Intel Icelake}\hspace{0.01\textwidth}%
\includegraphics[width=0.45\textwidth]{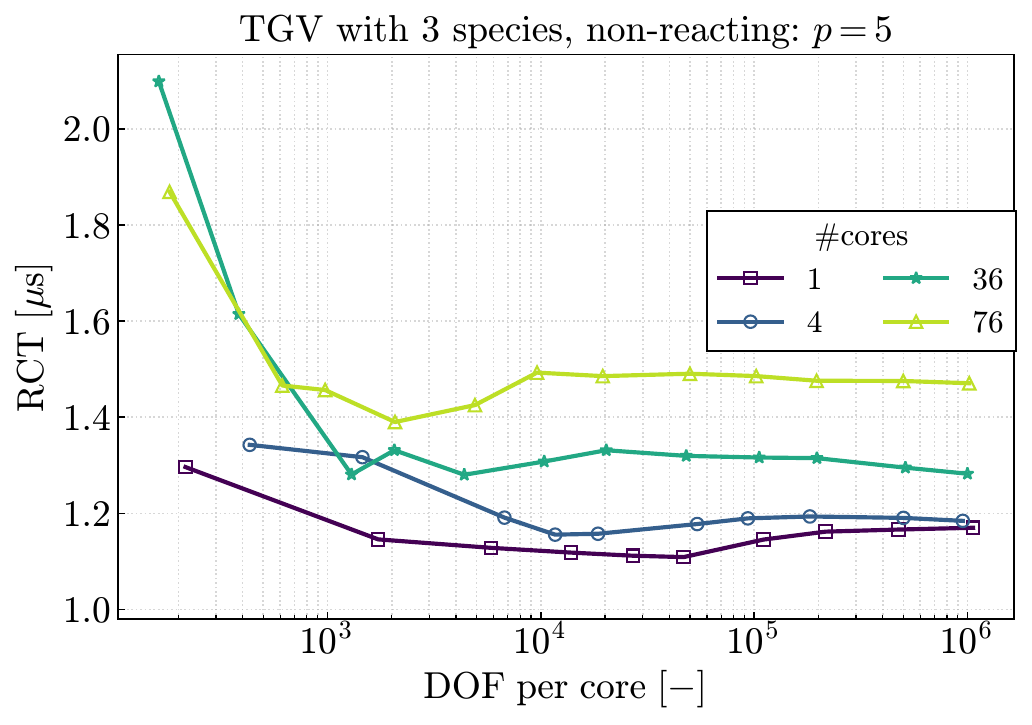}\hspace{0.01\textwidth}%
\includegraphics[width=0.45\textwidth]{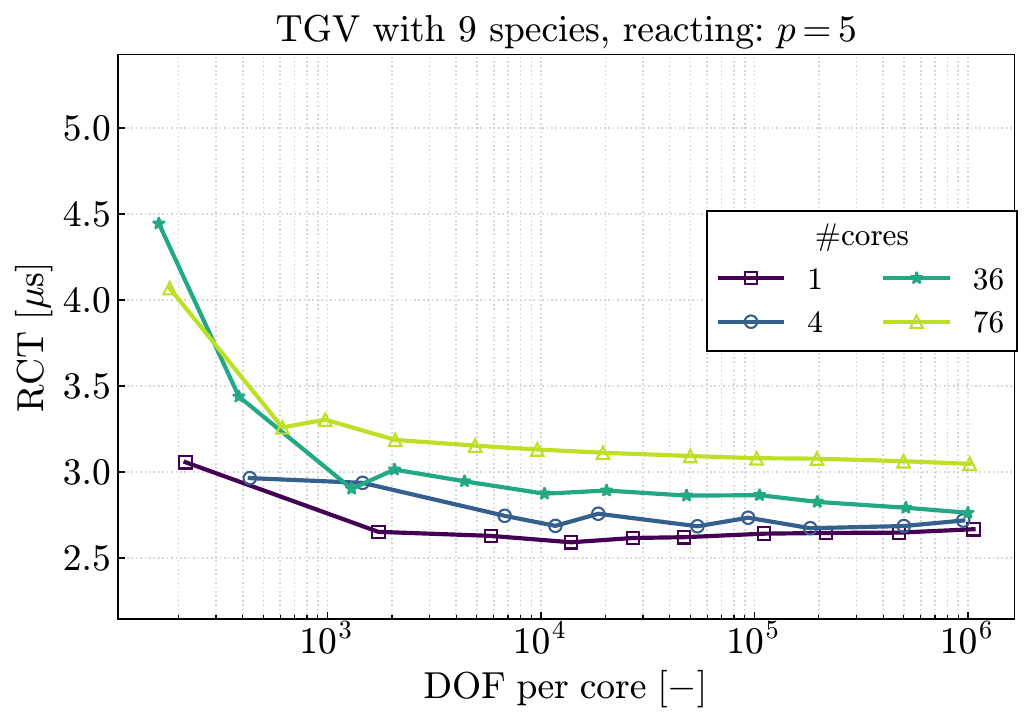}
  \end{subfigure}

  \begin{subfigure}[b]{\textwidth}\centering\phantomcaption\label{fig:rct_cpu_skx}%
\rotatebox{90}{\hspace{1.2cm}\small (c) Intel Skylake}\hspace{0.01\textwidth}%
\includegraphics[width=0.45\textwidth]{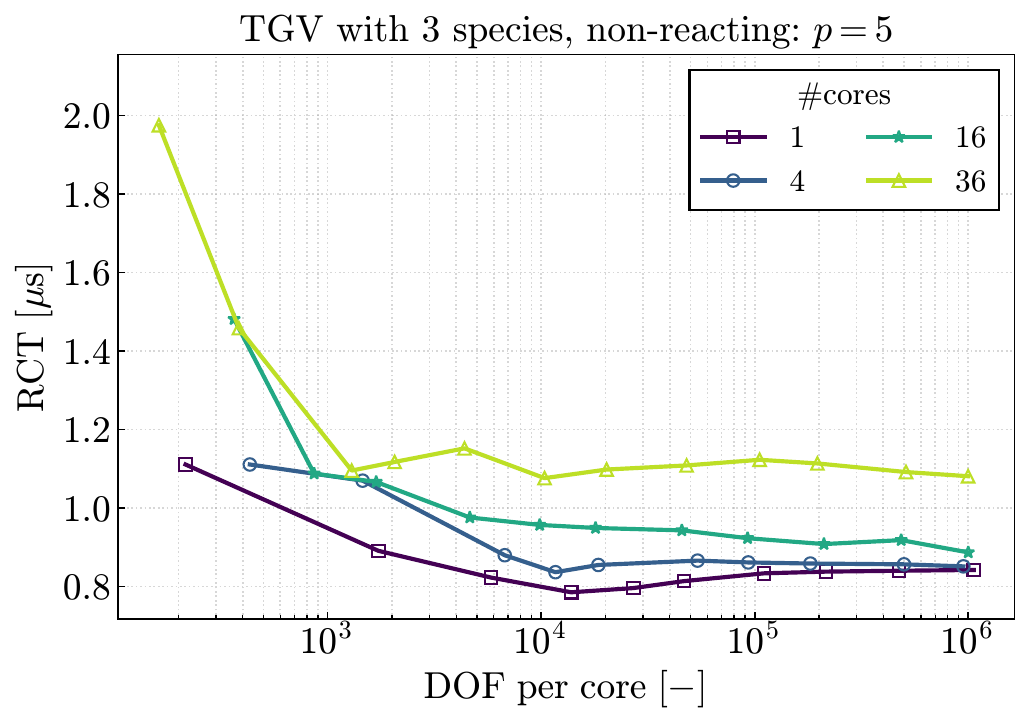}\hspace{0.01\textwidth}%
\includegraphics[width=0.45\textwidth]{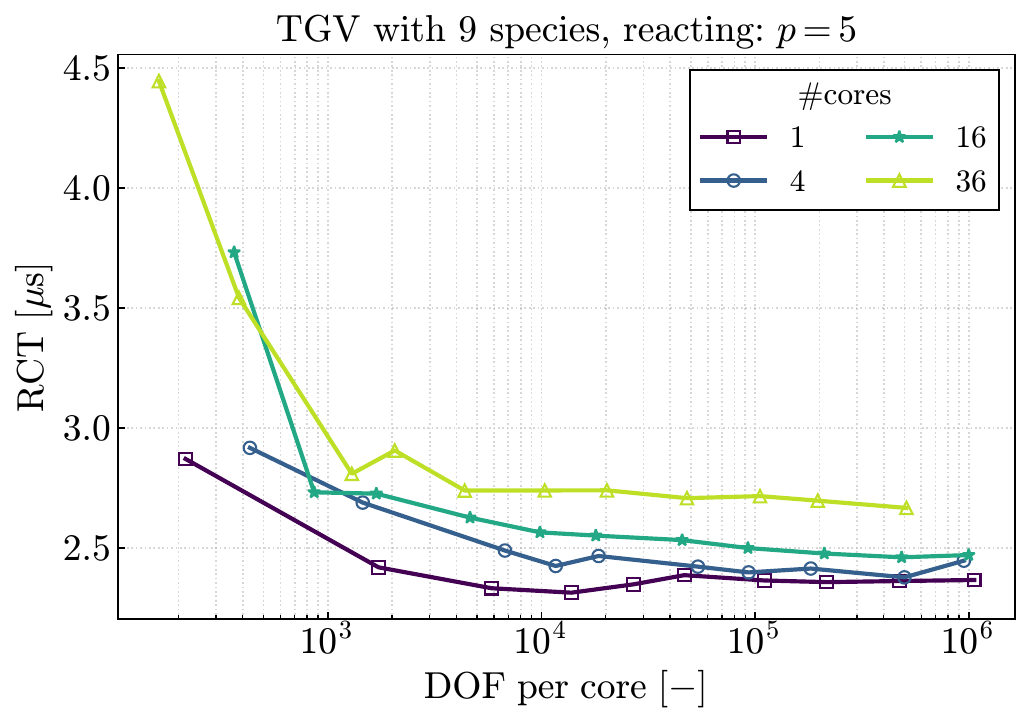}
  \end{subfigure}

  \begin{subfigure}[b]{\textwidth}\centering\phantomcaption\label{fig:rct_cpu_zen2}%
\rotatebox{90}{\hspace{1.2cm}\small (d) AMD Rome}\hspace{0.01\textwidth}%
\includegraphics[width=0.45\textwidth]{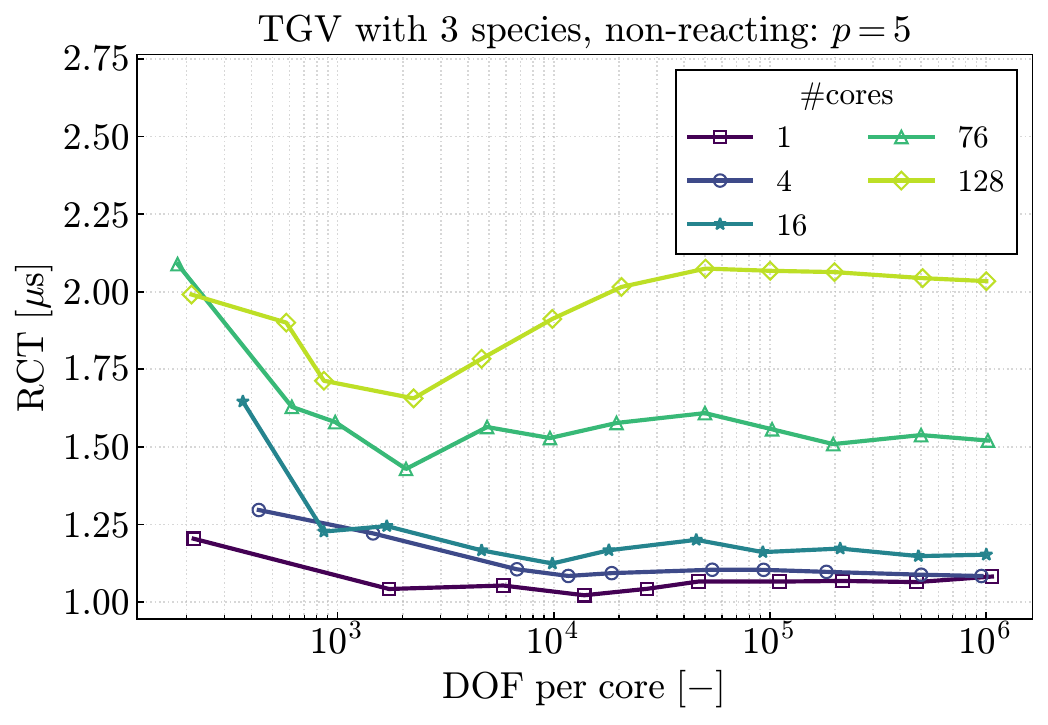}\hspace{0.01\textwidth}%
\includegraphics[width=0.45\textwidth]{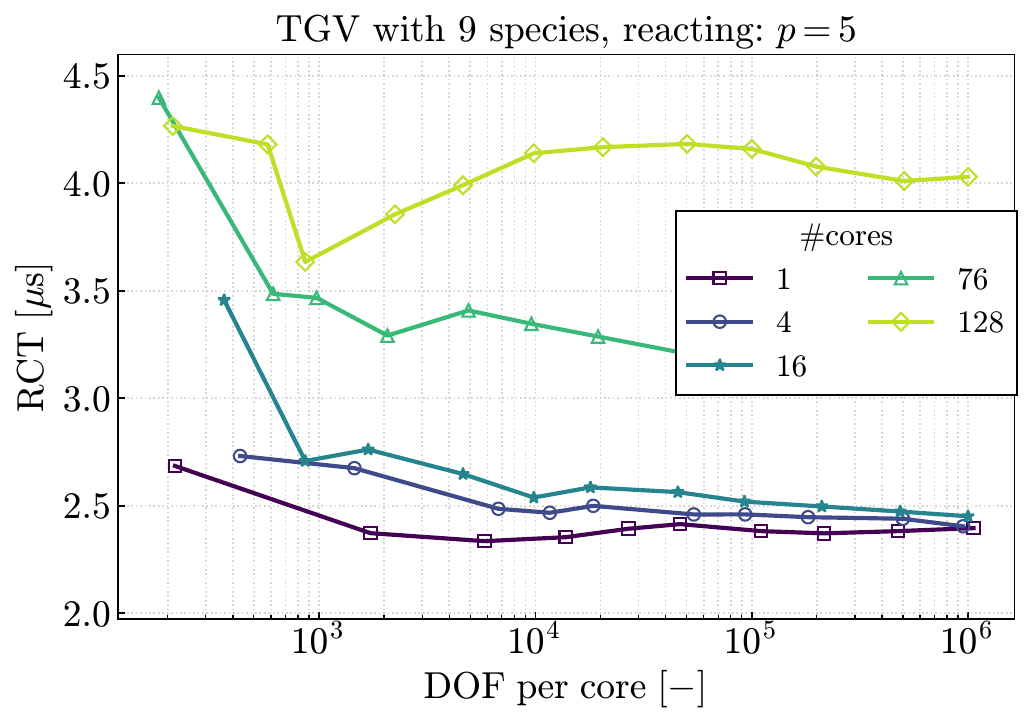}
  \end{subfigure}
\caption{RCT versus DOF per core on one node of each CPU type, $p=5$: (a) AMD Genoa, (b) Intel Icelake, (c) Intel Skylake, (d) AMD Rome. Left: non-reacting case. Right: reacting case.}
  \label{fig:rct_cpu_node}
\end{figure}

Figure~\ref{fig:rct_cpu_node} gives the RCT
on one node of each CPU type: AMD Genoa, Intel Icelake and Skylake, and AMD
Rome. Table~\ref{tab:rct-cpu-node} gathers the full-node values. On all four,
the RCT reaches a value within a few percent of its large-size value beyond about
$10^5$ DOF per core, and this value increases with the number of active cores.
On one Genoa node (Fig.~\ref{fig:rct_cpu_zen4_one_node}), in the non-reacting
case, it goes from about $0.88\,\mu\mathrm{s}$ on one core to $0.98\,\mu\mathrm{s}$
on 76 cores, $1.10\,\mu\mathrm{s}$ on 128 cores and $1.31\,\mu\mathrm{s}$ on the
full node, which is also the value of the multi-node runs up to 768 cores
(Fig.~\ref{fig:rct_cpu_zen4}). This rise with the number of active cores, at the
same load per core, points to the saturation of the memory bandwidth shared by
the cores of a node, rather than to the cost of MPI communications. At smaller
loads, from 128 cores onward, the RCT on Genoa reaches its minimum around
$5\times10^3$ DOF per core, where the data of the node fit in the last-level
cache. The same band, 3000 to 10\,000 DOF per
rank, is reported as the performance optimum of the CPU solver FLEXI
\cite{FLEXI,GALAEXI}.

\begin{table}[htbp]
\centering
\small
\resizebox{\textwidth}{!}{\begin{tabular}{@{}lrrrrrrr@{}}
\toprule
Node & Cores & \multicolumn{2}{c}{Bandwidth (GB/s)} & \multicolumn{2}{c}{$\mathcal{R}$ ($\mu$s)} & \multicolumn{2}{c}{$\mathcal{R}/N_\mathrm{cores}$ (ns)} \\
\cmidrule(lr){3-4}\cmidrule(lr){5-6}\cmidrule(l){7-8}
 & & node & per core & non-reacting & reacting & non-reacting & reacting \\
\midrule
Skylake & 36 & 256 & 7.1 & 1.08 & 2.67 & 30.0 & 74.1 \\
Icelake & 76 & 410 & 5.4 & 1.47 & 3.05 & 19.4 & 40.1 \\
Genoa & 192 & 920 & 4.8 & 1.31 & 2.85 & 6.8 & 14.8 \\
Rome & 128 & 410 & 3.2 & 2.03 & 4.03 & 15.9 & 31.5 \\
\bottomrule
\end{tabular}}

\vspace{2pt}
\caption{RCT on a full CPU node beyond $10^5$ DOF per core, $p=5$. Bandwidth: theoretical peak of the node (see Table~\ref{tab:cpu-hw}), and the same divided by the number of cores. $\mathcal{R}/N_\mathrm{cores}$: time to update one DOF for one RK stage on the whole node.}
\label{tab:rct-cpu-node}
\end{table}

For a given core count, Genoa has the lowest RCT in the reacting case, and in
the non-reacting case from 36 cores onward. Below 36 cores, Skylake is up to 5\,\% lower. On a
full node, the RCT is lowest on Skylake, which has the largest bandwidth per
core, and highest on Rome, which has the smallest. The RCT is normalized by the number of cores and therefore
favors the nodes with few cores. To compare the nodes as a whole, which matters
when the resource is counted in node hours, Table~\ref{tab:rct-cpu-node} also
gives $\mathcal{R}/N_\mathrm{cores}$, the time the whole node takes to update
one DOF for one RK stage. This cost follows the bandwidth of the node. In the
non-reacting case, a Genoa node is 2.3 times faster than a Rome node and 2.8
times faster than an Icelake node, for a bandwidth 2.2 times higher, and 4.4
times faster than a Skylake node, for a bandwidth 3.6 times higher. In the
reacting case, the full-node RCT is 2.0 to 2.5 times the non-reacting one, and
these ratios become 2.1, 2.7 and 5.0.

\subsection{Single-GPU performance}\label{sec:singlegpuperf}
\begin{figure}[htbp]
  \centering
  \begin{subfigure}[b]{0.48\textwidth}\centering
\includegraphics[width=\linewidth]{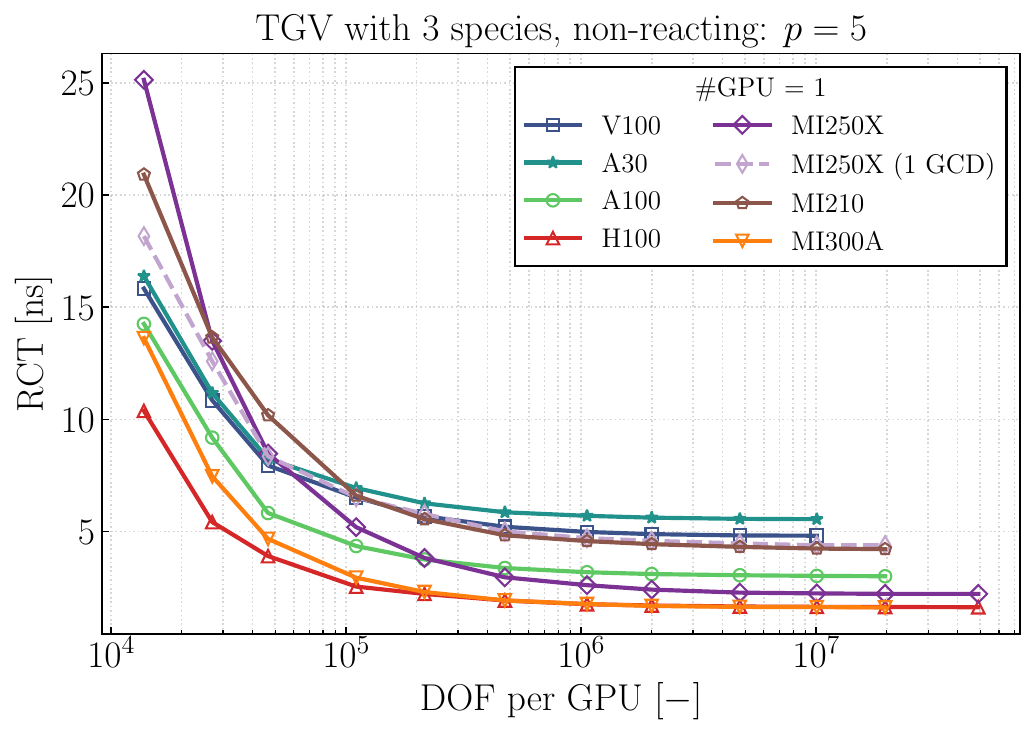}
\end{subfigure}
  \begin{subfigure}[b]{0.48\textwidth}\centering
\includegraphics[width=\linewidth]{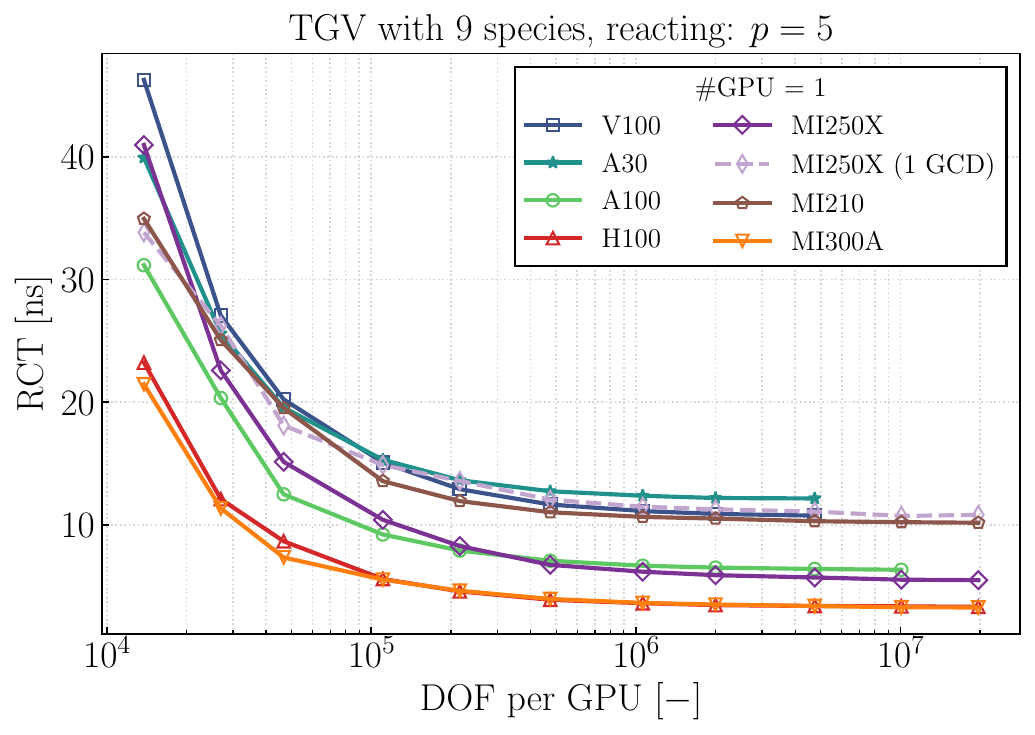}
\end{subfigure}
\caption{RCT versus DOF per GPU on four Nvidia GPU generations and three AMD GPUs, $p=5$. MI250X: full card (two GCD, solid line) and one GCD (dashed line). Left: non-reacting case. Right: reacting case.}
  \label{fig:rct_gpu}
\end{figure}

Figure~\ref{fig:rct_gpu} gives the RCT on four Nvidia GPUs from three
generations and on three AMD GPUs, and Table~\ref{tab:rct-gpu} gathers its values
beyond $10^6$ DOF. For the MI250X, the figure gives both one GCD and the full
card, whose two GCD are driven by two MPI ranks. The RCT of the card is computed
with $N_\mathrm{cu}=1$, so that it compares with that of the other GPUs. On all
GPUs, the RCT decreases with the load and reaches a plateau beyond about $10^6$
DOF. H100 and MI300A have the lowest values, within 2\,\% of each other at
$4.7\times10^6$ DOF, followed by the MI250X card and A100, then by MI210, one GCD
of MI250X, V100 and A30. Compared with the full-node values of
Table~\ref{tab:rct-cpu-node}, one H100 or one MI300A updates a DOF 3.8 to 4.5
times faster than a Genoa node with 192 cores, in both cases.

Among the Nvidia GPUs, this ranking follows the memory bandwidth. At
$4.7\times10^6$ DOF, A100 is 1.8 to 1.9 times faster than A30, for a bandwidth
1.7 times higher, and H100 is 1.8 to 1.9 times faster than A100, for a bandwidth
2.1 times higher. The AMD GPUs draw less from their bandwidth: one GCD of MI250X
is 1.1 to 1.3 times faster than an A30, for a bandwidth 1.7 times higher, and
MI300A matches H100 with a bandwidth 1.6 times higher. The two GCD of an MI250X
card give 1.94 to 1.96 times the throughput of one GCD, so that the card is 1.1
to 1.3 times faster than an A100. For the MI250X, the gap with the Nvidia GPUs is
larger in the reacting case.

The GPU kernels of Section~\ref{subsec:intragpu}, with the fields stored by
element and the gradients kept in shared memory, matter most on the AMD GPUs. The MI210 has one CDNA2 die, with the memory and the L2 cache of
one GCD of MI250X (see Table~\ref{tab:gpu-hw}), and its RCT at $4.7\times10^6$ DOF is 3 to 8\,\% lower than
that of one GCD. Between $10^6$ and $2\times10^7$ DOF, these kernels divide its
RCT by 2.3 to 2.7 in the non-reacting case and by 2.7 to 3.0 in the reacting
case, compared with the former kernels of the same commit, which store the
element index fastest and write the gradients to global memory. The measured
gain combines the two changes. At
$4.7\times10^6$ DOF, the time per iteration drops from 233 to 102\,ms and from
650 to 244\,ms. On an A100, the former and the new kernels remain within
4\,\% of each other. One GCD of MI250X has 8\,MB of L2 cache,
against 40\,MB on an A100 and 50\,MB on an H100 (see Table~\ref{tab:gpu-hw}), and
depends more on the reuse of data in shared memory.
The kernel profiles locate the larger gap of the MI250X in the reacting case
(see Section~\ref{sec:profiling}): from the non-reacting to the reacting case,
\texttt{GradViscIntConvFlx} costs 3.0 times more on one GCD of MI250X and 1.8
times more on an H100.

\begin{table}[htbp]
\centering
\small
\begin{tabular}{@{}lrrr@{}}
\toprule
GPU & Bandwidth & \multicolumn{2}{c}{$\mathcal{R}$ (ns)} \\
 & (TB/s) & non-reacting & reacting \\
\midrule
V100 & 0.90 & 4.8 to 5.0 & 10.8 to 11.1 \\
A30 & 0.93 & 5.6 to 5.7 & 12.2 to 12.4 \\
A100 & 1.56 & 3.0 to 3.2 & 6.4 to 6.7 \\
H100 & 3.35 & 1.6 to 1.8 & 3.3 to 3.6 \\
MI210 & 1.6 & 4.2 to 4.6 & 10.2 to 10.7 \\
MI250X, 1 GCD & 1.6 & 4.4 to 4.7 & 10.7 to 11.5 \\
MI250X, card & 3.2 & 2.2 to 2.6 & 5.5 to 6.2 \\
MI300A & 5.3 & 1.6 to 1.8 & 3.3 to 3.6 \\
\bottomrule
\end{tabular}

\vspace{2pt}
\caption{RCT on one GPU beyond $10^6$ DOF, $p=5$. Bandwidth: theoretical peak (see Table~\ref{tab:gpu-hw}). MI250X card: two GCD, two MPI ranks, RCT computed with $N_\mathrm{cu}=1$.}
\label{tab:rct-gpu}
\end{table}

\subsection{Weak and strong scaling on CPU}\label{sec:weakstrong-cpu}
\begin{figure}[htbp]
  \centering
  \begin{subfigure}[b]{0.48\textwidth}\centering
\includegraphics[width=\linewidth]{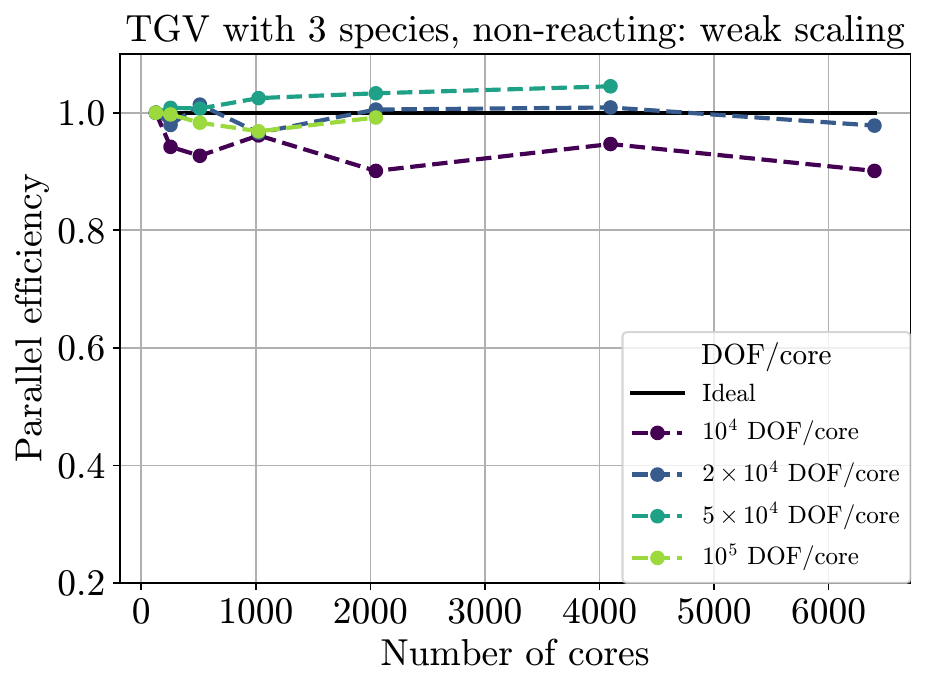}
\end{subfigure}
  \begin{subfigure}[b]{0.48\textwidth}\centering
\includegraphics[width=\linewidth]{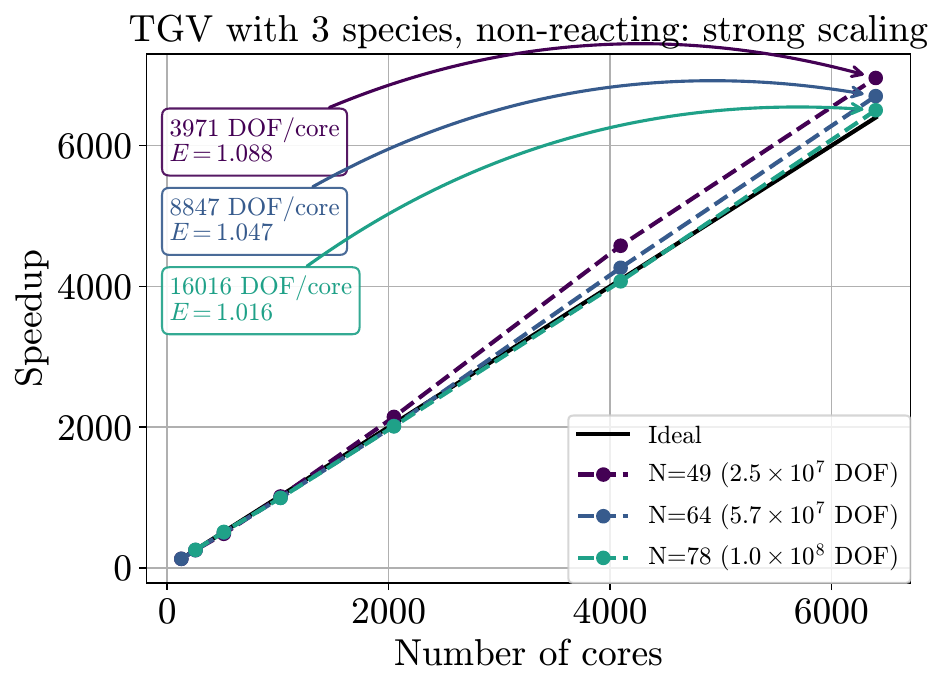}
\end{subfigure}
\caption{Non-reacting case on AMD Rome nodes, $p=5$: weak scaling efficiency with respect to 128 cores (left) and strong scaling speedup with respect to 128 cores, or to the smallest core count of the mesh when it does not fit on one node, 256 cores for $N=78$ (right).}
  \label{fig:weak_and_strong_scaling_nonreacting}
\end{figure}

\begin{figure}[htbp]
  \centering
  \begin{subfigure}[b]{0.48\textwidth}\centering
\includegraphics[width=\linewidth]{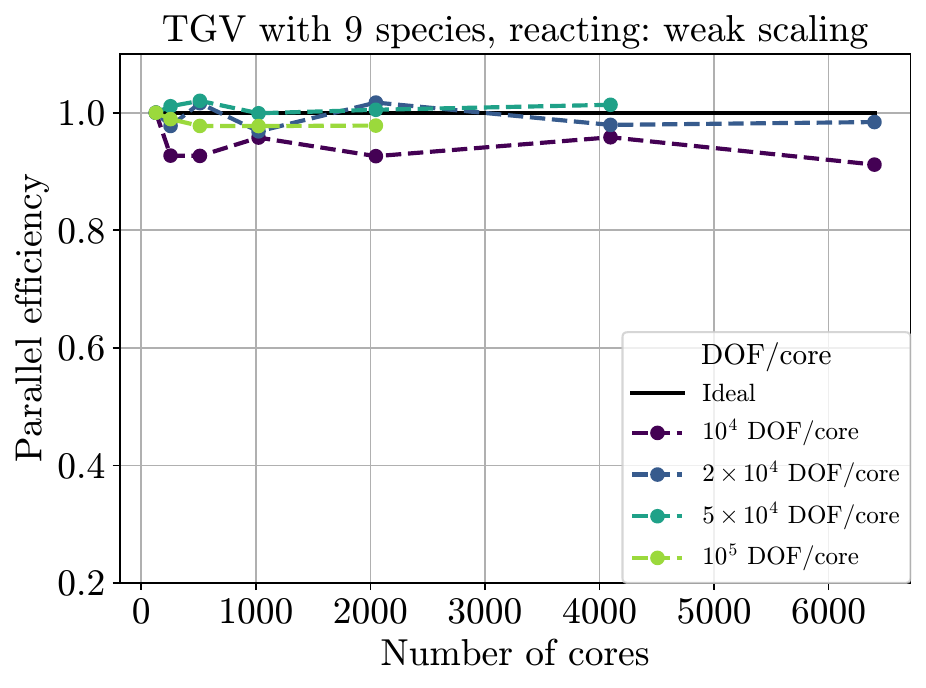}
\end{subfigure}
  \begin{subfigure}[b]{0.48\textwidth}\centering
\includegraphics[width=\linewidth]{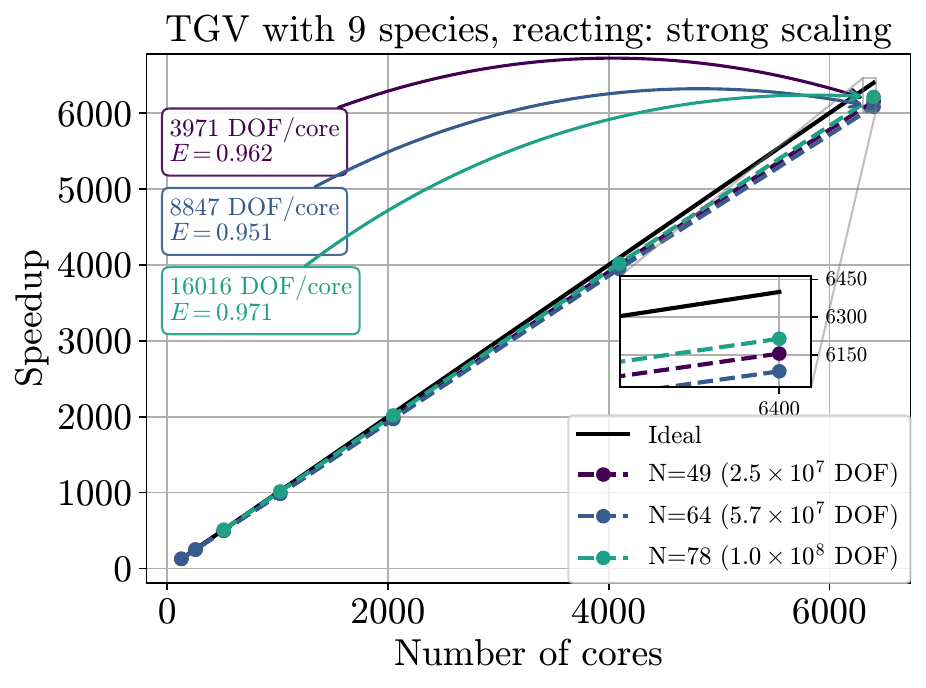}
\end{subfigure}
\caption{Reacting case on AMD Rome nodes, $p=5$: weak scaling efficiency with respect to 128 cores (left) and strong scaling speedup with respect to 128 cores, or to the smallest core count of the mesh when it does not fit on one node, 512 cores for $N=78$ (right).}
  \label{fig:weak_and_strong_scaling_reacting}
\end{figure}

Figures~\ref{fig:weak_and_strong_scaling_nonreacting} and
\ref{fig:weak_and_strong_scaling_reacting} give the weak scaling efficiency and
the strong scaling speedup on the AMD Rome scaling nodes (see Table~\ref{tab:cpu-hw}), up to 6400 cores. In weak scaling,
the efficiency stays between 0.97 and 1.05 at $5\times10^4$ and $10^5$ DOF per
core in both cases, up to the largest core count at which the mesh fits in
memory (4096 and 2048 cores). It decreases at lower densities, down to 0.90
(non-reacting) and 0.91 (reacting) at 6400 cores for $10^4$ DOF per core.
The weak scaling efficiency is computed with respect to one node (128 cores).
In strong scaling, the efficiency is also computed with respect to one node,
or to the smallest number of nodes that holds the mesh in memory when
one node is not enough: 2 and 4 nodes (256 and 512 cores) for $N=78$ in the
non-reacting and reacting cases. Within a node, the
RCT increases with the number of active cores
(see Section~\ref{sec:singlenodecpuperf} and \cite{HOPPS}): for $N=49$, it increases by 7\,\% in
the non-reacting case and by 9\,\% in the reacting case from 32 to 128 cores. A
reference below one node would attribute this intra-node effect to the parallel
scaling. For the largest problem ($N=78$, $1.0\times10^8$ DOF), the efficiency
at 6400 cores ($16\,016$ DOF per core) is 1.016 in the non-reacting case and
0.971 in the reacting case.

On CPU, the cubes are written by blocks rather than in the Morton order: the
domain is split by recursive bisection into one box per node, then one box per
core, and the elements are stored box after box, so that the contiguous range
of elements given to each rank (see Section~\ref{sec:mpi}) is exactly one box. With
the Morton order, whose blocks are irregular when $N$ is not a power of two
(see Section~\ref{sec:weakstrong-gpu}), the RCT of $N=49$ on 256 cores is 14\,\%
higher in the reacting case.

In the non-reacting case, the efficiency exceeds one at the largest core
counts: at 6400 cores, it is 1.09 for $N=49$, 1.05 for $N=64$ and 1.02 for
$N=78$. This superlinear
speedup comes from the load per core, which falls to 3971, 8847 and
$16\,016$ DOF at 6400 cores. On a full Rome node, the RCT decreases at such loads, because a growing
part of the data fits in the cache (Fig.~\ref{fig:rct_cpu_zen2}): in the
non-reacting case, it is 12\,\% lower at about $10^4$ DOF per core and 19\,\%
lower at about $5\times10^3$ DOF per core than beyond $10^5$ DOF per core. For
$N=49$, the RCT at 6400 cores is accordingly 8\,\% below its value on one node.
The same cache effect gives a performance peak at intermediate loads in the
first HOPPS article \cite{HOPPS}, and an optimum between 3000 and 10\,000 DOF
per rank for FLEXI \cite{FLEXI,GALAEXI}. On the reacting TGV, the low-Mach
solvers DINO and Nek5000 also show a superlinear strong scaling, with an optimum
at about 15\,000 points per core, which the authors relate, among other
possible causes, to a better use of the memory and caches at lower loads
\cite{AbdelsamieTGVReacting}.
The reacting case, whose cost depends less on the memory bandwidth
(see Section~\ref{sec:singlenodecpuperf}), benefits less from this effect: the RCT of
a full node decreases by only 8\,\% at about $5\times10^3$ DOF per core, and
its strong scaling efficiency stays between 0.95 and 0.99 for the three
problems up to 6400 cores.

\subsection{Weak and strong scaling on GPU}\label{sec:weakstrong-gpu}
\begin{figure}[htbp]
  \centering
  \begin{subfigure}[b]{0.48\textwidth}\centering
\includegraphics[width=\linewidth]{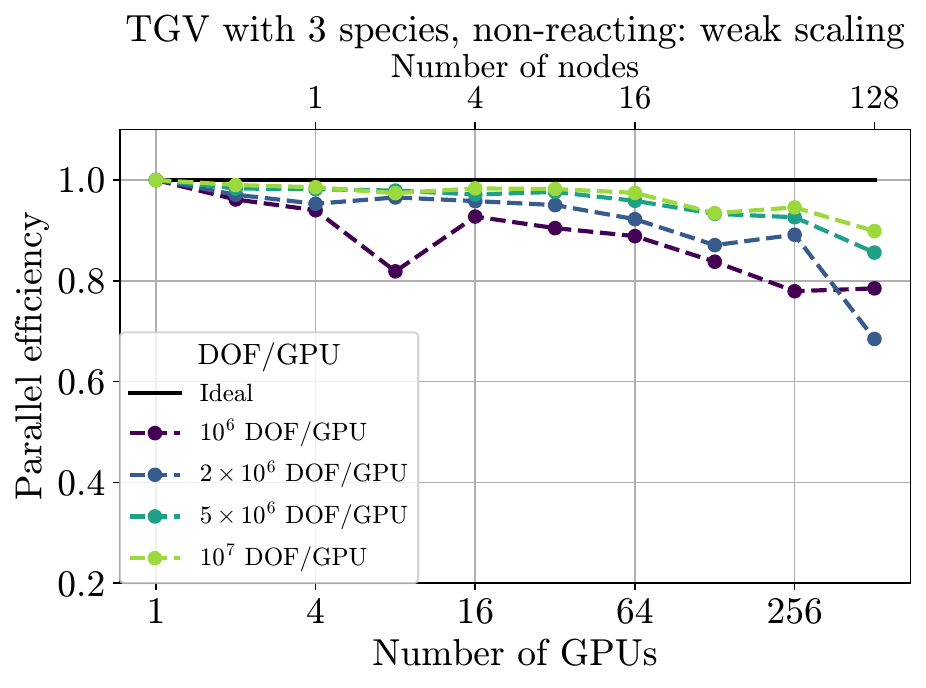}
\end{subfigure}
  \begin{subfigure}[b]{0.48\textwidth}\centering
\includegraphics[width=\linewidth]{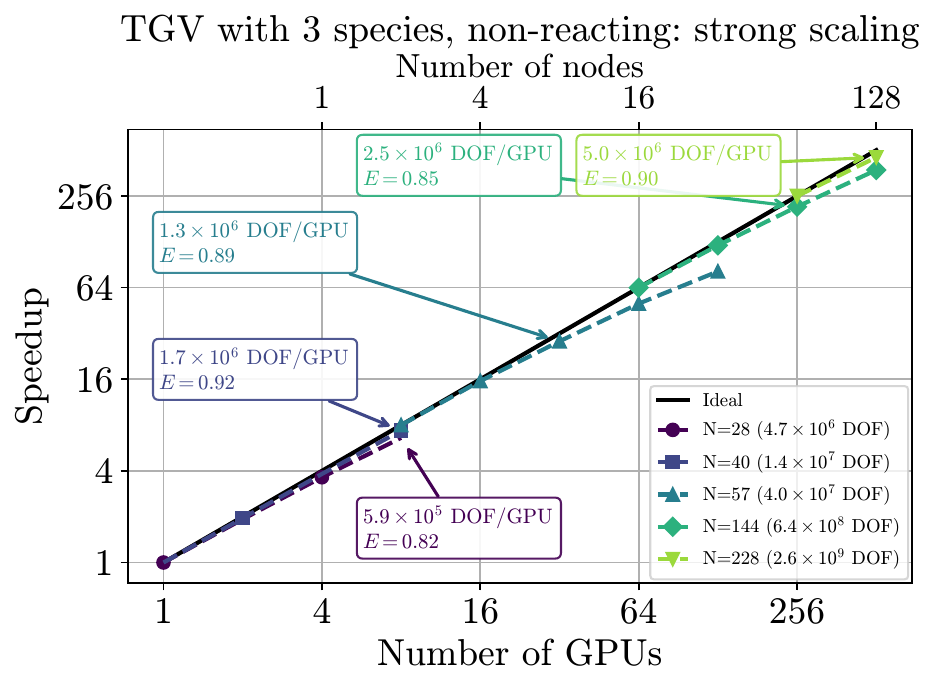}
\end{subfigure}

  \begin{subfigure}[b]{0.48\textwidth}\centering
\includegraphics[width=\linewidth]{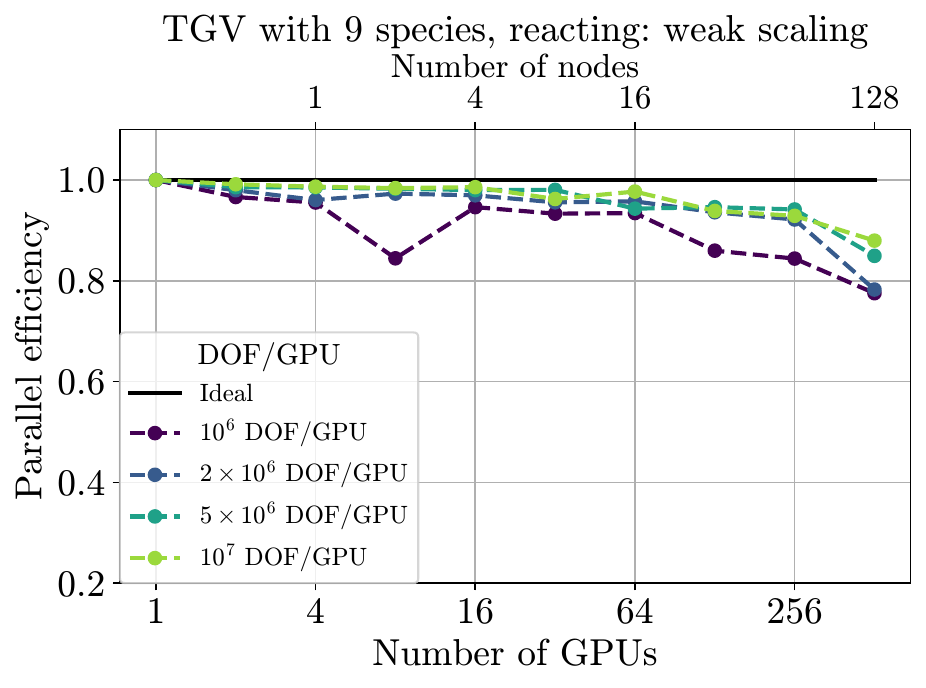}
\end{subfigure}
  \begin{subfigure}[b]{0.48\textwidth}\centering
\includegraphics[width=\linewidth]{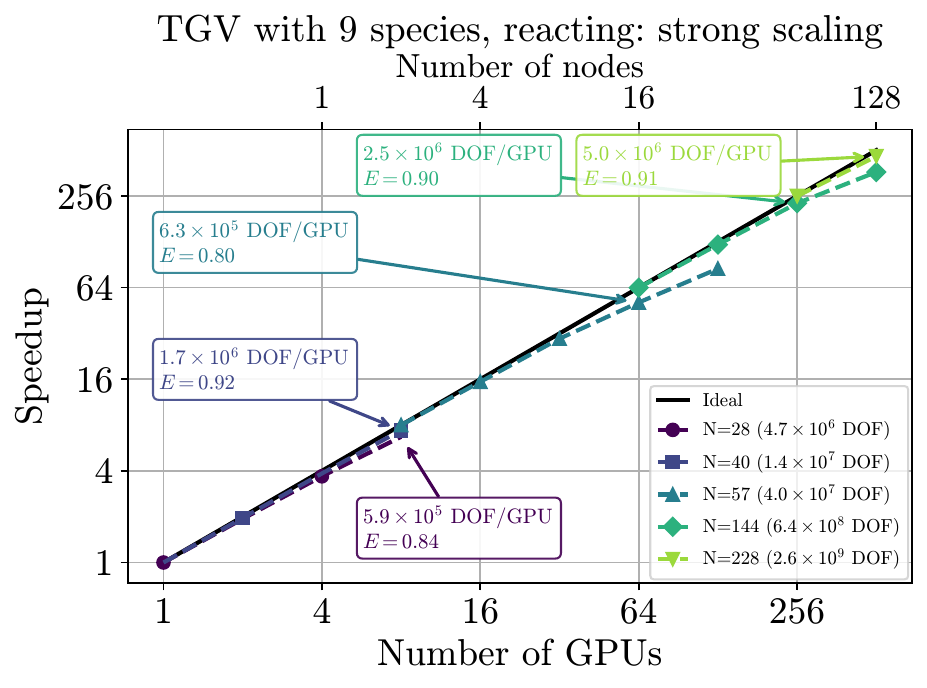}
\end{subfigure}
\caption{Nvidia H100 GPUs, $p=5$: weak scaling efficiency with respect to one GPU (left) and strong scaling speedup with respect to the smallest GPU count of each mesh (right). Top: non-reacting case. Bottom: reacting case.}
  \label{fig:scaling_h100}
\end{figure}

\begin{figure}[htbp]
  \centering
  \begin{subfigure}[b]{0.48\textwidth}\centering
\includegraphics[width=\linewidth]{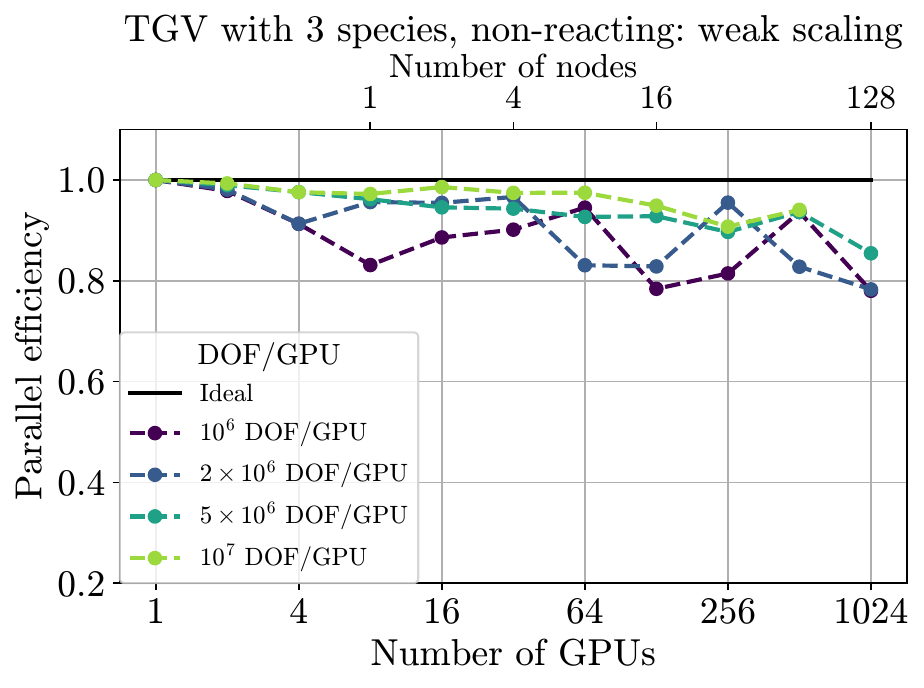}
\end{subfigure}
  \begin{subfigure}[b]{0.48\textwidth}\centering
\includegraphics[width=\linewidth]{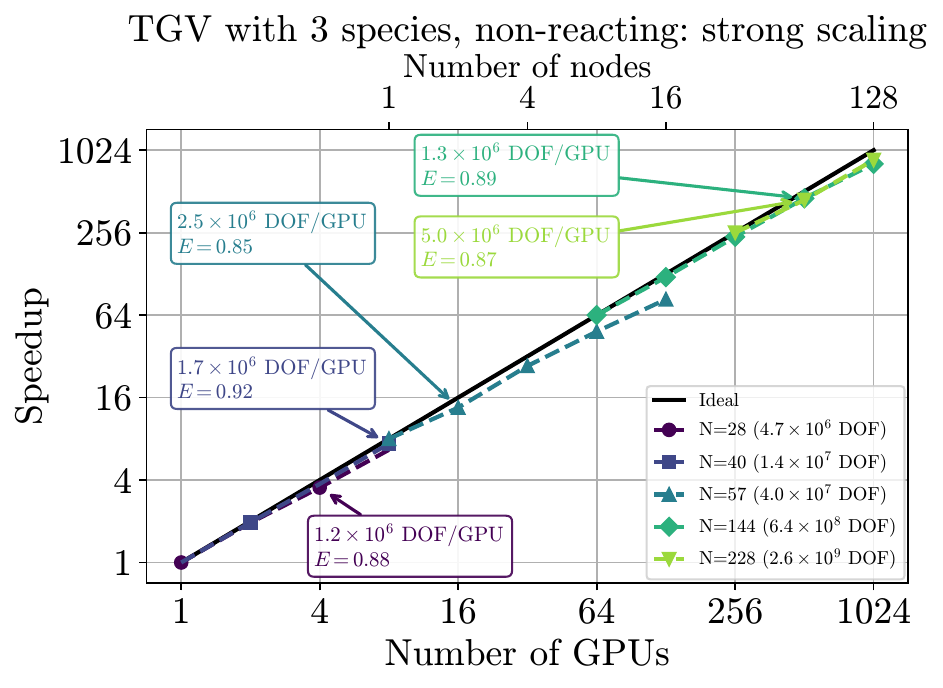}
\end{subfigure}

  \begin{subfigure}[b]{0.48\textwidth}\centering
\includegraphics[width=\linewidth]{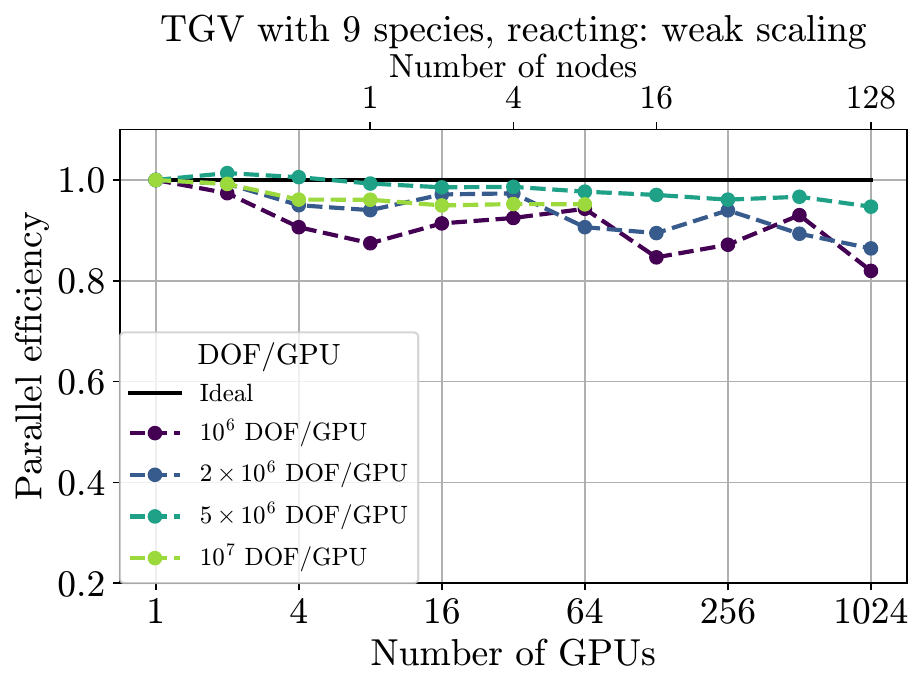}
\end{subfigure}
  \begin{subfigure}[b]{0.48\textwidth}\centering
\includegraphics[width=\linewidth]{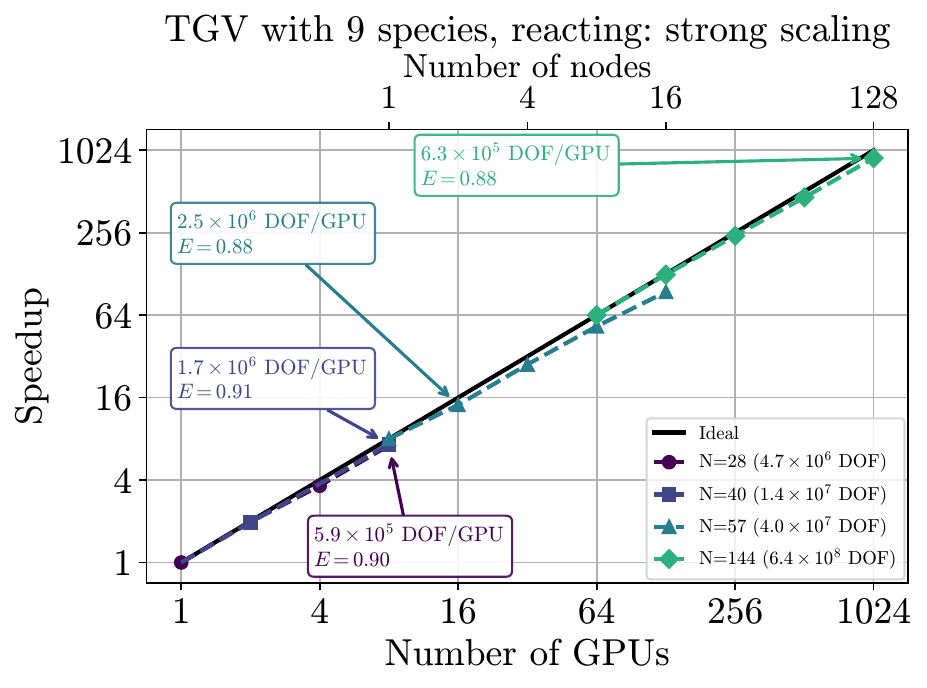}
\end{subfigure}
\caption{AMD MI250X GPUs, one GCD per MPI rank, $p=5$: weak scaling efficiency with respect to one GCD (left) and strong scaling speedup with respect to the smallest GCD count of each mesh (right). Top: non-reacting case. Bottom: reacting case.}
  \label{fig:scaling_mi250x}
\end{figure}

\begin{figure}[htbp]
  \centering
  \begin{subfigure}[b]{0.48\textwidth}\centering
\includegraphics[width=\linewidth]{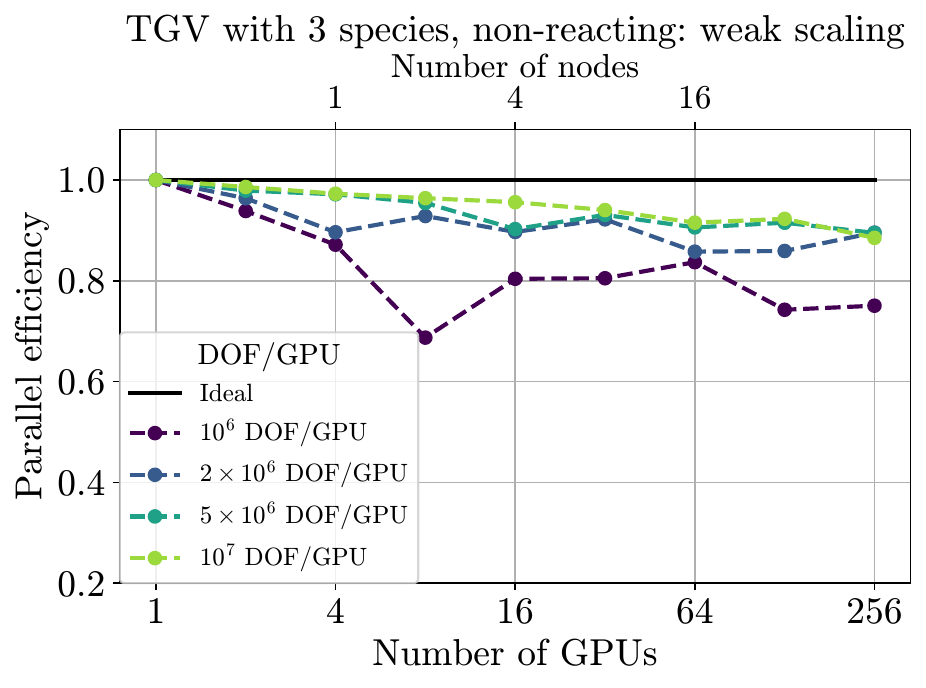}
\end{subfigure}
  \begin{subfigure}[b]{0.48\textwidth}\centering
\includegraphics[width=\linewidth]{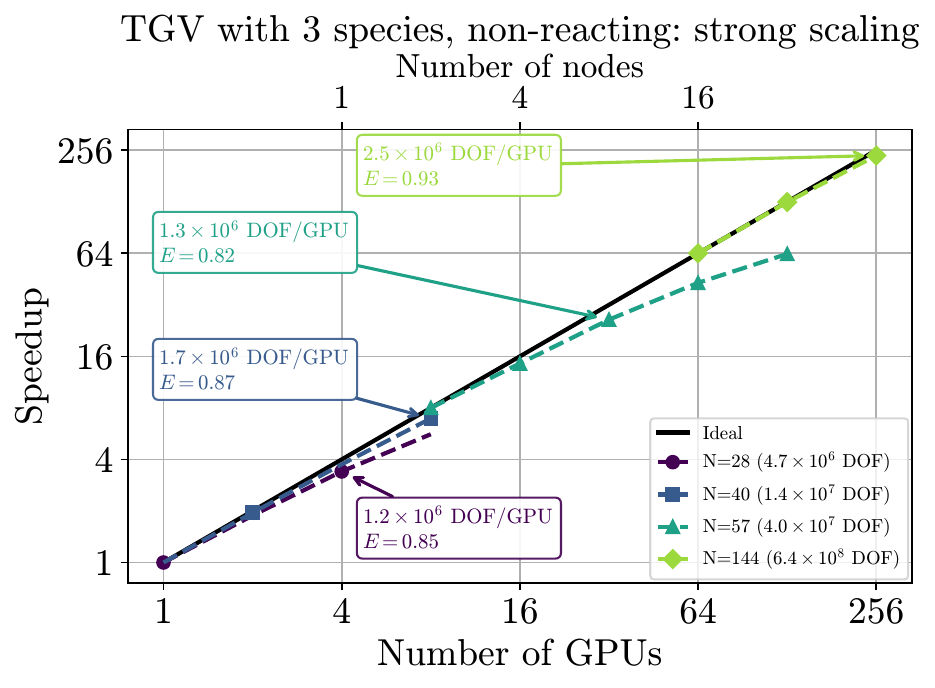}
\end{subfigure}

  \begin{subfigure}[b]{0.48\textwidth}\centering
\includegraphics[width=\linewidth]{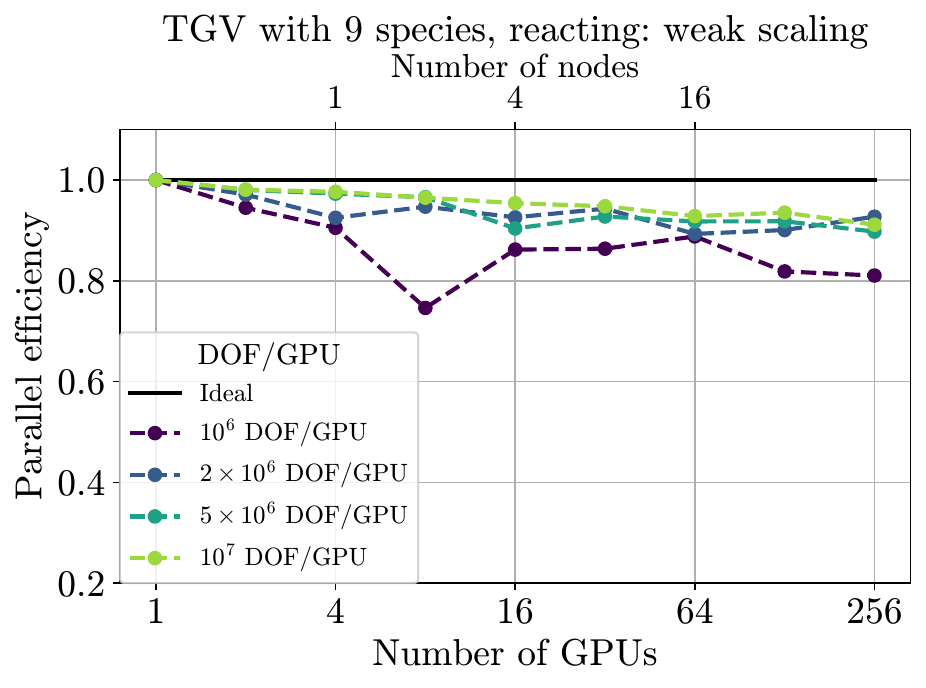}
\end{subfigure}
  \begin{subfigure}[b]{0.48\textwidth}\centering
\includegraphics[width=\linewidth]{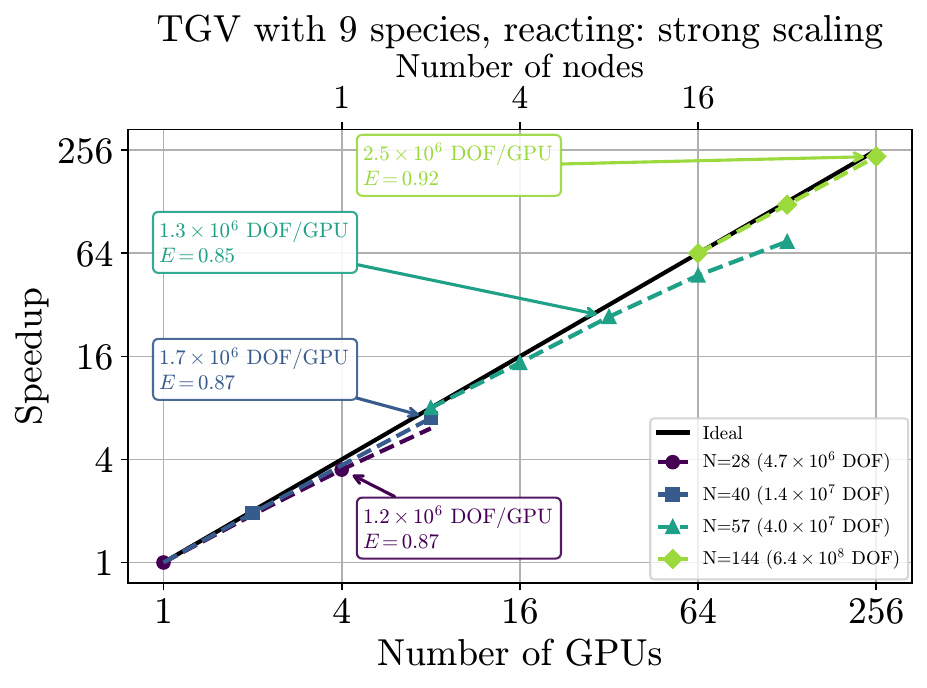}
\end{subfigure}
\caption{AMD MI300A APUs, $p=5$: weak scaling efficiency with respect to one APU (left) and strong scaling speedup with respect to the smallest APU count of each mesh (right). Top: non-reacting case. Bottom: reacting case.}
  \label{fig:scaling_mi300a}
\end{figure}

The scaling studies on GPU use three machines (see Table~\ref{tab:gpu-hw}): the H100
nodes of Jean Zay, up to 512 GPUs (128 nodes), and the MI250X and MI300A nodes
of Adastra, up to 1024 GCD (128 nodes) and 256 APUs (64 nodes). On the three
machines, each GPU, or each MI250X card, has its own network adapter.
Figures~\ref{fig:scaling_h100} to \ref{fig:scaling_mi300a} give the weak scaling
efficiency for four loads per GPU, $10^6$, $2\times10^6$, $5\times10^6$ and
$10^7$ DOF, and the strong scaling speedup for five meshes, from $N=28$
($4.7\times10^6$ DOF) to $N=228$ ($2.6\times10^9$ DOF), the last one on H100
and MI250X only. On MI250X, the reacting case at $10^7$ DOF per GCD stops at 64
GCD: beyond, the creation of the reacting solver exceeds the host memory of the
node (256\,GB for eight GCD). The weak scaling takes one GPU as
reference, so that the efficiencies include the exchanges within a node. Within
a weak scaling family, the load per GPU varies slightly from one run to the
next, since $N$ is an integer, and the weak scaling efficiency is computed from
the RCT, $E=\mathcal{R}(1)/\mathcal{R}(N_\mathrm{cu})$. The load per GPU differs
from its target by at most 12\,\% at $10^6$ and $2\times10^6$ DOF per GPU, 5\,\% at
$5\times10^6$ and 2\,\% at $10^7$. From the single-GPU RCT curves
(Fig.~\ref{fig:rct_gpu}), this variation changes the efficiency by at most
2\,\% at $10^6$ DOF per GPU, 1\,\% at $2\times10^6$ and 0.3\,\% at the two
higher loads. In strong scaling, each
mesh starts from the smallest GPU count of its series: one GPU for $N=28$ and
40, 8 for $N=57$, 64 for $N=144$, and 256 for $N=228$ (512 in the reacting case
on MI250X).

The cubes are written in the order of a Morton space-filling curve, so that the
contiguous ranges of elements given to the ranks (see Section~\ref{sec:mpi}) form
compact blocks. In the natural, lexicographic order, each rank would receive a
slice of the domain, and the number of faces exchanged per element would grow
with the number of ranks: on MI250X, the weak scaling efficiency at
$5\times10^6$ DOF per GCD then falls to 0.85 at 64 GCD in the non-reacting case,
against 0.93 with the Morton order. When $N$ is a power of two, the Morton
blocks are cubes or boxes. Otherwise, the curve is cut by the boundary of the
domain and some blocks become irregular, with more faces to exchange. This
gives the dips of the weak efficiency at $N=33$, 84 and 105, compared with
$N=42$, 64 and 128 at the neighboring GPU counts. They are most visible at
$10^6$ DOF per GPU, where the exchanges weigh most, and on MI250X.

At $5\times10^6$ DOF per GPU, the weak scaling efficiency remains between 0.93
and 0.98 up to 256 H100 GPUs, in both cases, and falls to 0.86 (non-reacting)
and 0.85 (reacting) at 512 GPUs. On MI250X, it remains at or above 0.90 up to
512 GCD, and 0.85 (non-reacting) and 0.95 (reacting) at 1024 GCD. At $10^7$ DOF
per GPU, it is 0.90 (non-reacting) and 0.88 (reacting) at 512 H100 GPUs, 0.94
at 512 GCD in the non-reacting case, and 0.95 at 64 GCD in the reacting case.
At $10^6$ DOF per GPU, it is 0.78 at 512 H100 GPUs in both cases, and 0.78
(non-reacting) and 0.82 (reacting) at 1024 GCD. On MI300A, at $5\times10^6$ DOF
per APU, it remains at or above 0.90 up to 256 APUs in both cases. At $10^7$ DOF
per APU, it is 0.89 (non-reacting) and 0.91 (reacting) at 256 APUs, and at
$10^6$ DOF per APU, 0.75 and 0.81.

In strong scaling, for $N=57$ ($4.0\times10^7$ DOF), the efficiency at 64 GPUs,
that is, $6.3\times10^5$ DOF per GPU, is 0.79 (non-reacting) and 0.80
(reacting) on H100, 0.76 and 0.83 on MI250X, and 0.67 and 0.74 on MI300A. For
$N=144$ ($6.4\times10^8$ DOF), from 64 to 512 GPUs, it is 0.74 and 0.72 on H100,
and 0.89 and 0.91 on MI250X, which still reaches 0.80 and 0.88 at 1024 GCD. From
64 to 256 GPUs, it is 0.85 and 0.90 on H100 (mean of three runs), 0.93 and 0.95 on MI250X, and 0.93
and 0.92 on MI300A. For
$N=228$, from 256 to 512 GPUs, it is 0.90 and 0.91 on H100, and 0.87 on MI250X
in the non-reacting case. On MI250X and MI300A, the reacting case keeps higher
efficiencies, consistent with more computation per exchanged value, as on CPU.
On H100, the two cases are close, without a systematic order.

The load a GPU needs follows from the ratio between its computing speed and the
bandwidth of the network. The parallel efficiency compares the time per DOF on
$N_\mathrm{cu}$ GPUs with that on one GPU, and it falls as soon as the time of
the exchanges is no longer small compared with that of the computation. A GPU
computes much faster than a CPU core, while the data to exchange per DOF are the
same: it therefore needs a much larger load for the same efficiency. The MI300A, as
fast as an H100 (see Section~\ref{sec:singlegpuperf}) with half its network
bandwidth per GPU, has the lowest strong scaling efficiency at small loads
($N=57$ at 64 GPUs). One GCD of MI250X
computes 2.7 to 3.3 times slower than an H100, and eight GCD share a node, so that a
larger part of the exchanges stays within the node: it keeps the highest strong
scaling efficiency at 512 GPUs. NekRS reaches its strong scaling limit, defined
as 80\,\% efficiency, at 2000 to 4000 points per MPI rank on a BlueGene/Q
system, against 2 to 4 million points per V100 GPU on Summit \cite{NekRS}. For
high-order benchmark problems, Fischer et al.\ find about $10^7$ points per
node at 80\,\% efficiency on Summit, against $10^5$ on BlueGene/Q, and they
compare the platforms through the time per step reached at this efficiency
rather than through the efficiency itself \cite{Fischer2020}. For the same
reason, faster kernels on one GPU lower the efficiency at a given load, while
they shorten the time to solution.

The load thresholds reported for other high-order solvers thus depend on the
network of their machines as much as on the codes. GALÆXI shows almost perfect
weak and strong scaling above $10^6$ DOF per GPU, up to 1024 GPUs, on JUWELS
Booster, where each A100 GPU has its own 200\,Gb/s InfiniBand adapter, and it
hides the communications with parallel GPU streams \cite{GALAEXI}. HORSES3D
requires at least 16\,000 elements per GPU at $p=5$, that is, about
$3.5\times10^6$ DOF, on the H100 GPUs of MareNostrum~5, which have four
200\,Gb/s InfiniBand adapters per node of four GPUs \cite{HORSES3DGPU}. PyFR
shows a similar loss in strong scaling on the GH200 GPUs of Alps, where its
efficiency drops from 0.93 at 512 GPUs to 0.47 at 2048 GPUs on a fixed mesh. It
reaches about the same efficiency at 2048 ranks on the MI250X GPUs of Frontier,
although a GH200 is about 3.7 times faster than one GCD of an MI250X, and
transfers 3.7 times more data over the network \cite{PyFR}: a faster GPU does
not lose more efficiency when the network keeps up with it. With one network
adapter per GPU, HOPPS reaches weak scaling efficiencies above 0.9 at
$5\times10^6$ DOF per GPU up to 256 GPUs on H100 and 512 GCD on MI250X, in line
with these solvers. The first HOPPS article found the communications within an
H100 node nearly free and a jump of the RCT beyond one node \cite{HOPPS}. With
the kernels of Section~\ref{subsec:intragpu} and the Morton order, the weak scaling
efficiency at $5\times10^6$ DOF per GPU is still 0.98 at 8 H100 GPUs, on two
nodes. On nodes with a single network adapter for four GPUs, the exchanges
between nodes limit the efficiency from the second node on
(\ref{app:a30}).

\section{Conclusion}\label{sec:conclusion}

This article extends the SD solver HOPPS to compressible multi-species reacting
flows. The physical models are selected at compile time, so that the compute
kernels contain no runtime branch on the configuration, and the shared-memory
parallelism is written once with Kokkos and compiled for CPUs and GPUs. An
auxiliary variable built at the FP from the interpolated primitive
variables performs the inversion of
the tabulated thermochemistry once per SP and removes it from the
interface operations. On a three-dimensional TGV
carrying hydrogen, in a non-reacting three-species and a reacting nine-species
configuration, HOPPS on CPU, HOPPS on GPU and the Fortran solver JAGUAR give
superimposed velocity, temperature, species and heat release rate profiles, and the
same history of the maximum temperature, including its non-monotonic evolution
in the reacting case. This cross-verification shows that the multi-species
extension reproduces JAGUAR, and that the CPU and GPU back-ends give the same
results on an unsteady, three-dimensional reacting case.

On AMD Genoa nodes, HOPPS is 1.7 to 2.8 times faster than JAGUAR over four core
counts and four polynomial degrees.
On AMD Rome nodes, the strong scaling efficiency at 6400 cores is 1.016 and 0.971
for the non-reacting and reacting cases. On GPU, kernels organized around the
elements, with the gradients kept in shared memory, make the same source
efficient on Nvidia and AMD devices: one MI300A reaches the RCT of one H100, and
one MI250X card is faster than one A100. At $5\times10^6$ DOF per GPU, the weak
scaling efficiency is 0.86 and 0.85 at 512 H100 GPUs, for the non-reacting and
reacting cases, 0.85 and 0.95 at 1024 GCD of MI250X, and 0.90 in both cases at
256 MI300A APUs, and it stays at or above 0.90 up to 256 H100 GPUs and 512 GCD.
A Morton ordering of the mesh keeps the
subdomains compact, and the efficiency depends on the network bandwidth
available per GPU: on nodes with one adapter for four A30 GPUs, it is only 0.77
and 0.81 at 24 GPUs for the same load (\ref{app:a30}). On both architectures, the
reacting case costs two to 2.5 times as much per DOF as the non-reacting one,
with nearly the same or a higher parallel efficiency.
The kernel profiles show
that this extra cost comes mostly from the transport of more variables on GPU,
whereas the evaluation of the reaction rates remains a large part of it on a
CPU core.

The performance study relies on a periodic Cartesian benchmark. The next steps
follow from this scope. Wall-bounded and complex reacting configurations, with curved
elements and the load imbalance of realistic geometries, can build on the
adaptive refinement of t8code. Finally, stiffer mechanisms with more species
will reduce the chemical time step of the explicit integration, and call for an
implicit or split treatment of the chemical source terms.

\section*{CRediT authorship contribution statement}
\textbf{Romaric Simo Tamou:} Conceptualization, Methodology, Software, Validation, Formal analysis, Investigation, Writing -- original draft, Writing -- review \& editing, Visualization.
\textbf{Thomas Marchal:} Conceptualization, Methodology, Software, Validation, Formal analysis, Investigation, Writing -- review \& editing, Supervision, Project administration.
\textbf{Carlos Montilla:} Conceptualization, Methodology, Software, Validation, Formal analysis, Investigation, Writing -- review \& editing, Supervision, Project administration.
\textbf{Jean-Fran\c{c}ois Boussuge:} Conceptualization, Methodology, Software, Validation, Formal analysis, Investigation, Writing -- review \& editing, Supervision, Project administration.

\section*{Declaration of competing interest}
The authors declare that they have no known competing financial interests or
personal relationships that could have appeared to influence the work reported
in this paper.

\section*{Funding}
This research did not receive any specific grant from funding agencies in the
public, commercial, or not-for-profit sectors.

\section*{Data availability}
HOPPS is developed at CERFACS and its source code is not publicly available.
Access to the code can be granted for research purposes under a collaboration
agreement with CERFACS. The input files of the TGV cases, the
scripts used to produce the performance figures and the data that support the
findings of this study are available from the corresponding author upon
reasonable request.

\section*{Declaration of generative AI and AI-assisted technologies in the manuscript preparation process}
During the preparation of this work the authors used Claude (Anthropic) in order to edit the language of the manuscript and to write the scripts that
produce some of the figures. After using
this tool, the authors reviewed and edited the content as needed and take full
responsibility for the content of the published article.

\section*{Acknowledgments}
This work was granted access to the HPC resources of CINES (Adastra) under the
allocation AD012B18459 made by GENCI, and to the HPC resources of IDRIS (Jean
Zay) under the same allocation. The authors also used the Kraken and Calypso clusters of
CERFACS, and the Belenos supercomputer of Météo-France.

\appendix
\section{Initial profiles of the Taylor--Green vortex}
\label{app:tgv-ic}

\paragraph{Velocity and pressure} With $L = 2\pi L_0$ the domain size, $L_0 = 1\,\mathrm{mm}$, and $u_0 = 4\,\mathrm{m/s}$ the reference velocity, the velocity field is
\begin{subequations}
\begin{align}
        u(x,y,z) &= u_0 \sin\!\Big(\frac{x}{L_0}\Big)\cos\!\Big(\frac{y}{L_0}\Big)\cos\!\Big(\frac{z}{L_0}\Big),\\
        v(x,y,z) &= -u_0 \cos\!\Big(\frac{x}{L_0}\Big)\sin\!\Big(\frac{y}{L_0}\Big)\cos\!\Big(\frac{z}{L_0}\Big),\\
        w &= 0,
\end{align}
\end{subequations}
giving a reference time $\tau_\mathrm{ref} = L_0/u_0 = 0.25\,\mathrm{ms}$. The pressure field, consistent with the compressible extension of the TGV analytical solution, is
\begin{equation}
P(x,y,z) = P_0 + \frac{\rho_0 u_0^2}{16}\left[\cos\!\Big(\frac{2x}{L_0}\Big)+\cos\!\Big(\frac{2y}{L_0}\Big)\right]\left[\cos\!\Big(\frac{2z}{L_0}\Big)+2\right],
\qquad \rho_0 = \frac{P_0}{R\,T_0},
\end{equation}
with $P_0 = 101325\,\mathrm{Pa}$, $T_0 = 300\,\mathrm{K}$, and $R$ the local gas constant of the mixture, computed from the initial mass fractions given below. In both cases, the density follows from the ideal-gas law, $\rho = P/(R\,T)$, with the temperature $T$ given below.

\paragraph{Smoothing function} Temperature and composition are built from a single smoothed-step building block, applied to the distance to the domain mid-plane $R_d(x) = |x - 0.5L|$:
\begin{equation}
\psi(R_d; \delta, \kappa) = \frac{1}{2}\left[1+\tanh\!\Big(\kappa\,\frac{R_d-\delta}{\delta}\Big)\right],
\end{equation}
where $\delta$ sets the half-width of the transition and $\kappa$ its steepness. This block is reused, with different $\delta$ and $\kappa$, for every scalar profile below.

\paragraph{Temperature (both configurations)} With $\delta_T = 0.875\,L_0$ and $\kappa_T = 3.2$,
\begin{equation}
T(x) = \begin{cases}
T_0 + 2\,\psi(R_d;\delta_T,\kappa_T)\,\big(T_\mathrm{ad}-T_0\big), & R_d \le \delta_T,\\[2pt]
T_0 + 2\,\big[1-\psi(R_d;\delta_T,\kappa_T)\big]\,\big(T_\mathrm{ad}-T_0\big), & R_d > \delta_T,
\end{cases}
\end{equation}
with $T_\mathrm{ad} = 1945.5\,\mathrm{K}$. This profile is continuous but not differentiable at $R_d=\delta_T$, where it reaches $T_\mathrm{ad}$.

\paragraph{Non-reacting species ($\mathrm{H_2}$, $\mathrm{O_2}$, $\mathrm{N_2}$)} With $\delta_Y = \pi L_0/4$, reusing $\kappa_T$ as the transition steepness,
\begin{subequations}
\begin{align}
        Y_{\mathrm{H_2}}(x) &= Y_{\mathrm{H_2}}^0\big[1-\psi(R_d;\delta_Y,\kappa_T)\big],\\
        Y_{\mathrm{O_2}}(x) &= Y_{\mathrm{O_2}}^0\,\psi(R_d;\delta_Y,\kappa_T),\\
        Y_{\mathrm{N_2}}(x) &= 1 - Y_{\mathrm{H_2}}(x) - Y_{\mathrm{O_2}}(x),
\end{align}
\end{subequations}
with $Y_{\mathrm{H_2}}^0 = 0.0556$ and $Y_{\mathrm{O_2}}^0 = 0.233$ the fuel-side and oxidizer-side mass fractions.
 \begin{figure}[htbp]
  \centering
  \begin{subfigure}[b]{0.48\textwidth}\centering
\includegraphics[width=\linewidth]{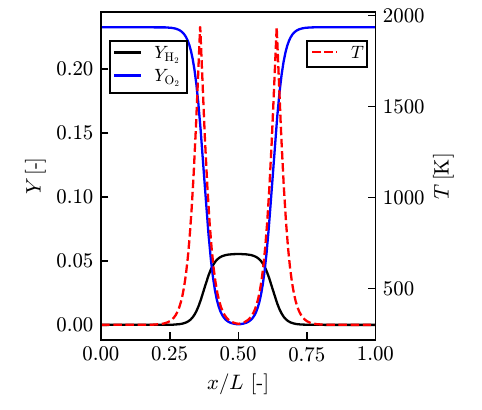}
\end{subfigure}
  \begin{subfigure}[b]{0.48\textwidth}\centering
\includegraphics[width=\linewidth]{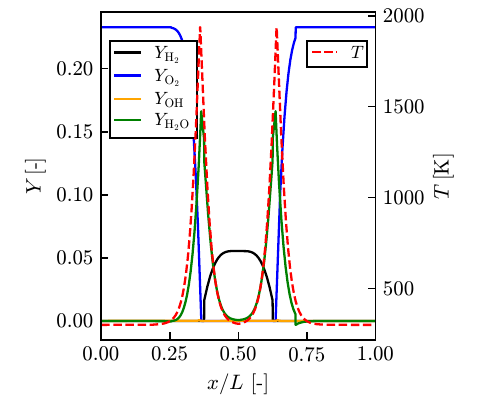}
\end{subfigure}
\caption{Initial profiles of temperature and mass fractions along the line $y = z = 0.5 L$. Left: non-reacting case. Right: reacting case.}
\label{fig:bench_nonreact_init_1D}
\end{figure}
\paragraph{Reacting species (9-species mechanism)} The temperature profile is the same as in the non-reacting case. $\mathrm{N_2}$ and $\mathrm{OH}$ use the same hyperbolic tangent form, with their own width and steepness,
\begin{subequations}
\begin{align}
        Y_{\mathrm{N_2}}(x) &= 0.9444 - 0.1773\,\psi(R_d;\,0.785\,\mathrm{mm},\,3.0),\\
        Y_{\mathrm{OH}}(x) &= \begin{cases} 2\,\psi(R_d;\,0.9\,\mathrm{mm},\,30)\;Y_{\mathrm{OH}}^\mathrm{ad}, & R_d \le 0.9\,\mathrm{mm} \\ 2\,[1-\psi(R_d;\,0.9\,\mathrm{mm},\,30)]\;Y_{\mathrm{OH}}^\mathrm{ad}, & R_d > 0.9\,\mathrm{mm} \end{cases}
\end{align}
\end{subequations}
with $Y_{\mathrm{OH}}^\mathrm{ad} = 7\times10^{-4}$. $\mathrm{H_2}$ and $\mathrm{O_2}$, whose equilibrium profiles are not simple sigmoids, are instead fitted by polynomials of $\tilde{x} = x/(1\,\mathrm{mm})$ to the reference equilibrium profile. Each polynomial is fitted on one half of the domain and mirrored about the mid-plane, $\tilde{x} \to 2\pi - \tilde{x}$, on the other half: for $\tilde{x} < \pi$ for $\mathrm{H_2}$, and for $\tilde{x} \ge \pi$ for $\mathrm{O_2}$. The polynomials are therefore evaluated at
\begin{equation}
\xi_{\mathrm{H_2}} = \begin{cases} 2\pi - \tilde{x}, & \tilde{x} < \pi,\\ \tilde{x}, & \tilde{x} \ge \pi, \end{cases}
\qquad
\xi_{\mathrm{O_2}} = \begin{cases} \tilde{x}, & \tilde{x} < \pi,\\ 2\pi - \tilde{x}, & \tilde{x} \ge \pi, \end{cases}
\end{equation}
and
\begin{subequations}
\begin{align}
        Y_{\mathrm{H_2}}(\tilde{x}) &= \begin{cases} \max\!\Big(0,\ \textstyle\sum_{i=0}^{6} a_i\,\xi_{\mathrm{H_2}}^{\,i}\Big), & R_d \le 0.785\,\mathrm{mm},\\ 0, & R_d > 0.785\,\mathrm{mm}, \end{cases}\\
        Y_{\mathrm{O_2}}(\tilde{x}) &= \begin{cases} 0, & R_d \le 0.785\,\mathrm{mm},\\ \max\!\Big(0,\ \textstyle\sum_{i=0}^{8} b_i\,\xi_{\mathrm{O_2}}^{\,i}\Big), & R_d > 0.785\,\mathrm{mm}, \end{cases}
\end{align}
\end{subequations}
with $Y_{\mathrm{O_2}}$ set to its far-field value $0.2329$ for $\tilde{x} \le 1.6$ and $\tilde{x} \ge 4.45$. The coefficients $a_i$ and $b_i$ are given in Table \ref{tab:polyCoeffTGV}. Finally, $\mathrm{H_2O}$ closes the mixture by conservation, $Y_{\mathrm{H_2O}} = 1 - Y_{\mathrm{H_2}} - Y_{\mathrm{O_2}} - Y_{\mathrm{N_2}} - Y_{\mathrm{OH}}$, and the remaining minor radicals (H, O, HO$_2$, H$_2$O$_2$) are initialized to zero.
\begin{table}[htbp]
    \centering

    \begin{tabular}{@{}ccccc@{}}
        \toprule
        $i$ & 0 & 1 & 2 & 3 \\
        \midrule
        $a_i$ ($\mathrm{H_2}$)
        & 61.8667
        & $-127.3234$
        & 107.0607
        & $-47.1849$ \\
        $b_i$ ($\mathrm{O_2}$)
        & 0.23528
        & $-0.043358$
        & 0.38877
        & $-1.3896$ \\
        \bottomrule
    \end{tabular}
    \qquad
    \begin{tabular}{@{}cccccc@{}}
        \toprule
        $i$ & 4 & 5 & 6 & 7 & 8 \\
        \midrule
        $a_i$ ($\mathrm{H_2}$)
        & 11.5188
        & $-1.4785$
        & 0.077994
        & & \\
        $b_i$ ($\mathrm{O_2}$)
        & 2.4697
        & $-2.3804$
        & 1.2579
        & $-0.33928$
        & 0.03589 \\
        \bottomrule
    \end{tabular}
\caption{Polynomial coefficients for the fitted $\mathrm{H_2}$ and $\mathrm{O_2}$ initial profiles.}\label{tab:polyCoeffTGV}
\end{table}

\FloatBarrier
\section{Weak and strong scaling on nodes with one network adapter for four GPUs}
\label{app:a30}

This appendix gives the scaling of HOPPS on Nvidia A30 GPUs, up to 24 GPUs (six
nodes of four GPUs), with the same GPU kernels as in
Section~\ref{sec:weakstrong-gpu}. The A30 nodes have a single 100\,Gb/s
Omni-Path adapter for four GPUs, without GPUDirect RDMA, and they show how a
network that does not keep up with the GPUs limits the parallel efficiency.
The cubes are written in the natural order, so that each node holds a slice of
the domain. Both studies take one node (4 GPUs) as reference, and the weak
scaling efficiency is computed from the RCT, as in
Section~\ref{sec:weakstrong-gpu}.

\begin{figure}[htbp]
  \centering
  \begin{subfigure}[b]{0.48\textwidth}\centering
\includegraphics[width=\linewidth]{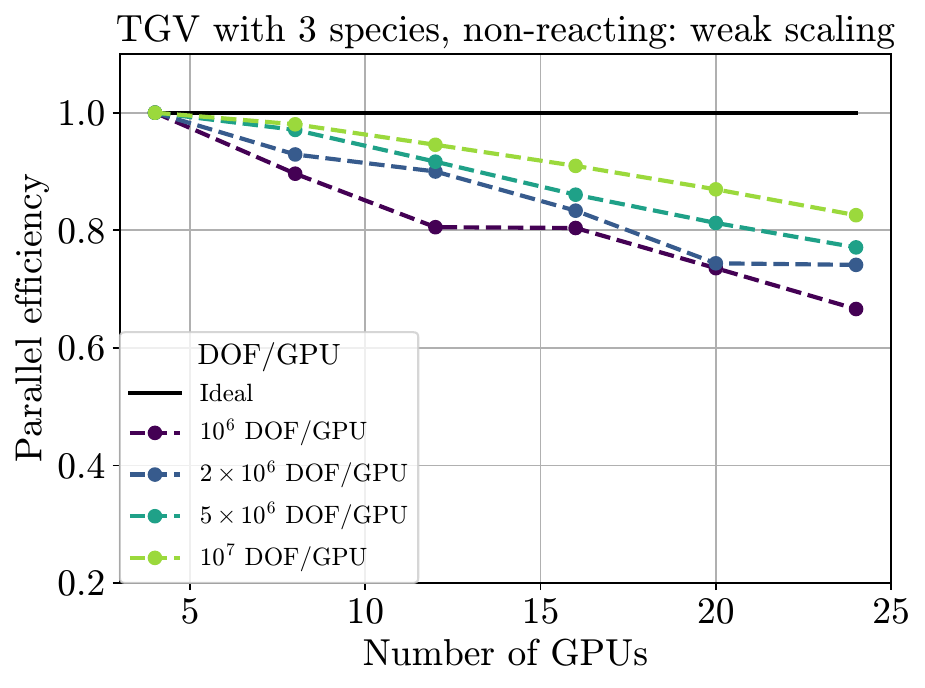}
\end{subfigure}
  \begin{subfigure}[b]{0.48\textwidth}\centering
\includegraphics[width=\linewidth]{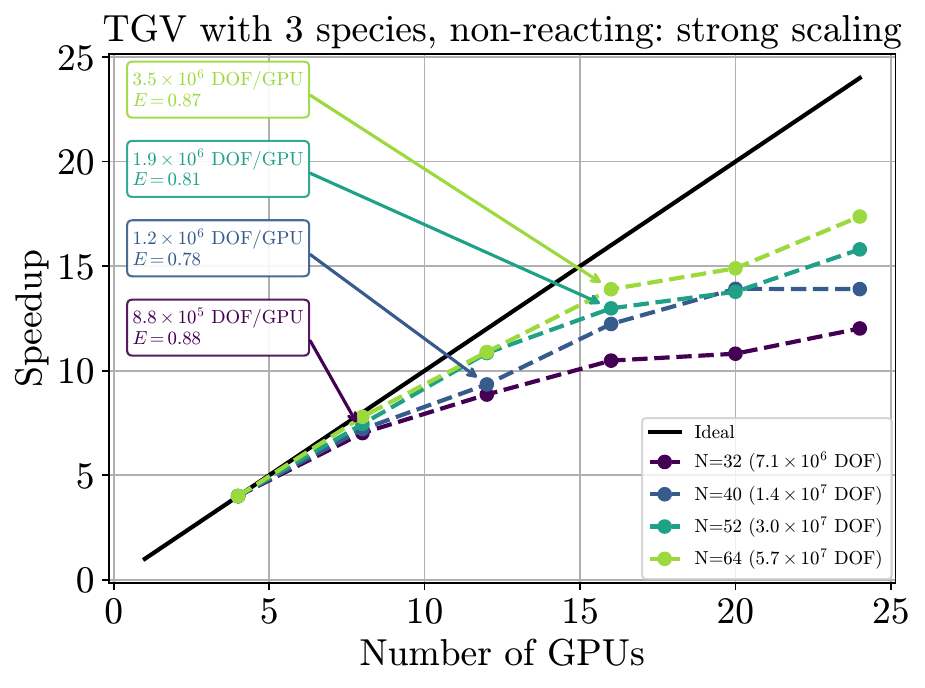}
\end{subfigure}
\caption{Non-reacting case on Nvidia A30 GPUs, $p=5$: weak scaling efficiency (left) and strong scaling speedup (right), with respect to one node (4 GPUs).}
  \label{fig:scaling_a30_nonreacting}
\end{figure}

\begin{figure}[htbp]
  \centering
  \begin{subfigure}[b]{0.48\textwidth}\centering
\includegraphics[width=\linewidth]{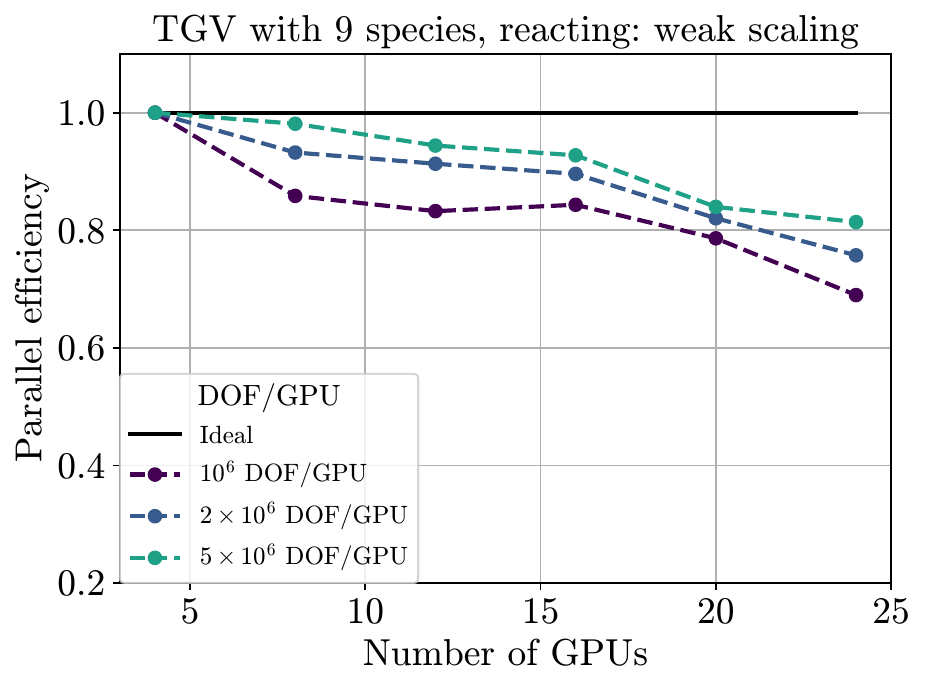}
\end{subfigure}
  \begin{subfigure}[b]{0.48\textwidth}\centering
\includegraphics[width=\linewidth]{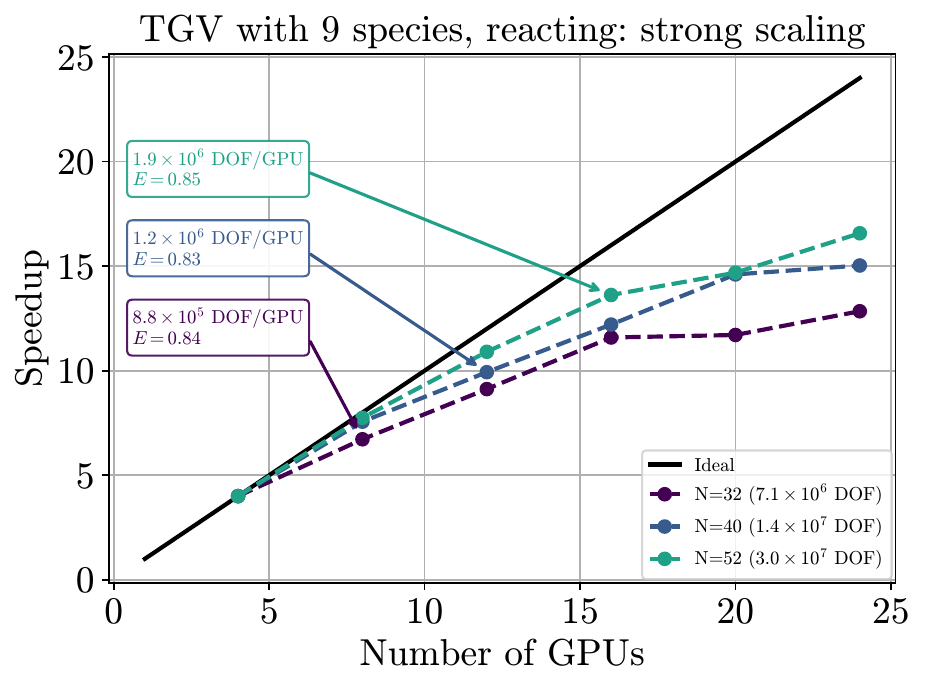}
\end{subfigure}
\caption{Reacting case on Nvidia A30 GPUs, $p=5$: weak scaling efficiency (left) and strong scaling speedup (right), with respect to one node (4 GPUs).}
  \label{fig:scaling_a30_reacting}
\end{figure}

At 24 GPUs, the weak scaling efficiency is 0.67 in the non-reacting case and
0.69 in the reacting case for about $10^6$ DOF per GPU, 0.77 and 0.81 for
$5\times10^6$ DOF per GPU, and 0.83 for $10^7$ DOF per GPU in the non-reacting
case (Figs.~\ref{fig:scaling_a30_nonreacting} and
\ref{fig:scaling_a30_reacting}). For $5\times10^6$ DOF per GPU, it stays at or
above 0.92 up to 12 GPUs. In strong scaling, for $N=32$ ($7.1\times10^6$ DOF),
the efficiency is 0.88 (non-reacting) and 0.84 (reacting) at 8 GPUs, and 0.50
and 0.54 at 24 GPUs. For $N=52$ ($3.0\times10^7$ DOF), it is 0.93 and 0.97 at 8
GPUs, and 0.66 and 0.69 at 24 GPUs. For $N=64$ ($5.7\times10^7$ DOF), in the
non-reacting case, it is 0.97 at 8 GPUs and 0.72 at 24 GPUs.

The loss of efficiency comes from the exchanges between nodes. At equal load
per GPU, the RCT on one node is within 2\,\% of the single-GPU RCT, whereas for
$N=52$ in the non-reacting case it exceeds the RCT on one node by 11\,\% at 12
GPUs and by 52\,\% at 24 GPUs. Measured between two nodes, the Omni-Path adapter
delivers about 11\,GB/s in total, that is, 2.7\,GB/s per GPU when the four GPUs
of a node send at the same time, whether the buffers are in device or host
memory. With slices of the domain, each node exchanges two planes of $N^2$
faces with its neighboring nodes, whatever the number of GPUs. At each RK
stage, the two halos of Section~\ref{sec:mpi} cross these faces, at each of
their $(p+1)^2$ FP. The first one is the auxiliary variable, $(T, u, v, w, P,
Y_1,\dots,Y_{N_s}, R, C_p, \rho E)$, that is, $N_s+8$ values (see
Section~\ref{sec:Auxvariable}). The second one is the averaged flux, with one
value per conservative variable $(\rho, \rho u, \rho v, \rho w, \rho E, \rho
Y_k)$, that is, $N_s+5$ values. Each FP thus carries $2N_s+13$ values, 19 for
three species and 31 for nine species. For $N=52$, a node thus sends 30\,MB per
stage in the non-reacting case and 48\,MB in the reacting case, which takes at
least 13.5 and 22\,ms per time step at the bandwidth of the adapter. The
overhead measured at 20 and 24 GPUs is about 19 and 35\,ms per time step: the
network bandwidth accounts for two thirds to three quarters of it, which the
overlap with the internal interfaces cannot hide. Each A30 has a nominal network
bandwidth of 25\,Gb/s, sixteen times less than an H100 of
Section~\ref{sec:weakstrong-gpu}, and HOPPS needs several $10^6$ DOF per GPU on
these nodes to keep the exchanges small compared with the computation.

\FloatBarrier
\section{Single-species performance on one GPU}
\label{app:mono}

\begin{figure}[htbp]
  \centering
\includegraphics[width=0.6\linewidth]{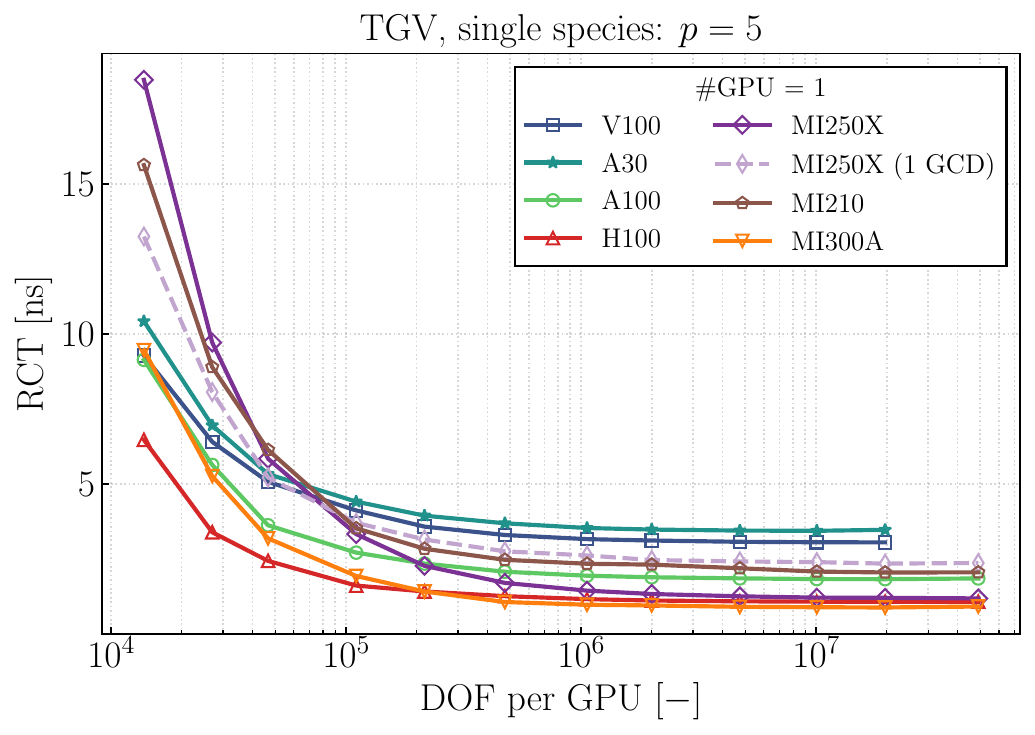}
\caption{Single-species RCT versus DOF per GPU on four Nvidia GPU generations and three AMD GPUs, $p=5$. MI250X: full card (two GCD, solid line) and one GCD (dashed line).}
  \label{fig:rct_gpu_mono}
\end{figure}

This appendix gives the RCT of the single-species solver on one GPU, with the
same GPU kernels and the same protocol as in
Section~\ref{sec:singlegpuperf}. HOPPS is built without species, and the TGV
is computed with a single ideal gas ($\gamma=1.4$) on the same meshes and with the same numerical options as the
multi-species cases. Figure~\ref{fig:rct_gpu_mono} gives
the RCT versus the DOF per GPU and Table~\ref{tab:rct-gpu-mono} its values
beyond $10^6$ DOF.

The curves have the same shape as in Fig.~\ref{fig:rct_gpu}, with a plateau
beyond about $10^6$ DOF, and the same ranking: MI300A and H100 have the lowest
RCT, followed by the MI250X card and A100. On the plateau, the single-species
RCT is 0.47 to 0.64 times the non-reacting multi-species RCT of
Table~\ref{tab:rct-gpu}. If the cost per DOF were proportional to the number
of conservative variables, five without species against eight with three
species, this ratio would be $5/8=0.625$. The measured ratio is close to or below this
value, since the single-species solver is also simpler: with a constant ratio
of specific heats, the temperature follows in closed form from the internal
energy, without the search in the tabulated energies (see
Eq.~\eqref{energyInterval}), and there are no species diffusion fluxes.

The first HOPPS article \cite{HOPPS} normalizes the RCT by time step of a
three-stage Runge--Kutta scheme: its values are therefore divided by three in
Table~\ref{tab:rct-gpu-mono} to compare costs per stage. At $10^7$ DOF, the
single-species RCT is 3.9, 3.4 and 4.1 times lower on V100, A30 and A100, 2.7
times lower on H100 and about 6 times lower on MI210. For the H100, the
comparison uses the H100 of the same machine in \cite{HOPPS}. Against the GH200
superchip of \cite{HOPPS}, whose memory bandwidth is higher, the gain is 2.4.
The first HOPPS article does not report MI250X and MI300A results.

These gains come from changes made since the first HOPPS article to parts of
the solver that the single-species and multi-species cases share. The
conservative state is converted to primitive variables once per SP, and the
temperature and the pressure are passed to the Riemann solver and to the flux
functions instead of being recomputed (see Section~\ref{sec:Auxvariable}). The
HLLC solver no longer uses matrix-vector products. The element kernels store the
fields by element and compute the gradients and the fluxes in a single kernel,
with the gradients kept in scratch memory, and only the gradients needed by the
viscous fluxes are computed (see Section~\ref{subsec:intragpu}). The interface
kernels process the FP pairs sorted by memory address and write the Riemann
fluxes through shared memory (see Section~\ref{subsec:intragpu}). The gain is
largest on the MI210, on which the storage by element matters most (see
Section~\ref{sec:singlegpuperf}).

\begin{table}[htbp]
\centering
\small
\begin{tabular}{@{}lrrr@{}}
\toprule
GPU & \multicolumn{2}{c}{$\mathcal{R}$ (ns)} & Gain at $10^7$ DOF \\
 & this work & \cite{HOPPS}, divided by 3 & \\
\midrule
V100 & 3.1 to 3.2 & 11.7 to 12.0 & 3.9 \\
A30 & 3.4 to 3.5 & 11.7 & 3.4 \\
A100 & 1.8 to 1.9 & 7.3 to 7.7 & 4.1 \\
H100 & 1.1 to 1.2 & 2.8 to 3.0 & 2.7 \\
MI210 & 2.1 to 2.3 & 10 to 12.3 & 5.9 \\
MI250X, 1 GCD & 2.3 to 2.6 & -- & -- \\
MI250X, card & 1.2 to 1.5 & -- & -- \\
MI300A & 0.9 to 1.0 & -- & -- \\
\bottomrule
\end{tabular}

\vspace{2pt}
\caption{Single-species RCT on one GPU beyond $10^6$ DOF, $p=5$, compared with the first HOPPS article \cite{HOPPS}, whose values, read from its figures, are given per Runge--Kutta stage. MI250X card: two GCD, two MPI ranks, RCT computed with $N_\mathrm{cu}=1$.}
\label{tab:rct-gpu-mono}
\end{table}

\end{document}